%% file: main.tex
\documentclass[twocolumn]{aastex631}

\usepackage{natbib}
\usepackage{amsmath}
\usepackage{blindtext, rotating}

\usepackage{graphicx}
\graphicspath{{./}{figures/}}
\usepackage{array,xcolor,colortbl}
\usepackage{multirow}
\renewcommand{\arraystretch}{1.2}
\usepackage[normalem]{ulem}

\usepackage{xcolor}
\usepackage{xspace}
\newcommand{\redmagic}{redMaGiC\xspace}
\newcommand{\redmapper}{redMaPPer\xspace}
\newcommand{\maglim}{MagLim\xspace}
\newcommand{\degree}{$^{\circ}$\xspace}

\newcommand{\dv}[1]{\mathrm{d} #1}
\newcommand{\hMpc}{{\ifmmode{\,h^{-1}{\rm Mpc}}\else{$h^{-1}$Mpc}\fi}}
\newcommand{\hkpc}{{\ifmmode{\,h^{-1}{\rm kpc}}\else{$h^{-1}$kpc}\fi}}
\newcommand{\hMsun}{{\ifmmode{\,h^{-1}{\rm {M_{\odot}}}}\else{$h^{-1}{\rm{M_{\odot}}}$}\fi}}
\newcommand{\Msun}{{\ifmmode{{\rm {\,M_{\odot}}}}\else{${\rm{M_{\odot}}}$}\fi}}

\newcommand{\Mstar}{{\ifmmode{\,M_{*}}\else{$M_{*}$}\fi}}
\newcommand{\Mhalo}{{\ifmmode{\,M_{\rm halo}}\else{$M_{\rm halo}$}\fi}}
\renewcommand{\tilde}[1]{\stackrel{\sim}{\smash{#1}\rule{0pt}{1.1ex}}}
\newcommand{\theth}{\textsc{The Three Hundred\,}}

\usepackage[absolute,overlay]{textpos}

\begin{document}

\title{\large Superclustering with the Atacama Cosmology Telescope and Dark Energy Survey \\ \normalsize II. Anisotropic large-scale  coherence in hot gas, galaxies, and dark matter}

\input{authors_numbered}

\begin{textblock*}{10cm}(7.5cm,11.5cm) 
   \small{{\em \textcolor{black}{Author affiliations are listed at the end of this paper.}}}
\end{textblock*}

\begin{abstract}

Statistics that capture the directional dependence of the baryon distribution in the cosmic web enable unique tests of cosmology and astrophysical feedback. We use \textit{constrained oriented stacking} of thermal Sunyaev-Zel'dovich (tSZ) maps to measure the anisotropic distribution of hot gas $2.5-40$~Mpc away from galaxy clusters embedded in massive filaments and superclusters. The cluster selection and orientation (at a scale of $\sim15$ Mpc) use Dark Energy Survey (DES) Year 3 data, while expanded tSZ maps from the Atacama Cosmology Telescope Data Release 6 enable a $\sim3\times$ more significant measurement of the extended gas compared to the technique's proof-of-concept. Decomposing stacks into cosine multipoles of order $m$, we detect a dipole ($m=1$) and quadrupole ($m=2$) at $8-10\sigma$, as well as evidence for $m=4$ signal at up to $6\sigma$, indicating sensitivity to late-time non-Gaussianity. We compare to the Cardinal simulations with spherical gas models pasted onto dark matter halos. The fiducial tSZ data can discriminate between two models that deplete pressure differently in low-mass halos (mimicking astrophysical feedback), preferring higher average pressure in extended structures. However, uncertainty in the amount of cosmic infrared background contamination reduces the constraining power. Additionally, we apply the technique to DES galaxy density and weak lensing to study for the first time their oriented relationships with tSZ. In the tSZ-to-lensing relation, averaged on 7.5~Mpc (transverse) scales, we observe dependence on redshift but not shape or radial distance. Thus, on large scales, the superclustering of gas pressure, galaxies, and total matter is coherent in shape and extent.

\end{abstract}


\section{Introduction}\label{Sec:Intro} 
Although baryons make up only about one-sixth of the matter in the universe \citep{Planck2018}, as tracers of the total large-scale structure (dark matter included), they are essential components for cosmological prediction and measurement. The distribution of baryons with respect to the dark cosmic web is complex: galaxies are known to align with their surrounding large-scale tidal fields \citep[known as intrinsic alignment; overviewed in][]{Hirata2004PhRvD..70f3526H, Troxel2015PhR...558....1T}, and galaxy mass and color is environmentally-dependent \citep[see, e.g.][]{Bamford2009MNRAS.393.1324B, AragonCalvo2010, Codis2018}. Moreover, the matter in galaxies represents only a small fraction ($\lesssim$10\%) of the baryonic matter in the universe, with another $\lesssim10$\% split by gas in the circumgalactic medium and intracluster medium; the remainder ($\gtrsim80\%$) is expected to be distributed in the gas of the diffuse intergalactic medium (IGM) \citep[e.g.,][]{read2005, Shull2012, Nicastro2017, Sorini2022}. The distribution of the density and temperature of the diffuse baryons in low-mass clusters, galaxy groups, halo outskirts, and intergalactic filaments is poorly understood. The general picture is of a repeating cycle of shock-heating during infall onto filaments and into halos and subsequent outflow and heating from feedback, but details remain murky \citep{McQuinn2016, Donahue2022PhR...973....1D}. As with any other tracer, tidal forces play an important role in shaping the gas in the cosmic web into its characteristic filamentary network \citep{vdWetal:2008b}, although the extent to which tidal forces are important in shaping feedback inflows and outflows in the late ($z<2$) universe is an open question.

These details of the relationships among gas, galaxies, and dark matter in their cosmic web environment are becoming increasingly important as we receive larger amounts of multi-wavelength survey data, enabling analyses that combine information from vastly different length and energy scales. A promising area is the study of cosmic microwave background (CMB) secondary effects combined with galaxy survey data. CMB surveys are sensitive to late-time scattering signatures from both concentrated and diffuse gas. In particular, the thermal Sunyaev-Zel'dovich effect \citep[tSZ,][]{SZ1970, SZ1972} is sensitive to inverse Compton scatterings of cold CMB photons off hot ionized gas along the line-of-sight. It is parametrized by Compton-$y$:
\begin{equation}\label{eq:tSZ}
    y =  \int  n_\mathrm{e} \frac{k_\mathrm{B} T_\mathrm{e}}{m_\mathrm{e} c^2} \sigma_\mathrm{T} \dv{\ell},
\end{equation}
describing a line-of-sight integral over proper distance $\ell$ of the number density of electrons $n_\mathrm{e}$ and the electron temperature $T_\mathrm{e}$, taking into account the Thomson cross section for scattering ($\sigma_\mathrm{T}$) and normalized by the electron rest mass energy $m_\mathrm{e} c^2$. Since $n_\mathrm{e} T_\mathrm{e}$ is the electron pressure $P_\mathrm{e}$, the $y$ signal is related to thermal energy per volume. Massive clusters emerge by eye in modern $y$ maps, but gas from low-mass halos and diffuse regions is challenging to observe, as the signals lie well beneath the noise \citep{Planck2016tSZ, Coulton2023arXiv230701258C}.

To overcome the signal-to-noise challenges, some details of the distribution of matter in the lower-density universe can be inferred through the technique of cross-correlations between maps of $y$ and maps of the positions of low-mass objects, such as galaxies, groups, or low-mass clusters \citep[e.g.,][]{Pandey2019, Yan2021, Sanchez2023MNRAS.522.3163S}. An essentially equivalent technique is stacking, where one takes the average of $y$ map cutouts at positions of many such objects \citep[e.g.,][]{Vikram2017, Schaan2021}. By construction, these methods capture the average tSZ signal as a function of \textit{directionless} distance from the sample of objects. However, in reality, this signal originates from a combination of concentrated and diffuse gas that clusters in an \textit{oriented} manner, in accordance with the local cosmic web surrounding each point in the sample. By incorporating large-scale orientation information to make aligned stacks of filaments and/or superclusters, we can increase the signal-to-noise of extended measurements \citep[as in][]{deGraaff2019, Tanimura2019, Kraljic2020, Lokken2022PaperI} and learn about the tidal dependence of the gas pressure and its relationships.

Because full hydrodynamic simulations are computationally expensive to run, predictions for gas signatures in large sky surveys are typically made using approximate halo models. These models can be pasted onto dark matter simulations to generate mock maps. Because gas in large (cluster- and group-hosting) halos is hot ($\sim10^6-10^8$ K), and because simple isotropic halo-centric models have been developed which can account for some of the effects of feedback on massive halos \citep{Battaglia2012b, OsatoNagai2023MNRAS.519.2069O}, the halo approach functions reasonably well for directionally-independent tSZ predictions in which the largest halos dominate \citep[e.g.,][]{Greco2015, Planck2016tSZ, Horowitz2017MNRAS.469..394H, Gong2019MNRAS.486.4904G}. This is promising for the future of cosmological inference using the tSZ effect.

However, several factors motivate us to test these models in an oriented manner. First, most galaxy clusters are themselves non-spherical, as reviewed by \citet{Limousin2013SSRv..177..155L} and further demonstrated by more recent observational studies such as \citet{Chiu2018ApJ...860..126C, gouin2020A&A...635A.195G}. Second, there is evidence in simulations and observations that cosmic web environment impacts the properties of galaxies within halos \citep[e.g.][]{Hahn2009MNRAS.398.1742H, Yuan2021MNRAS.502.3582Y}; it is reasonable to question whether environment also impacts the state of the gas pressure in halos such that it cannot be predicted by mass and redshift alone. Third, when examining the tSZ signal between pairs of galaxies or clusters, isotropic halo models for the massive pairs do not fit well the signal from the intervening region \citep{deGraaff2019, Tanimura2019}, and neither do elliptical models \citep{Hincks2022}, indicating the presence of diffuse filament gas bridging the gap (as expected from hydrodynamic simulations). As directionally-dependent statistics become more popular as ways to extract greater detail from cosmic web structure \citep[e.g.,][]{Takada2004, Valogiannis2022PhRvD.105j3534V, Naidoo2022MNRAS.513.3596N}, it is important to probe where and when simple halo models begin to deviate from reality.

In \citet[][hereafter referred to as Paper I]{Lokken2022PaperI}, we presented a framework to measure the oriented superclustering of gas that makes use of photometric galaxy data, includes scale flexibility, and incorporates the possibility of environmental selection. In this framework, the projected galaxy position data is used to select locations of interest in the cosmic web, e.g., regions of strong superclustering. The smoothed galaxy maps are also used to determine the orientation of large-scale structure (LSS) around the selected locations at a chosen scale. The selection and orientation information is then applied to a (much noisier) $y$ map. Cutouts are taken from the $y$ map at each location, then rotated and stacked (averaged) to align the LSS orientation axis. This process augments the signal-to-noise ratio (SNR) from the noisier tracer both at the selected centers of the stack and along the axis of superclustering. In Paper I, we studied this method using simulations and presented a proof-of-concept by combining the Dark Energy Survey (DES) Year 3 (Y3) galaxy data with a Compton-$y$ map from the Atacama Cosmology Telescope (ACT) Data Release 4 (DR4) which incorporated both ACT and \textit{Planck} data over an overlapping area of $\sim430$~sq.~deg.

This follow-up Paper II makes use of the expanded ACT DR6 $y$ map \citep{Coulton2023arXiv230701258C}, which now has near-complete overlap ($\sim3,900$ sq. deg.) with DES Y3 \citep{DESY3+CMB2023PhRvD.107b3531A}. We also apply oriented stacking to DES galaxy number density and weak lensing data. Through this combination, we probe the large-scale anisotropic relationships between hot gas, galaxies, and the (dark-matter-dominated) total matter content at redshifts $0.2<z<0.92$. In other words, this work probes anisotropic bias and measures for the first time the evolution in the anisotropic bias of the gas pressure.

We then compare the ACT$\times$DES observations with the corresponding distribution in the Buzzard and Cardinal simulations \citep{DeRose2019, To2024ApJ...961...59T}, using halo models for the galaxies and gas pressure. The pressure is pasted isotropically around each halo and we do not model any diffuse gas, such that anisotropy in the oriented stacks arises solely from the halo distribution. This simplified modeling was fully consistent with observed data in Paper I. In the present work, the key goals are to determine whether (1) the model continues to be sufficient given SNR improvements from the newly expanded data, and (2) our measurements of extended structure can distinguish between variations in the gas and galaxy prescriptions within the halo framework. (We do not attempt to quantify the amount of diffuse gas.)

The paper is structured as follows. We begin with an overview of the constrained oriented stacking methodology in Section~\ref{sec:methods_overview}. In Sec.~\ref{sec:Data} we describe our ACT$\times$DES data and corresponding simulations. In Sec.~\ref{sec:methods} we discuss further details of the orientation methods, focusing on updates compared to Paper I. Sec.~\ref{sec:uncertainties} describes our approach to calculating uncertainties and covariance matrices. In Sec.~\ref{sec:tests} we describe various tests of the results and uncertainties. Our results are presented in Sec.~\ref{sec:res_indiv} (for individual tracers) and Sec.~\ref{sec:res_multiple} (for relationships between different tracers) and we summarize and conclude the discussion in Section~\ref{sec:conclusion}.

Throughout this work, we use a flat $\Lambda$CDM cosmology with parameters from \citet{Planck2018} for distance--redshift conversions   ($\Omega_{\mathrm{M}}=0.3097$, $H_0= 67.66\,\mathrm{km}\, \mathrm{s^{-1}}$) implemented through \texttt{astropy.cosmology} \citep{astropy2022ApJ...935..167A}. All distances are in comoving units unless otherwise specified.

\section{Method Overview} \label{sec:methods_overview}

We begin by selecting clusters above a mass threshold from a catalog. We then limit the sample to those located within large-scale regions of strong superclustering, which will serve as the stacking locations. The motivation behind this selection is discussed at length in Paper I; briefly, cosmologically these regions are both overdense and therefore expected to be sensitive to the nature of dark matter, while elongated (and non-virialized) on large scales, thus sensitive also to dark energy slowing and preventing their collapse. Furthermore, they are rare and highly non-Gaussian and therefore unique regions to search for deviations from the standard cosmological model. In terms of their gas signatures, superclusters are expected to induce a significant extended tSZ signal due to the correlation and alignment of multiple massive clusters, galaxy groups and galaxies, and the shock-heated and/or feedback-heated gas beyond member halos \citep[as demonstrated in Paper I and also discussed in][]{FloresCacho2009MNRAS.400.1868F, Tanimura2019,  Lokken2023MNRAS.523.1346L}. 

For such regions, we require the matter field to be highly elongated and sufficiently overdense with respect to the typical matter fluctuations at that redshift. Generally these features could be determined from any high SNR tracer, $F_1$, of the total matter distribution. Here, we define these features in terms of a map of galaxy overdensity limited to a narrow range in redshift. It is important to limit the extent of the redshift bin to span no more than a couple hundred comoving Mpc along the line of sight, as we are particularly interested in structure shaped by late-time formation processes that induce non-Gaussianity into matter tracer fields, and projections much wider than the correlation length of structure suppress non-Gaussianity due to the central limit theorem \citep{Jeong2011}. We smooth the resulting projected galaxy overdensity map $\delta_\mathrm{g}$ with a kernel of real-space comoving size $R$ to create a map of the large-scale galaxy fluctuations $F_{\mathrm{g},R}$. In this work we will smooth all $\delta_\mathrm{g}$ maps with a Gaussian filter with a FWHM of $20~\mathrm{Mpc}$, making our analysis sensitive to structures of $\sim8-10\hMpc$ separations; details of this choice and conversion are in Section~\ref{subsec:smoothing}.

On the smoothed map, we define the rarity of the overdensity (which we shall refer to as the `field excursion') as
\begin{equation}
    \nu = F_{\mathrm{g},R}/\sigma_R,
\end{equation}
where $\sigma_R$ is the variance of that smoothed map. Henceforth we will drop the $R$ subscript as we do not vary the smoothing scale in this work. We define the ellipticity, a normalized measure of the elongation of the same field, to be:
 \begin{equation} \label{eq:e}
     e = \frac{\lambda_1 - \lambda_2}{2 (\lambda_1 + \lambda_2)},
 \end{equation}
where the $\lambda_i$ are the eigenvalues of the Hessian matrix $H$; $H_{ij}=\frac{\partial^2 F_\mathrm{g}}{\partial x_i \partial x_j}$.
As in Paper I, we enforce $\mid{\lambda_1}\mid>\mid{\lambda_2}\mid$, such that $\lambda_1$ is the eigenvalue corresponding to the `short axis' of curvature and $\lambda_2$ describes the long axis.

We define high-superclustering regions as the galaxy cluster locations at positions $\boldsymbol{\hat{n}}_i$=(RA$_i$,dec$_i$) in $F_\mathrm{g}$ which satisfy both $\nu>2$ and $e>0.3$, isolating $\sim20\%$ of clusters. Thus there are constraints on the matter-tracer field at two scales: at small scales, the condition of a massive halo, and at large scales the conditions of $\nu$ and $e$. We will refer to these multi-scale constraints as $\xi$. The particular choice of threshold is motivated by maximizing the SNR of extended oriented tSZ emission in simulations (Paper I).

Having selected a sample of supercluster locations, we then cut out square regions around those locations from a second, separate map of a matter tracer $F_2$: the map to be stacked, which ideally has independent noise. The maps we will use as $F_2$ are Compton-$y$, projected galaxy number density $n_\mathrm{g}$, and lensing convergence $\kappa$. The cutouts are 80~Mpc per side, 4$\times$ the FWHM of the Gaussian filter applied to $F_\mathrm{g}$, sized so as to encompass the signal well beyond the region where the selections are defined. In this work, we leave each $F_2$ map in its native state without additional smoothing (but will perform averaging at a later stage by binning the stacked $F_2$ data).

We rotate each cutout such that the long-axis of structure is aligned among all the cutouts. To calculate the rotation angle, we use features of the first map $F_\mathrm{g}$: the Hessian eigenvector $\boldsymbol{v_2}$ pointing along the long-axis (i.e., the filament direction in a simplified straight-filament picture) defines an angle $\gamma$ from the R.A. axis. Each cutout from $F_2$ is rotated by $-\gamma$, i.e. into the basis of the eigenvectors such that the long-axis is the $x$ axis of the rotated image. In this new basis, we use polar coordinates on the image, with $r$ being the (projected) distance from the central halo and $\theta$ the polar angle measured from the new $x$ axis. This technique is very similar to measuring an \textit{alignment correlation function} as done in intrinsic alignment (IA) studies \citep[e.g.,][]{Faltenbacher2009RAA.....9...41F, Blazek2011}, which examine the projected correlation function of galaxies as a function of the projected separation (called $r_p$ in IA work) and angle from the galaxy's major axis. The galaxy major and minor axis in IA models are analogous to the large-scale Hessian eigenvector axes of $F_\mathrm{g}$ in our work.

If the cutouts are stacked at this point to create the average $\langle F_2 | \xi \rangle$, in the limit of an infinite number of combined cutouts, this process naturally results in a stack with reflective symmetry about the $x$ and $y$ axis (as in Paper I). However, the picture of filaments connected symmetrically to either side of each halo is an oversimplification: in reality there is a distribution of 2D cluster connectivity (number of connected filaments) which includes even and odd numbers, and odd-numbered connectivity necessarily means reflection asymmetry. A cluster with an even number of connections may also have high asymmetry if there is significantly more material on one side than the other. To propagate this information through the stacking process, we make use of asymmetry along the primary filament axis in the galaxy distribution as described below.

In the eigenvector basis for each massive halo, the sign of the gradient of the smoothed galaxy map $\partial F_\mathrm{g}/\partial x$ indicates whether LSS is becoming denser (positive) in the $+x$ direction or less dense (negative). We record this polarity information. If the gradient is negative, after rotating the cutouts from $F_2$ into the same basis, we flip that cutout horizontally. The same process is repeated for the $y$ direction. In other words, we enforce that the positive gradient directions from $F_\mathrm{g}$ (the first map) must be oriented towards the right and upwards ($+y$) in each $F_2$ cutout.

Finally, all $N$ cutout images are simply averaged (i.e. stacked), so the final stacked image $I$ is:
\begin{equation} \label{eq:simple_stack}
    I(r,\theta)\equiv \langle F_{2}(r,\theta)|\xi \rangle = \frac{\sum_{\mathrm{i}=1}^{N} F_2(\boldsymbol{\hat{n}}_\mathrm{i}, r, \theta)}{N}.
\end{equation}
The benefit of using polar coordinates will become evident shortly. Although in this work we apply equal weights to each cutout, a weighting scheme based on the large-scale density or ellipticity may be an interesting avenue to pursue to enhance the signal in future work.

Neither the rotation nor gradient-flip procedure automatically enforces that the stack of $F_2$ will display anisotropic structure or positive gradients oriented upwards and to the right: for completely uncorrelated $F_1$ and $F_2$, the stack would be isotropic. The technique relies on the assumption that structure is highly correlated between $F_1$ and $F_2$, while noise is not, as is true for the LSS tracers used in this work. Individual cutouts from the same location in correlated maps $F_1$ and $F_2$ may appear very different due to differing noise properties, but when averaged the oriented stacking technique brings out signal-to-noise from the three-point function of the map of constrained clusters, the first LSS map $F_1$ (specifically $F_\mathrm{g}$ in our case), and the second LSS map $F_2$.

To encapsulate the anisotropic information of the oriented and constrained stack, we decompose each image into a 2D multipole series (in other words, a Fourier series in $\theta$ for each $r$):
\begin{equation} \label{eq:image_composition_series}
    I(r,\theta)  = \sum_m \left (  C_m(r) \cos(m \theta)+S_m(r) \sin(m \theta) \right ),
\end{equation}
where $C_m$ and $S_m$ are the (real) Fourier coefficients. The cosine moments $C_\mathrm{m}(r)$ are calculated by
\begin{equation}\label{eq:multipole_moments}
    C_m(r) = \frac{1}{X\pi}\int_0^{2\pi}  I(r,\theta)  \cos{(m \theta)} \dv{\theta},
\end{equation}
where $X=2$ for $m=0$ and $X=1$ for all other $m$. The sine coefficients $S_m(r)$ are calculated by replacing $\cos{(m\theta)}$ with $\sin{(m\theta)}$ in Eq.~\ref{eq:multipole_moments}. We measure the $C_m(r)$ and $S_m(r)$ profiles out to a maximum $r$ of half a side length of the square stack (i.e., with the maximum $r$ corresponding to the circle inscribed in the square), in our case 40~Mpc.

For $m=0$, the $C_0(r)$ profile contains equivalent information to an unoriented stack, describing the isotropically-averaged signal as it decreases outwardly from a massive halo embedded in a large-scale overdensity. Meanwhile, higher-order moments like the dipole $(m=1)$ and quadrupole ($m=2$) incorporate shape information as a function of $r$. By construction, most of the oriented structure will fall along the $+x$ axis (at $\theta=0$, where $\cos(\theta)=1$), and therefore most of the shape information will be encapsulated by the $C_m(r)$ coefficients. Unlike in Paper I, where only the even cosine moments contained signal due to left-right symmetry, the addition of the gradient-flip towards $+x$ is expected to induce signal into the odd cosine multipoles. Moreover, in the case that the LSS is also asymmetric along the short-axis in the eigenvector basis, the $+y$ flip will induce asymmetry, bringing signal into the sine terms of Eq.~\ref{eq:image_composition_series}. As previously explained, the reflective asymmetries partially relate to the connectivity of halos, and such connectivity has been explored in, e.g., \citet{Codis2018MNRAS.479..973C, Gouin2022}.

As a function of $r$, as previously mentioned, the $m=0$ profiles will fall from high values as $r$ increases, as they represent a cross-correlation between $F_2$ and the constrained cluster sample $\boldsymbol{\hat{n}}_\mathrm{i} | \xi$. However, as shown in Paper I, the $m\neq0$ profiles will typically exhibit a rise and fall as a consequence of the large-scale orientation procedure. Specifically, they tend to begin small near $r=0$ and rise to a peak at $r_\mathrm{pk}\sim 0.5-0.75\times$ the Gaussian FWHM, if a Gaussian is used to smooth $F_1$. This is because the method explicitly aligns structure at that chosen scale, such that in the stack the anisotropy is enhanced at that characteristic distance. Meanwhile, structure at both $r<r_\mathrm{pk}$ and $r>r_\mathrm{pk}$ is in general somewhat misaligned from the orientation axis and thus averages to a more isotropic distribution in the stack.

\section{Observational Data}\label{sec:Data}

Following a similar prescription as Paper I, we use the \redmapper cluster catalog from three years of DES data (Y3) for the stacking locations due to its precise photometric redshifts (see Sec.~\ref{subsec:cluster_data} for references and description). We use the  DES Y3 \maglim galaxy catalogue (Sec.~\ref{subsec:galaxy_orientation_data}) to define the large-scale superclustering constraints and the direction of structure along which to stack. \maglim has been optimized for cosmological analysis given the trade-off between number density and redshift uncertainty. 
Using the above two catalogs, we stack on the ACT DR6 Compton-$y$ map (Sec.~\ref{subsec:ACT-y-map}); maps of the DES \maglim galaxy number density (Sec.~\ref{subsec:DES-nd-maps}); and a map of the weak lensing convergence from DES galaxy shears (Sec.~\ref{subsec:DES-kappa-map}).

\subsection{Galaxy cluster data} \label{subsec:cluster_data}
To define the sample of stacking locations, we begin by using the DES Y3 \redmapper cluster catalog (version \texttt{v6.4.22+2}
Full). DES \citep{DES2005} finished its six-year observing run of the southern sky in 2019. Using five optical filters (\textit{grizY}) on the Dark Energy Camera \citep{Flaugher2015}, fitted on the 4-meter Blanco Telescope at the Cerro Tololo Inter-American Observatory (CTIO) in Chile, DES mapped out 5,000 square degrees of the sky. The \redmapper algorithm \citep{Rykoff2014} was applied to Y3 \texttt{GOLD} data \citep{SevillaNoarbe2021ApJS..254...24S} to identify clusters as overdensities of red-sequence galaxies, assigning a richness $\lambda$ to each cluster  based on the probabilistic number of members within a defined radius from the center. Y3 \redmapper achieved precise photo-$z$ estimates of $\sigma_z/(1+z)\sim0.01-0.02$ \citep{McClintock2019}. In Paper I, we used all clusters with $\lambda>10$ due to the limited overlap of ACT $y$ with the DES footprint. Now with a much expanded survey overlap, for our main analysis, we restrict the sample more conservatively to $\lambda>20$, or $ M\gtrsim10^{14} \Msun$ at $z=0.35$ \citep{McClintock2019}. This excludes lower-mass clusters that are more commonly impacted by mis-centering and contaminated by line-of-sight projections \citep{Rykoff2016}. We also limit the cluster sample to the overlapping region of the DES and ACT sky footprints, including a buffer of 2 degrees from the survey edges such that clusters falling within the buffer zone are removed, in order to avoid edge effects that could impact both the orientation and stacking (validated in Section~\ref{subsec:mask_holes}). The reduction to this limited overlap region reduces the catalog by $\sim40\%$, from 53,610 to 32,176 clusters. The redshift extent is from $z=0.1$ to $z=0.95$.

\subsection{Galaxy orientation data}
\label{subsec:galaxy_orientation_data}

We make use of DES photometric galaxy data from the \maglim sample \citep{Porredon2021} to reduce the cluster sample by the large-scale $\nu>2$ and $e>0.3$  superclustering constraints as well as to determine, at the location of each cluster, the axis of strongest superclustering. Both are done in projection.
 
Here we briefly discuss the motivation to choose \maglim from the multiple DES Y3 galaxy samples. All DES Y3 galaxy samples derive from the core dataset selected for cosmological analysis, \texttt{Y3 GOLD} \citep{DESY32021photodata}. The impetus to consider \maglim rather than red-sequence galaxies selected by the \redmagic algorithm \citep{Rozo2016} (which we used in Paper I and has higher photometric redshift accuracy) came from the uncovering of selection and calibration problems for \redmagic that lead to internal inconsistencies in the DES Y3 cosmology analysis \citep{RodriguezMonroy2022MNRAS.511.2665R}. The problems arose in an analysis much more sensitive than that which we conduct here, so while they may not be an obstacle to our study, they motivated us to consider \maglim, which was specifically optimized for cosmological inference and showed no inconsistencies in the Y3 analysis \citep{Porredon2021_cosmoconstraints}. Meanwhile, an improved version of \redmagic is available in which many of the calibration concerns have been addressed, so we make tests to decide which is optimal for our analysis.

The \maglim sample has a redshift-dependent magnitude limit of $i <
4 z_\mathrm{phot} + 18$ \citep{Porredon2021}. It has poorer photometric redshift uncertainties by a typical factor of a few ($0.02<\sigma_z/(1+z)<0.05$ compared to \redmagic's $\sigma_z/(1+z)<0.0126$ \citep{Porredon2021}. However, the \maglim number density is $\sim3-5\times$ higher than \redmagic's, with a difference of $\sim7\times$ at low redshifts. Therefore, while the increased redshift uncertainty smears out the structure, limiting the accuracy of orientation, the boost in number density may compensate. We test the accuracy of orientation using simulated mocks of each sample, in comparison with a true-redshift simulated sample. The test is described in Appendix.~\ref{appdx:maglim_v_redmagic}. We find that \maglim produces more accurate orientations, so we choose to exclusively use \maglim.

The \maglim catalog includes per-galaxy weights which compensate for the survey-dependent angular selection function which would otherwise bias the sample. When projecting these galaxies into number density maps in tomographic redshift bins (Section~\ref{sec:methods}), we apply these weights. The catalog with non-zero weights extends over a redshift range of $0.2<z<1.05$, but we only make use of the range overlapping with \redmapper.

\begin{figure}
    \centering
    \includegraphics[width=\columnwidth, trim={0cm 4.5cm 0cm 4.5cm},clip]{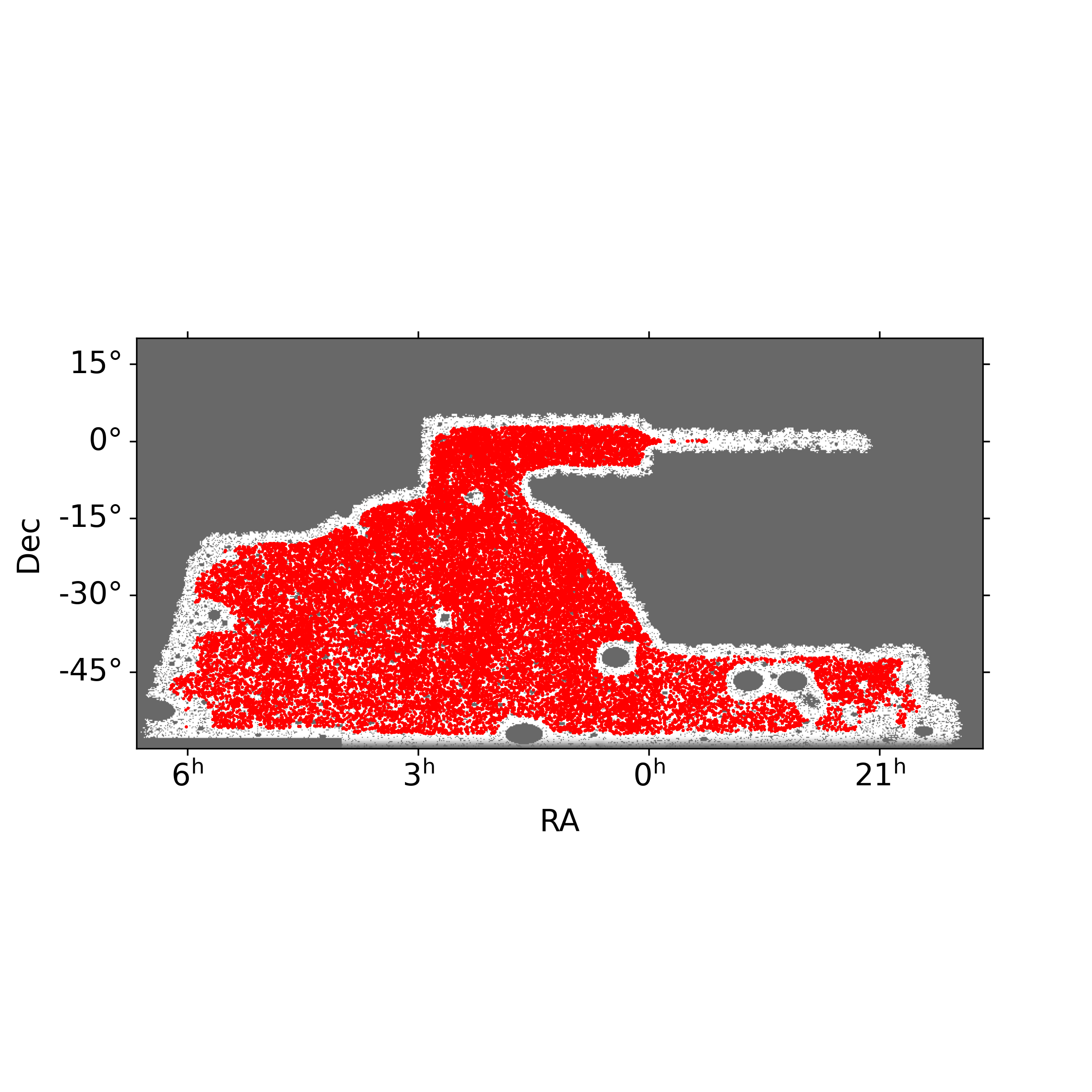}
    \caption{\redmapper clusters (red) overlaid on the DES survey footprint mask in Cartesian projection. Grey represents regions where the mask is 0, while white (obscured in most of the footprint by the over-plotted red points) represents where it is 1. Each of the white pixels have associated \maglim galaxies. The cluster catalog has been cut to include only clusters $>2$ degrees away from the mask edges and holes.}
    \label{fig:clusters_mask}
\end{figure}
Figure~\ref{fig:clusters_mask} shows the binary DES footprint mask in grey and white (where white regions are unmasked, i.e., observed, while grey are masked). \maglim galaxies densely populate the unmasked footprint, so the white region can be interpreted as the \maglim galaxy coverage. The reduced \redmapper cluster catalog is overlaid in red points, such that the white region is mostly obscured, except for the buffer zone near the footprint edges.

\subsection{ACT DR6+Planck Compton-y map}\label{subsec:ACT-y-map}
The first projected map we stack, and the primary focus of our analysis, is the recent Compton $y$ map described in \citet{Coulton2023arXiv230701258C} which was created using a combination of ACT data, up to and including Data Release 6 (DR6), with \textit{Planck} satellite data.

ACT recently completed its final observing run in Fall 2022, rounding out a 15-year run as a major large-sky, ground-based CMB telescope. ACT was a 6-meter telescope that mapped out the microwave sky from its vantage point 5190~m high in the Atacama Desert, Chile \citep{Fowler2007, Swetz2011}. From 2008-2013, ACT was only sensitive to temperature fluctuations, but after the ACTPol receiver was deployed in 2013, it began to take polarization data as well \citep{thornton/etal:2016}. The receiver was subsequently upgraded to Advanced ACTPol \citep{Henderson2016, Shuay2017, Choi2018, Li2018}. In Data Release 4, ACT produced and analyzed data primarily from 2013-2016 in frequency bands centered on 98 and 150 GHz, covering 18,000 square degrees of the sky \citep{ACTDR52020, Aiola2020}. Data Release 5 \citep{ACTDR52020} extended the maps to include 2017-2018 data and added a band centered on $\nu\simeq220$~GHz. The latest and ongoing DR6 includes data up to the end of ACT's lifetime in 2022 \citep[as briefly described in][and further detailed in Naess et al, in prep]{Qu2023arXiv230405202Q}.

Prior to ACT, the \textit{Planck} satellite made lower-resolution observations in nine frequencies of the entire microwave sky from space, thus avoiding atmospheric noise. The \textit{Planck} team released their first results in 2013 and final results in 2018 \citep{Planck2014A&A...571A..16P, Planck2018}.

The ACT DR6 $y$ map covers 1/3 of the sky ($\sim$13,000 square degrees) and includes 93, 148 and 220 GHz data from ACT as well as data from nine \textit{Planck} frequencies ranging from $30$\,GHz to $545$\,GHz \citep{Coulton2023arXiv230701258C}. The map expands to a much wider footprint than the initially constructed ACT $y$ map, which covered 456 sq. deg. of sky using 2014-2015 data from only two ACT frequency channels and similar \textit{Planck} data \citep{Madhavacheril2020}, which we analyzed in Paper I. The input maps for DR6 have a wide range of resolutions (at worst, 32 arcmin for the lowest \textit{Planck} frequency and at best, 1 arcmin for the highest ACT frequency) but the final map is convolved with a 1.6 arcmin beam. The map is noise-dominated at all scales \citep{Coulton2023arXiv230701258C}. Thus the detail of the thermal energy of the cosmic web is drowned under a sea of noise, with only the highest peaks (the massive galaxy clusters) easily distinguishable by eye; statistical techniques must be applied to extract information buried below the noise.

The Compton $y$ signal in this map has been extracted from other frequency-dependent microwave signatures by performing an internal linear combination (ILC): adding together multi-frequency maps with different weights such that the combined map has optimal response to tSZ. The emission from dusty galaxies, called the cosmic infrared background (CIB), is a source of residual contamination due to its similar spectral energy distribution (SED) \citep{Madhavacheril2020}. It is the most important residual for cross-correlations of the tSZ with galaxies and galaxy clusters, as it is highly correlated with those locations. In this work, we test several additional $y$ maps with different models of the deprojected CIB SED. For such maps, in addition to requiring the ILC algorithm to have unit response to Compton-$y$, the final map must also have null response to the CIB. This second requirement introduces additional noise into the map but (ideally) nulls the contribution from CIB. We discuss the tests in Section~\ref{sec:tests}. We ultimately report results from both the non-deprojected map and one CIB-deprojected map.

In reality, modern microwave background experiments such as ACT and \textit{Planck} are insensitive to the monopole signal in $y$. In other words, the mean intensity is not recorded in any frequency, only the deviations from the mean. Therefore, a $y$ map is truly displaying $\Delta y$, the difference from the sky-averaged mean. We will generally refer to $\Delta y$ as simply $y$ in figures and results, but in cases in which the mean subtraction is relevant will be explicit in the notation.

Before performing any operations on the $y$ map, we first use the code \texttt{pixell}\footnote{\url{https://pixell.readthedocs.io/en/latest/readme.html}} to reproject it from its native CAR format (equally-shaped pixels) to \textsc{HEALPix} format \citep[equal-area pixels;][]{Gorski2005} with NSIDE=4096.

\begin{figure*}[htbp!]
    \centering
    \includegraphics[width=\textwidth, trim={6cm 0cm 5.5cm 0cm},clip]{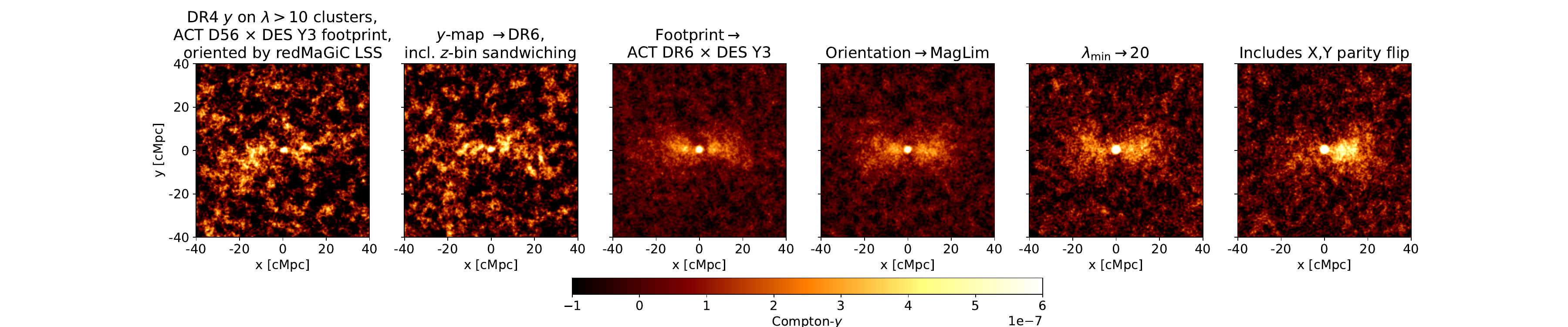}
    \caption{The impact of data combinations and methods on $y$ map stacks of clusters combined over $0.26<z<0.70$ (all except leftmost plot) and $0.25<z<0.72$ (leftmost plot, reproduced from Paper I). From left to right, the successive updates are described above each stack, with the largest impact to SNR coming from the expanded overlap of the two surveys (third plot). Note that successive stacks which contain similar structure do not necessarily place that structure in the same location to the left/right side of the image, as there is a random $\pm \pi$ flip added to each cutout prior to stacking. In the final right-most stack, the direction along the filament axis in which the galaxy density gradient is higher has been flipped to the right in each cutout prior to stacking.}
    \label{fig:ystack_impvts}
\end{figure*}

\subsection{Increases in signal-to-noise in Compton-y stacks}
The impacts of the expanded DR6 footprint, as well as the changes in cluster and galaxy data from Paper I, are presented in Figure~\ref{fig:ystack_impvts}. For simplicity, this figure does not yet incorporate any of the environmental constraints $\xi$ on cluster data. At left is the Paper I result for an oriented, symmetric stack of 975 \redmapper clusters with $\lambda>10$, oriented by all \redmagic galaxies on a scale of 18~Mpc, combined over the redshift range $0.25<z<0.72$. The clusters were restricted to only those which fell within the ACT D56 region; the cutouts were from the map released in \citet{Madhavacheril2020} made up of 2014-2015 ACT data in combination with \textit{Planck} data. At second-from-left, a nearly-identical cluster sample is stacked with a more optimal method of overlapping cluster bins sandwiched within galaxy bins along the line of sight, which will be presented in Sec.~\ref{subsubsec:slicewidth}. The redshift range must be slightly adjusted in the overlapping-bins pipeline, but we produce a very similar cluster sample spanning $0.26<z<0.7$ and containing 819 clusters. The map has also been updated to include DR6 data. These updates have relatively minor impacts on the signal-to-noise in the stack. Note that the first two stacks are statistically similar, but appear to show different distributions of structure. This is caused by a default, but unnecessary, component of our pipeline (when asymmetry is not enforced) which randomly adds $\pm \pi$ to the rotation angle of each cutout prior to stacking. Therefore, some $y$ signal which appears on the left in the first image is mirrored at right in the second. 

In the third stack from left, the cluster footprint has been expanded to have near-full overlap with DES Y3. This change causes the most significant increase in SNR as the cluster sample is expanded by $\sim 10 \times$ to 8,669. The quantitative SNR improvements to the multipole profiles for this particular stack will be compared to the Paper I results later, in Sec.~\ref{sec:res_indiv}. 

In the three images to the right, the galaxies used for orientation are updated to \maglim, yielding a marginal improvement due to the higher number density; then the clusters are reduced to the high-richness $\lambda>20$ sample, and finally the $x$ and $y$ parity flips are included. The adjustment to $\lambda>20$ increases signal in the extended regions due to the (weak) correlation between richness and extended environment as explored in Paper I. However, the noise also increases due to the $2.6\times$ reduction in sample size to 3,286 clusters. In the final image, the $x$-axis shows a notable asymmetry, while asymmetry in the $y$ direction appears subtler (further discussed later in Sec.~\ref{subsec:all_tracers}).

\subsection{Projected galaxy number density maps} \label{subsec:DES-nd-maps}
To compare the superclustering of thermal energy with that of galaxies, we use the DES \maglim data not only for large-scale constraints and orientation (as $F_1$ from Section~\ref{sec:methods_overview}), but also for stacking ($F_2$). Here it is useful to note that, due to the redshift-dependent magnitude limit, the galaxy bias varies per redshift \citep[see Table I in][]{Porredon2021_cosmoconstraints}.

To use the galaxy data for both purposes, it is necessary to split the data. This is because it is ideal for the noise in the map being stacked to be fully independent of the noise in the data used for orientation. When stacking the ACT $y$ map with orientations given by DES galaxies, the dominant noise in the former (instrumental and atmospheric) is expected to be independent from the DES noise. However, when stacking maps of DES galaxies with orientation given by the same galaxy data, the same Poisson (shot) noise that impacts the orientation also appears in the maps. For example, for a cluster at a node in the cosmic web where the surrounding structure is relatively isotropic, the orientation may be dominated by shot noise, which will also generate an apparent signal along the preferential axis of the stacked figure causing a misleading excess in the averaged filamentary structure.

To remove the bias due to this correlated shot noise, we split the \maglim catalog into two random subsets: $3/4$ for orientation and $1/4$ for stacking. The choice of fractions came from testing several different splits with mock data to measure the deterioration in orientation accuracy; using less than $3/4$ for orientation caused a significant downward bias in oriented signal. We divide the $1/4$ sample into four broad redshift bins, with edges [0.20, 0.36, 0.53, 0.72, 0.94]. The bins are larger than the typical spread in \maglim photo-$z$s: given that $\sigma_z$ for \maglim is typically $0.02<\sigma_z/(1+z)<0.05$ and maximally $\sim0.07$ at $z\sim0.4$ \citep[see Figure 1 from][]{Porredon2021}, the bins are larger than $2\sigma_z$ even in the case of poorest photo-$z$. The specific choice of bin edges is different than in the DES Y3 main cosmology analyses and arises from details of the orientation techniques described in Section~\ref{sec:methods}. In each bin, the galaxies are mapped in projection, weighted by the \maglim weights. The mapping is done in \textsc{Healpix} format \citep{Gorski2005} using \texttt{Healpy} \footnote{\href{https://healpy.readthedocs.io/en/latest/}{https://healpy.readthedocs.io/en/latest/}}, with NSIDE=4096. For each \textsc{Healpix} pixel we divide the weighted number of member galaxies by the coverage fraction for that pixel (what fraction of the pixel was observed), called `FRACGOOD' in the DES Y3 data catalogs.

\subsection{DES Y3 convergence map} \label{subsec:DES-kappa-map}
To make a comparison between the $y$ signal, representative of the gas thermal energy, and the total matter content of superclusters, we also make use of one of the DES Y3 lensing convergence maps \citep{Jeffrey2021}. The convergence ($\kappa$) maps, also called mass maps, reconstruct the matter in the foreground of source galaxies using the source galaxy shear data. In total, Y3 includes shear data from over 100 million galaxies. We use only the $\kappa$ map from source bin 4 which was reconstructed using the Kaiser-Squires (KS) method. The photometric redshift bin edges are [0.872, 2.0], and the estimated true redshift distribution is discussed in \citet{Myles2021MNRAS.505.4249M}. This range lies beyond all but one of the four broad redshift bins we use for analysis, presented later in Sec.~\ref{sec:redshift_ranges}. The map is in \textsc{HEALPix} format with NSIDE=1024.

This simplistic approach, using only one map derived from high-$z$ shear galaxies, is certainly non-optimal given the additional available DES lensing data at lower redshifts. However, the goal of including the convergence map in our study is to make a preliminary measurement of the levels of coherence in the large-scale superclustering of hot gas, galaxies, and total matter. Towards that end, this simplified approach is sufficient, and using the fourth source bin -- which includes, as much as possible, the foreground objects from \redmapper and \maglim that we seek to study -- represents a conservative choice. In future work it would be interesting to apply these techniques with a more robust algorithm combining multiple lensing tomographic bins.

\subsection{Buzzard and Cardinal mocks}
For direct comparison with data, we examine both the Buzzard simulations \citep{DeRose2019} and the updated version of this suite, called Cardinal \citep{To2024ApJ...961...59T}. We examine both because they each have advantages and disadvantages for the comparison with DES Y3 observations: Buzzard was specifically created to mimic the same DES Y3 survey effects, while Cardinal is tailored for Y6; however, Cardinal benefits from improvements in galaxy modeling. Furthermore,  using both allows us to examine the impact of improved halo modeling on the anisotropic superclustering signal. Both simulations share the same underlying halo catalog, produced by an N-body simulation run with a $\Lambda$CDM cosmology with $\Omega_\mathrm{M}=0.286, h=0.7, \sigma_8=0.82$, referred to hereafter as the Buzzard cosmology. Comparing to the DES Y3 best-fit $\Lambda$CDM cosmology from their $3\times2$pt analysis \citep{DESY32022galaxy_clustering_weak_lensing_PhRvD.105b3520A}, Buzzard has a 1.4$\sigma$ lower value of $\Omega_\mathrm{M}$ and 2.2$\sigma$ higher value of $\sigma_8$. Though DES clustering doesn't independently constrain $h$, the combined DES + baryon acoustic oscillations + big bang nucleosynthesis or DES + \textit{Planck} constraints on $h$ are $\sim2-3\sigma$ lower than Buzzard's $h=0.7$. In other words, the Buzzard cosmology is not inconsistent with DES, but also not in strong agreement.

There are also some differences between the Buzzard cosmology and the \textit{Planck} 2018 cosmology \citep{Planck2018} used in this work for radial and transverse distance calculations; these differences contribute error only in the rescaling of cutouts from different redshifts into the same comoving size. They contribute at the few-percent level to the rescaling, a discrepancy well within the bin size for the radial profiles which will constitute our final results; therefore this mismatch is expected to contribute negligible error. 

Both Buzzard and Cardinal apply galaxies with given spectral energy distributions (SEDs) to halos using the \texttt{ADDGALS} algorithm \citep{Weschler2021}.
Cardinal presents significant improvements to the models for subhalo abundance matching and the color assignment to galaxies that enter into \texttt{ADDGALS}. The resulting galaxies have improved color-dependent clustering, which in particular impacts the color-selected \redmapper and \redmagic samples \citep{To2024ApJ...961...59T}. The \redmapper samples for both Buzzard and Cardinal were run with somewhat different algorithms than that which was applied to DES Y3. The Buzzard run produced far fewer \redmapper clusters than observed in DES Y3 \citep{DeRose2019}, and the large-scale bias of cluster-galaxy cross-correlations (both \redmapper-\maglim and \redmapper-\redmagic) was strongly inconsistent. Both these features show strong improvement in Cardinal; overall the \redmapper sample is far more consistent with DES Y3, despite the algorithmic differences \citep{To2024ApJ...961...59T}.

The Buzzard \maglim sample was selected from one realization of the Buzzard v2.0 simulations as described in section IV. of \citet{Porredon2021_cosmoconstraints}. Because of differences in the color-magnitude relation in Buzzard galaxies vs. DES, the authors re-defined the magnitude limits to reproduce more accurately the DES Y3 $n(z)$ distributions. Meanwhile, the Cardinal \maglim sample has been produced with the standard \maglim algorithm but with DES Y6-like survey conditions. However, as \maglim has undergone negligible changes from Y3 to Y6 in the redshift bins used in this work, we do not expect this to generate major inconsistencies with the data.

The simulations are not produced with accompanying CMB foreground maps. To produce a correlated Compton-$y$ map, we input the common underlying halo catalog  to the Websky code package \texttt{XGPaint}\footnote{\url{https://github.com/WebSky-CITA/XGPaint.jl}} \citep[adapted from][]{Stein2019, Stein2020} which paints tSZ profiles (and other foreground signatures, if desired) onto halos based on their mass and redshift. Each profile originates from a pressure prescription that is an isotropic function of radial distance from the halo center. The advantage of this rapid painting approach is that it allows us to perform the same exact oriented stacking procedure on the simulations as on observational data, using the mock DES galaxy and cluster samples to perform selection and orientation on the $y$ map with a very similar redshift distribution. 

To connect the pasting approach to insights from hydrodynamic simulations, we consider several related studies. Many hydrodynamic simulations have shown that AGN feedback can explain SZ observations which show pressure depletion in lower halo mass halos compared to a self-similar mass-thermal energy relation \citep[e.g.,][]{Battaglia2012a, McCarthy2010MNRAS.406..822M, LeBrun2017MNRAS.466.4442L, Yang2022}. Moreover, hydrodynamic simulations also show us that it is not only the near-halo regime that matters: in some simulations with AGN feedback, the amount and pressure of gas beyond $2R_{200}$ of halos has a major impact on the mock observed SZ signal for extended structures like filaments and superclusters \citep{Lokken2023MNRAS.523.1346L}. Furthermore, the relative amount of gas to dark matter beyond these halos varies drastically from one simulation to the next, depending mostly on the AGN feedback prescription \citep{Ayromlou2023MNRAS.524.5391A, Yang2022}. The AGN prescription not only affects the distribution of gas density but also the temperature (the combination of which sets the tSZ strength), causing anisotropic impacts on the gas far beyond halos \citep[up to $\sim4R_{200c}$ and possibly beyond,][]{Yang2024MNRAS.527.1612Y}.

In this work, we create two mock Compton-$y$ maps representing two distinct gas pressure prescriptions mimicking different ways in which AGN feedback may affect lower-mass halos. In a separate work focusing on simulations, we will test further variations to the pressure profiles prescribed to each halo to mimic a broader range of the possibilities presented by various hydro sims (Lokken et al., in prep).

The first map mimics results from hydrodynamic simulations from \citet[][hereafter BBPS]{Battaglia2012b} which simulated halos with $M>10^{13}\Msun$ and included AGN feedback. The authors fit a model for the pressure profile as a function of $x=r/R_{200c}$ from the halo center:
\begin{eqnarray}\label{eq:BBPS}
    P_\mathrm{e}^{\mathrm{B12}} &=& G M_{200} 200 \rho_{\rm{cr}}(z) (f_{\rm{b}}/2)  \nonumber \\
    &&\times R_{200} P_0 (x/x_\mathrm{c})^\gamma [1+(x/x_\mathrm{c})^\alpha]^{-\beta},
\end{eqnarray}
where $\alpha=-1$ and $\gamma=-0.3$. The parameters $P_0$, $x_\mathrm{c}$ and $\beta$ are all functions of redshift; for each of these parameters generically referred to as $A$,
\begin{equation}
    A = A_0\bigg(\frac{M_{200}}{10^{14}\Msun}\bigg)^{\alpha_m} (1+z)^{\alpha_z}.
\end{equation}
The amplitude $A_0$, mass dependence $\alpha_m$, and redshift dependence $\alpha_z$ were varied separately for $P_0$, $x_c$ and $\beta$ to find the nine best-fit values for the simulated cluster sample. This model, with the BBPS best-fit parameters (listed in their Table 1), constitutes our fiducial pressure-pasting model. It implies a relationship between mass and integrated Compton-$y$ signal of $Y \propto M^{1.72}$, steeper than the self-similar relation of $Y \propto M^{5/3}=M^{1.67}$ \citep{Kaiser1986, Nagai2006ApJ...650..538N}. The parameters were originally calibrated to simulated clusters of mass $M_{200c}=5\times10^{13}-1.6\times10^{15}\Msun$ and out to radius $2R_\mathrm{200c}$, and the $\alpha_z$ redshift dependence parameter was fit to a more restricted mass range of $1.1\times10^{14}\Msun <
M_\mathrm{200c} < 1.7\times10^{14}\Msun$. Nevertheless, we apply the model to all halos down to $M>10^{10}\Msun$ and extend the profiles out to $4R_\mathrm{200c}$ to model a wider range of structures and account for more of the intergalactic gas. We find that the radial extension of the profiles has a minor impact on the superclustering signal because of the low pressure in the outskirts of the profiles---a smaller impact, for instance, than switching to the \textit{break model} discussed below (as will be demonstrated in Lokken et al., in prep).

For our second mock $y$ map, we refer to several observational studies which have shown evidence of a feedback-driven departure from a simple power-law relationship for the $Y-M$ relation for lower mass halos. In \citet{Hill2018}, the \textit{Planck} $y$ map stacked on SDSS galaxies was found to be best-fit with models for the gas profile that diverged from BBPS through either a mass-dependent scaling in the power law or a break in the power law. \citet[][hereafter P22]{Pandey2021} also tested a break in the power law slope at lower halo masses, equivalent to the `uncompensated break' model in \citet{Hill2018}, namely:
\begin{equation} \label{eq:break_model}
    P_\mathrm{e}^{\mathrm{B12},\mathrm{br}}(r|M,z) =
    \begin{cases}
        P_\mathrm{e}^{B12}(r|M,z), & \text{$M \ge  M_{\mathrm{br}}$}\\
        P_\mathrm{e}^{B12}(r|M,z) \left(\frac{M}{M_\mathrm{br}}\right)^{\alpha_\mathrm{m}^{\mathrm{br}}}, & \text{$M \le M_{\mathrm{br}}$}\\
    \end{cases}
\end{equation}
In that paper, the authors fixed $M_\mathrm{br}=2\times10^{14}$~\hMsun\ based on the steepening in the $Y-M$ relationship seen in the cosmo-OWLs simulations (with AGN feedback) at roughly that mass \citep{LeBrun2017MNRAS.466.4442L}. Allowing $\alpha_\mathrm{m}^{\mathrm{br}}$ to vary, they found a preference for a non-zero $\alpha_\mathrm{m}^{\mathrm{br}}$ in both the \textit{Planck}-only and \textit{Planck}+ACT results. In our second mock $y$ map, we implement the uncompensated \textit{break model} using the best-fit value result from that work, $\alpha_{\mathrm{m}}^{\mathrm{br}}=0.972$, and the same value for $M_\mathrm{br}$.

Implementing these two different gas pressure models will provide insight into the range that current uncertainties in gas physics contribute to the tSZ superclustering signatures under a single cosmological model. It is important to note that both of these prescriptions make the following assumptions about the gas: (1) that the pressure can be entirely described by an isotropic relationship to halos based on mass and redshift, and (2) that the cutoff radius of $4R_{200}$ allows for a full description of the gas beyond halos without causing excessive overlap. Neither assumption is expected to be true based on observations of filament and bridge gas \citep{ deGraaff2019, Tanimura2019, Hincks2022}, but the question we pose is whether the prescriptions are nevertheless sufficient to capture the binned signal in the oriented stacks.

\section{Detailed Methods} \label{sec:methods}

As broadly described in Sec.~\ref{sec:methods_overview} and introduced in Paper I, the oriented stacking approach that we take first divides the DES galaxy and cluster data into narrow tomographic bins of redshift. The projected information in each bin is then used in combination with independent maps (the fluctuations from the mean in $y$, galaxy number density $n_\mathrm{g}$, and surface mass density $\Sigma$) to isolate the signals from superclustering at that particular redshift. The algorithm uses \textsc{HEALPix} maps throughout \citep{Gorski2005} and performs curved-sky operations through the Cosmology Object Oriented Package \citep[\texttt{COOP}\footnote{\url{https://www.cita.utoronto.ca/~zqhuang/work/coop.php}},][]{Huang2016}. We refer the reader to full details in Paper I and in this work will focus on describing only the detailed updates to the methodology.

\subsection{Orientation}\label{sec:orientation_methods}
\subsubsection{Slice width} \label{subsubsec:slicewidth}
The accuracy of orienting each cluster by its local galaxy data depends on the true 3D correlation of the projected structures; for poor photometric redshifts and/or broad projection over the line of sight (l.o.s.), the orientation deteriorates due to contributions from uncorrelated foreground and background structures. To mitigate this effect, we first explore variations to the l.o.s. extent of the tomographic slices into which we divide the galaxy data, motivated by relatively poorer photometric redshift uncertainties of \maglim compared to \redmagic (which was used in Paper I with slices equivalent to 200~Mpc in comoving distance). Using the Websky simulations \citep{Stein2020}, we test the accuracy of orientation when galaxy and cluster data is divided into broader redshift slices of $\Delta z=0.15$ ($\Delta \chi \sim 500$ Mpc), the approximate bin sizes optimized for the DES Y3 cosmology results with \maglim  \citep{Porredon2021_cosmoconstraints}. These tests are described in detail in Appendix \ref{appdx:binsize_tests}. We find that projection effects dominate the orientation when such wide bins are used, and that narrower slices of 200~Mpc extent are significantly more effective at isolating local structure. We thus fix the extent of the galaxy slices at 200~Mpc as in Paper I.

\begin{figure}
    \centering
    \includegraphics[width=\columnwidth, trim={6.5cm 1.5cm 7cm 1.5cm},clip]{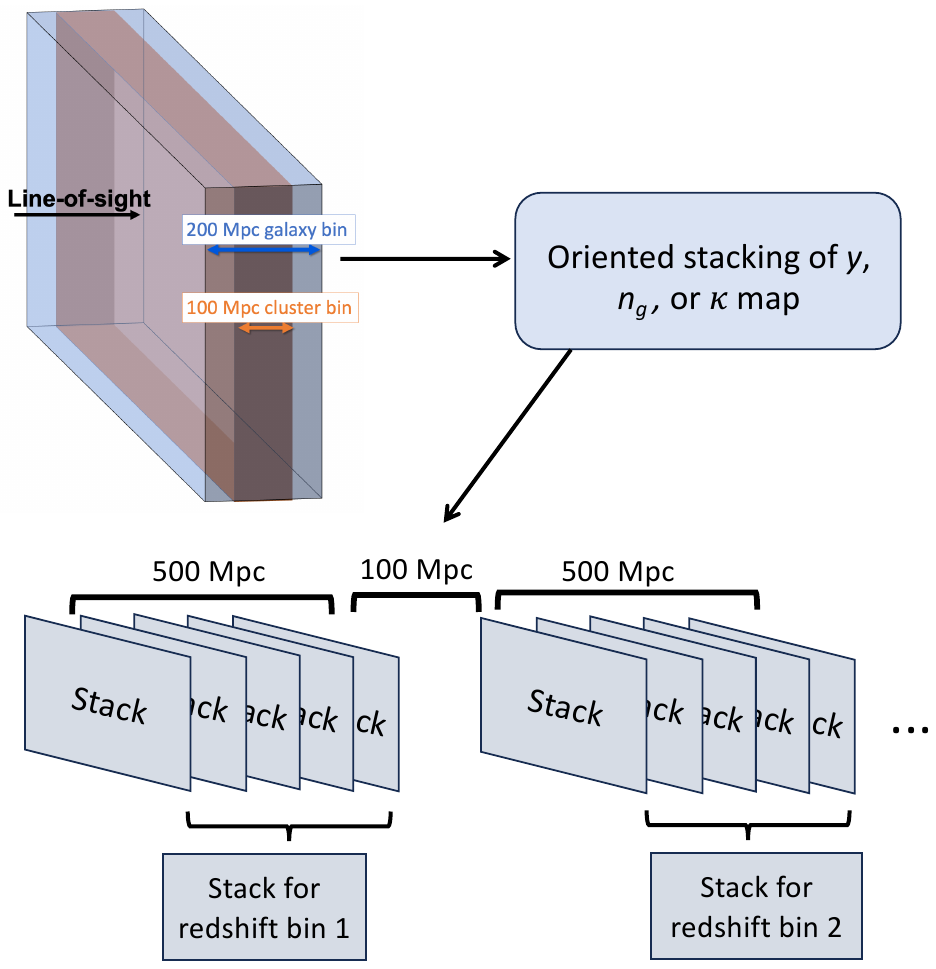}
    \caption{Redshift binning schema. From upper left, the image portrays the slicing of \redmapper galaxy clusters and \maglim galaxies for orientation. The galaxy slice encompasses the cluster slice to better account for extended structure around clusters near the bin edges. The schema also accounts for the lower photo-$z$ precision of the galaxies. After processing this data through the oriented stacking procedure, we then combine 5 stacks from successive 100 Mpc cluster slices to create each wide redshift bin result for analysis (lower). Between redshift bins, we omit 100 Mpc worth of cluster data as the galaxy data used for orienting this slice overlaps with the redshift bins on either side. In total we create 5 broad redshift bins.}
    \label{fig:bins_schematic}
\end{figure}

Through further tests with Websky, we also find that the orientation accuracy is improved when the cluster redshifts are restricted to a smaller range than the galaxy redshifts for each slice, as compared to the case in which they share the same extent. In this method, clusters are selected within only 100 Mpc, and the orientations are determined by galaxy information in a surrounding bin of 200 Mpc. The slices share the same center along the line-of-sight direction. This allows structures which extend both toward and away from the observer to be taken into account in the projected orientation (Fig.~\ref{fig:bins_schematic}, upper image).

In Appendix~\ref{appdx:binsize_tests} we also explore an alternative treatment of photometric redshifts that, rather than using only the mean, treats each galaxy as multiple entities spread through the $p(z)$ distribution. We find that this method does not improve the orientations and thus do not implement it.

For the sake of completeness, Appendix~\ref{appdx:ratios_tracers} presents an analytic description of how the narrow galaxy and cluster tomographic slice technique isolates signals from the $y$, $n_\mathrm{g}$, and $\kappa$ maps at the redshifts of interest.

\subsubsection{Smoothing choice}\label{subsec:smoothing}
We smooth each map with a Gaussian filter of FWHM 20~Mpc. Gaussian smoothing is distinct from top-hat smoothing in the way that it smoothly consolidates information from the central region but includes information from long tails; therefore the two cannot be exactly compared. However, a rough conversion from top-hat to Gaussian filter can be done by equalizing the volume under each non-normalized 2D function, i.e., a top-hat of height 1 and a Gaussian with amplitude 1. This yields FWHM$\sim1.67 R_{\mathrm{TH}}$, or $R_{\mathrm{TH}}\sim12$~Mpc ($\sim8 h^{-1}$~Mpc for $h=0.67)$ for our case. We test this conversion by comparing calculated orientations for several clusters embedded in a simulated galaxy field. In practice, we find that a slightly different conversion of FWHM$\sim1.3 R_{\mathrm{TH}}$ tends to produce similar orientations. With this conversion, 20~Mpc FWHM corresponds to $R_{\mathrm{TH}}=15.4=10.3 h^{-1}$~Mpc. In either case, the range of $8-10\hMpc$ separations is relatable to the range of structures covered in previous studies of inter-cluster / inter-galaxy bridges \citep{Tanimura2019, deGraaff2019}.

\subsubsection{Avoiding mask holes}\label{subsec:mask_holes}

\begin{figure}
    \centering
    \includegraphics[width=\columnwidth]{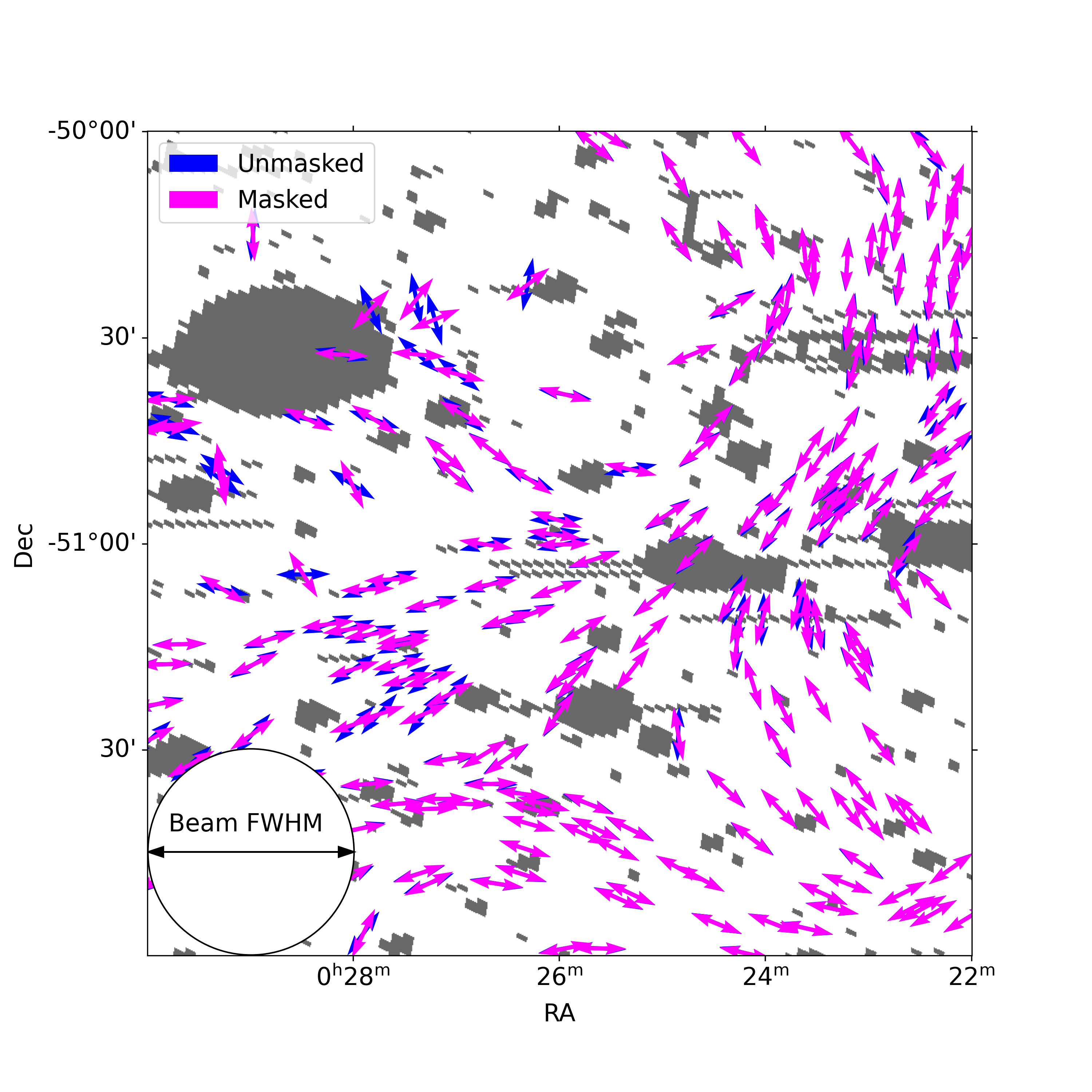}
    \caption{A test of the effects of the DES mask on determination of the LSS orientation. Each arrow is located on a massive halo in Websky and is oriented according to the surrounding smoothed halo density map. When masking the halo density map, the only orientations that are affected (going from pink to blue) are those positioned near the mask hole with a similar diameter as the FWHM of smoothing (the `beam', shown at lower-left). In most cases, the pink arrows overlap the blue well enough to hide them from view.}
    \label{fig:masking_test}
\end{figure}

We reduce the cluster sample to avoid not only the borders of the combined DES+ACT mask, but also holes in the DES mask which lie within the footprint. These holes exist at varying scales; Fig.~\ref{fig:masking_test} shows a 2$\times$2\degree region where the gray pixels represent fully-masked regions. Unphysical orientations may occur for objects near mask edges. To test this, we apply the mask to a mass-weighted overdensity map of Websky halos. We examine the orientations determined for the halos before and after masking. We find that, in general, holes smaller than the smoothing FWHM do not affect the orientation. For holes with a diameter approximately equal to the FWHM, the orientation determined for nearby points is affected. Points at a distance further than approximately one FWHM away from large holes are unaffected. 

Thus, we choose to cut clusters from the \redmapper catalog which lie within 2 degrees of the mask edge and holes with a diameter $\gtrsim$2 degrees. 2 degrees corresponds to $\sim30$~Mpc at $z=0.2$, the minimum redshift used in this work. 

To ensure that false orientation from the mask has not biased the results, we also perform oriented stacks of the mask itself as a null test. These are presented in section \ref{sec:tests}. We find that stacks of the mask have no anisotropy; in other words, the higher-order moments are consistent with zero.

\subsection{Stacking}
\subsubsection{Cutout sizes}
We fix the size of the cutouts from the secondary map ($F_2$) to correspond to 80 Mpc per side by updating the angular size of the cutouts for each 100 Mpc slice of clusters. In this way, stacks of structure in different redshift slices can be immediately compared out to 40 Mpc in radial distance. At the lowest redshift of $z=0.2$, 80 Mpc corresponds to 4.9\degree per side. Given that the clusters are cut 2 degrees within the combined DES+ACT mask edges, for this very lowest bin, there is a possibility of edge effects in the outer 0.45 degrees ($\sim7$ Mpc) of a stack in the lowest redshift bin. However, in the analysis there are no notable anomalies there, so we include it in our analysis. The procedure differs slightly from Paper I, in which we fixed the cutout sizes to 4\degree $\times4$\degree, then cut and scaled stacks from different redshifts to equal comoving size.
\subsubsection{Redshift ranges and combining multiple stacks}\label{sec:redshift_ranges}
Due to the thin cluster slicing, the stacked cutout sample in each slice is relatively small so each stack has a low SNR. To increase the SNR for analysis, we combine 5 successive stacks, i.e., 500 Mpc worth of cluster data, to create a single stack representing one broader redshift bin (lower part of Figure~\ref{fig:bins_schematic}). Each 100 Mpc stack has the same comoving size but different resolution, so we interpolate the four subsequent stacks to the pixel size of the first in that redshift bin.

We repeat this for four broad bins. To increase independence among the bins, we omit 100~Mpc of cluster data between each neighboring bin to remove correlations arising from the overlapping setup of the galaxy data for orientation. Nevertheless, the bins are unlikely to be completely independent due to the long photo-$z$ tails of the \maglim galaxy distribution.

\maglim begins at $z=0.2$, defining the lower-redshift end of our analysis. \redmapper extends out to $z\sim1$. However, to keep the number of combined stacks consistent for each redshift bin, we cut the sample slightly lower, such that each bin combines 5 stacks. The four final redshift bins used for analysis are: \begin{equation}    
z \in: \{[0.21, 0.34],[0.37, 0.51],[0.54, 0.70],[0.74, 0.92]\}.\end{equation}

\begin{figure*}
    \centering
    \includegraphics[width=\textwidth]{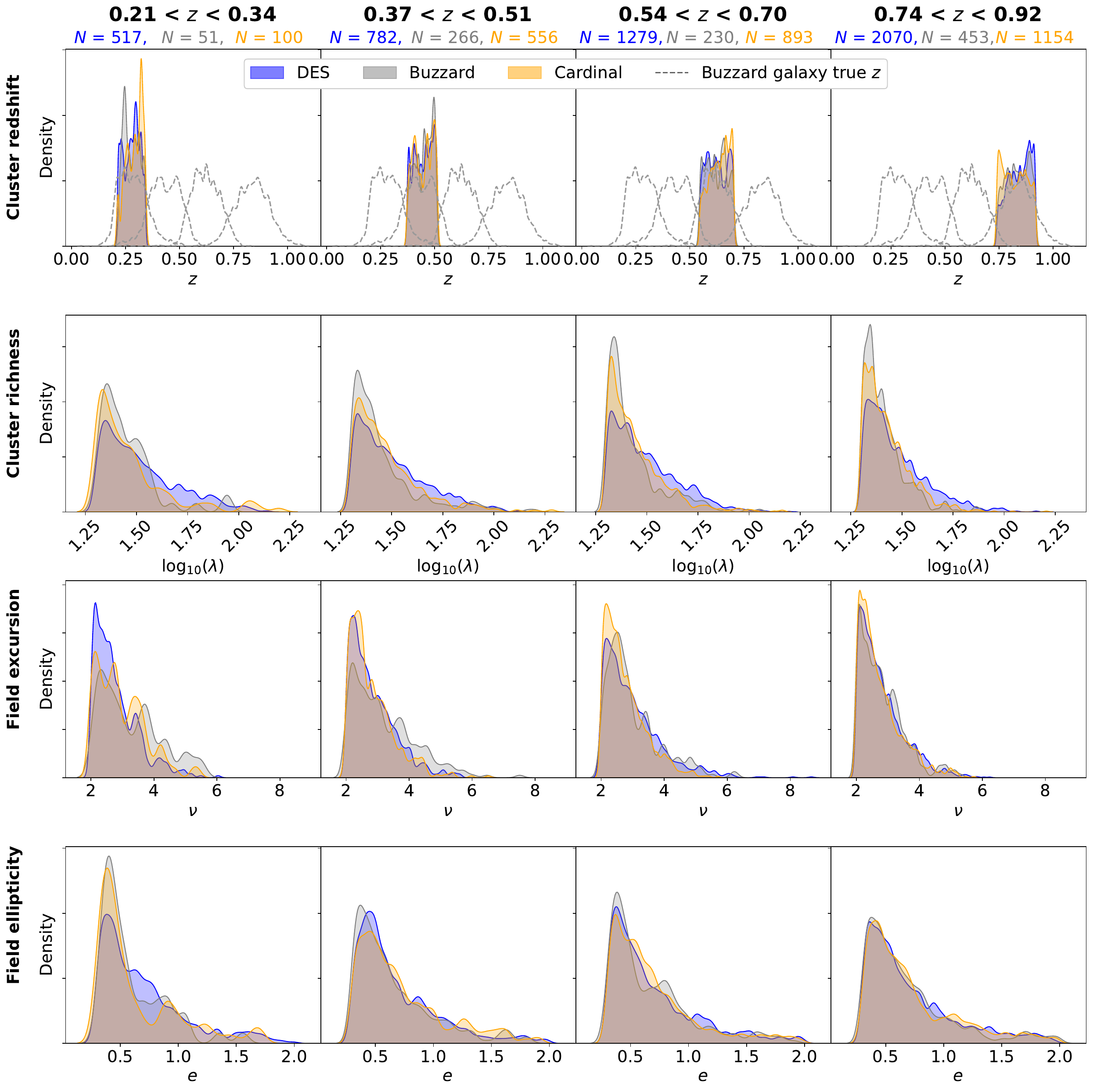}
    \caption{The kernel density estimate (KDE) for the distributions of cluster photometric redshift and richness (upper two rows), and field excursion $\nu$ and ellipticity $e$ as measured from the \maglim projected number density maps after 20~Mpc Gaussian smoothing (lower two rows). Each subplot shows DES in blue, Buzzard in gray, and Cardinal in orange. In the top row, in addition to the shaded photo-$z$ cluster distributions, the \textit{true} Buzzard \maglim redshift distributions are also shown in dotted lines. Only 0.1\% of the galaxy samples are plotted and the normalization is not to scale with the clusters. They are significantly broader than the mock cluster true-$z$s (not shown, but very similar to the cluster photo-$z$s) because of the lower galaxy photo-$z$ accuracy. The distribution of every property is normalized equally across the catalogs, hiding the stark difference in sample size; however, the numbers $N$ are shown at the top of each column. The imposed cuts of $\lambda>20$, $\nu>2$, and $e>0.3$ are smoothed slightly from the KDE. We clip the $e$ distribution at $e=2$ for the visualization as there is a long and sparse high-$e$ tail from clusters located at large-scale saddle points. Overall, the selections for all three samples are highly similar, with the most variation coming from the consistently larger high-$\lambda$ tail in DES.}
    \label{fig:multiscale_props}
\end{figure*}

The distributions of the \redmapper cluster richness, redshift, and the two large-scale field properties in these four bins are shown in Fig.~\ref{fig:multiscale_props} for DES, Buzzard, and Cardinal. The three datasets are in reasonably good agreement in the distributions. It is noteworthy how different, however, the number of (selected) clusters are between the observational data and the simulations, especially in the lowest redshift bin. The distribution for the low-redshift bin for Cardinal and Buzzard, therefore, has large Poisson noise fluctuations due to the small sample sizes. The most notable consistent difference in the samples is that DES data contains more high-richness clusters than in either Cardinal or Buzzard, a residual of the color-dependent clustering model which is not completely idealized even in Cardinal. As shown in Paper I, the large-scale $\nu$ and $e$ properties are more highly correlated than the cluster richness with the large-scale stacked signal, the focus of this analysis, and therefore the slight difference in richness distribution is not expected to significantly impact the comparison.

\subsubsection{Treatment of stack monopole} \label{subsec:treatment_stack_monopole}
The ACT+\textit{Planck} Compton-$y$ map has not explicitly been set to have a zero mean. In the outskirts of a stack, there is typically a residual background value of $\Delta y$ due to noise and contamination (e.g., from the impact of long-wavelength CMB modes in the ILC procedure); the amplitude and sign of the background varies for differently-sized stack samples. We remove this background for all stack image visuals by  subracting the average of an annulus spanning from 1/2 to 3/4 of the image. For a quantitative analysis of $m=0$ moments, we remove the average of the outer $30-40$~Mpc of the $y$ profiles (equivalent to subtracting a $30-40$~Mpc annulus from the image). By nature, the $m>0$ components of the image have zero mean and thus this subtraction is unnecessary for those profiles.

To treat all three matter tracers equivalently, we perform the same annulus subtraction on stacks of $\kappa$ and $n_\mathrm{g}$ as well. In all cases, it is possible that this removes some true structure from the outskirts of the stacks that in reality is still above the mean, especially because of the large-scale $\nu$ selection that has isolated a sample of rare large-scale overdensities. To make this explicit and caution against drawing conclusions from the overall amplitude of the  $m=0$ curves, we refer to them as `adjusted' or `adj.' for short. We do not work towards a more careful treatment of this background subtraction given that analysis of isotropic signal is not the focus of this work.

\subsection{Lensing conversion}\label{subsec:lensing_conversions}
In this section we present the approximate conversions to change our stacks of the lensing convergence---integrated over a broad kernel in redshift---to stacks that probe the surface mass density at a specific redshift of interest. The approximations and calculations herein will allow us to directly compare lensing results to $y$ and galaxy density for the same structures.

For gravitational weak lensing in a flat universe, as reviewed in \citet{Kilbinger2015}, the convergence $\kappa$ generally is
\begin{equation} \label{eq:kappa}
    \kappa(\boldsymbol{\hat{n}}) = \frac{3 H_0^2 \Omega_{\mathrm{M}}}{2c^2} \int_{0}^{\chi_{\mathrm{lim}}} \frac{\delta(\chi, \boldsymbol{\hat{n}}) }{a(\chi)} q(\chi) \chi \dv{\chi}. 
\end{equation}
It is a weighted projection of the intervening matter overdensities, $\delta$, defined as $(\rho-\overline{\rho})/\overline{\rho}$, where the weighting is a smooth function of comoving distance $\chi$ that depends on the source galaxy distribution, which extends to $\chi_{\mathrm{lim}}$. $q(\chi)$, the lens efficiency at a given comoving distance $\chi$, is:
\begin{equation}
    q(\chi) = \int_\chi^{\chi_{\mathrm{lim}}}  n(\chi')\frac{\chi'-\chi}{\chi'} \dv{\chi'},
\end{equation}
where $n(\chi')$ is the source distribution.

\begin{figure}
    \centering
    \includegraphics[width=\columnwidth]{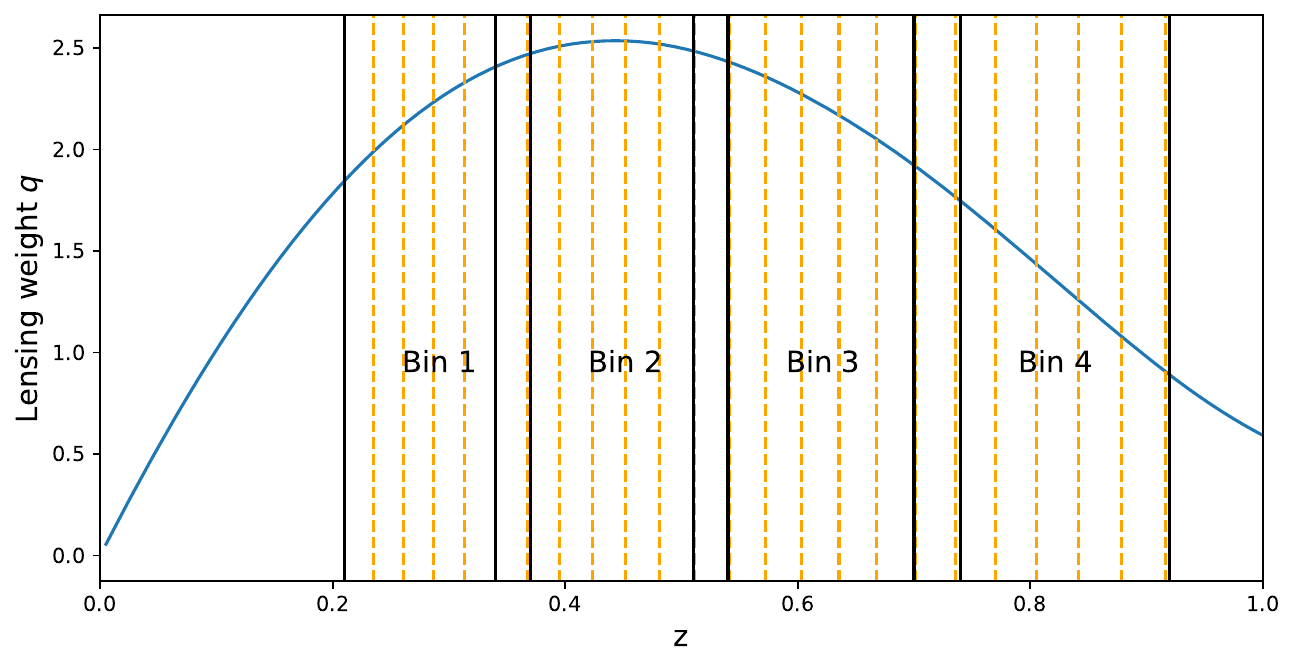}
    \caption{The lensing weight $q(z)$ for the DES Y3 $\kappa$ map created using the highest-redshift source bin, with photo-$z$s within $z=[0.87, 2.0]$ \citep{Myles2021MNRAS.505.4249M}. The blue curve shows the kernel, which peaks at $z\sim0.45$. The orange lines show the widths of our cluster slices, over which we approximate the kernel to be constant for the purposes of Eq.~\ref{eq:kappa_stack}. The black lines show the bin edges for the final stacks we will use for analysis of redshift-dependence -- we combine all stacks within the black bins.}
    \label{fig:lensing_kernel}
\end{figure}

For our purposes, Eq.~\ref{eq:kappa} can be divided into two parts, one correlated with the underlying structure of interest and the uncorrelated part. Given our restriction of the galaxy data into a narrow redshift slice corresponding to 200~Mpc, we assume that the component of $\kappa$ correlated with the constrained cluster at location $\boldsymbol{\hat{n}}_i$ spans a line-of-sight extent $\chi_1<\chi< \chi_2$, where $\chi_1$ and $\chi_2$ are within $\sim$100 Mpc of the center of the comoving distance bin to which that cluster belongs. Then the foreground integral $\int_{0}^{\chi_1}\delta(r,\theta)|\boldsymbol{\hat{n}}_i\, q(\chi)\chi/a \dv{\chi}$ can be considered a noise term, which we will call $\tilde{N_\kappa}$, as can the similar background integral from $\chi_2$ to $\chi_{\rm{lim}}$. This is because the integrand itself has mean-zero at any $\chi$, being uncorrelated with the stack locations, and the sum of mean-zero variables is itself also a mean-zero variable. The $\kappa$ image cutout around that cluster, oriented in the coordinate system defined by the eigenvectors of the galaxy field, is
\begin{multline}
    \langle \kappa | \boldsymbol{\hat{n}}_i \rangle (r, \theta) = \\ \int_{\chi_1}^{\chi_2} \frac{3 H_0^2 \Omega_{\mathrm{M}}}{2c^2} \frac{\big[\rho(r,\theta,\chi)|\boldsymbol{\hat{n}}_i-\overline{\rho}(\chi)\big]}{\overline{\rho}(\chi)} \frac{q(\chi)\,\chi}{a(\chi)}\,\dv{\chi} +\tilde{N}_{\rm{\kappa}} \\ 
    =  \int_{\chi_1}^{\chi_2} \frac{4\pi G}{c^2} \big[\rho(r,\theta,\chi)|\boldsymbol{\hat{n}}_i-\overline{\rho}(\chi)\big] q(\chi)\,\chi\, a^2(\chi)\, \dv{\chi} +\tilde{N}_{\rm{\kappa}} \\ =\int_{\ell({\chi_1})}^{\ell({\chi_2})} \frac{4\pi G}{c^2} \Delta\rho(r,\theta,\ell)|\boldsymbol{\hat{n}}_i \, q(\ell)\, \ell \, \dv{\ell} +\tilde{N}_{\rm{\kappa}} 
\end{multline}
where $\Omega_M=\rho_\mathrm{M,0}/\rho_\mathrm{cr,0}$ and in the second step we have rearranged by using the definition of the critical density, $\rho_\mathrm{cr,0}=\frac{3H_0^2}{8\pi G}$ and the fact that the average physical matter density at $\chi$ is $\overline{\rho}(\chi)=\rho_\mathrm{M,0}/a^3$. The last line changes the integration coordinates from comoving to proper distance, $\ell$. The \textit{stacked} image of a cluster slice selected under the set of constraints $\xi$, oriented by surrounding galaxies with both slices centered at  $\ell_{\rm{c}}$, is
\begin{multline} \label{eq:kappa_stack}
    \langle \kappa | \xi \rangle (r, \theta, \ell_{\rm{c}}) \approx 
    \frac{4\pi G}{c^2} \langle\Sigma_m(r,\theta,\ell_{\rm{c}})|\xi\rangle \, q(\ell_{\rm{c}})\, \ell_{\rm{c}} + \tilde{N}_{\rm{\kappa}},
\end{multline} 
where $\Sigma(\boldsymbol{\hat{n}}) = \int  \Delta \rho(\boldsymbol{\hat{n}},\ell)\,\dv{\ell}$ is the surface mass density. This makes the approximation that averaging over the integral of many structures in the distance slice is the same as integrating over the extent of a representative average structure which is located at the center of the slice. With the understanding that $q(\ell)$ and $\ell$ are a slowly-varying functions of distance compared to $\Delta \rho(\ell)$, we have also made the approximation that they can be pulled out of the integral as constants (their values at the bin center). To support this assumption for $q$, Fig.~\ref{fig:lensing_kernel} shows how the lensing kernel for the DES Y3 $\kappa$ map varies throughout our four broad redshift bins. In each narrow cluster slice of width $\Delta{\chi}=100$~Mpc (orange lines) and surrounding 200~Mpc galaxy slice, it is reasonable to approximate the lensing weight $q$ by a constant.

Rearranging Eq.~\ref{eq:kappa_stack}, what we actually seek to measure is
\begin{equation}
    \langle \Sigma | \xi \rangle (\ell_\mathrm{c}) = \langle \kappa | \xi \rangle (\ell_\mathrm{c}) \frac{c^2}{4\pi G} \frac{1}{\ell_{\rm{c}}\, q(\ell_{\rm{c}})},
\end{equation}
the average surface mass density (as a function of radial distance and polar angle, suppressed) given the constraints $\xi$ weighted by the lens efficiency and proper distance of structure.

As the lensing kernel is low in bin 4 (Fig.~\ref{fig:lensing_kernel}) due to the overlap between sources and lenses, the results therein are noisy, and therefore we do not present results from this bin in our final analyses which involve $\kappa$.

\section{Uncertainties and binning} \label{sec:uncertainties}
\subsection{Uncertainties}
For each tracer (galaxies, lensing, and $y$), we implement the same sub-sampling method to calculate uncertainties as was applied to the Buzzard mock data in Paper I. We split the cluster data into 24 approximately equal-sized regions on the sky using the \texttt{kmeans\_radec} algorithm\footnote{https://github.com/esheldon/kmeans\_radec}. Each region covers approximately 160 square degrees, much larger than the 24 square degree size of the largest-sized cutouts (at $z=0.2$). Since many cutouts fit within one region, a stack in one region is mostly independent from its neighbor. The motivation for such spatial splits is to get independent estimates of the sample variance. The largest contributor to this is the variance from coherent large-scale structures which influence the orientation over 20 Mpc scales. Spatial splits have the added benefit of sampling regions which may have varying instrumental noise, survey systematics, or other sources of contamination; e.g., in the $y$ map the spatial splits sample different amplitudes of the long-wavelength CMB contamination.

The number of clusters in each region varies slightly. The total stack is an unweighted average of all clusters (Eq.~\ref{eq:simple_stack}),
so to get the equivalent signal with a combination of separate stacks in $N_\mathrm{reg}$ regions, each stack $\langle F_{2,p} |\xi \rangle$ being an average over all $N_{\mathrm{cl},p}$ clusters satisfying the constraints $\xi$ in the $p^\mathrm{{th}}$ region, each stack must be weighted by $N_{\mathrm{cl},p}$:
\begin{equation}
    \langle F_2 | \xi \rangle  = \frac{\sum_{p=1}^{N_{\mathrm{reg}}} \langle F_{2,p} | \xi \rangle N_{\mathrm{cl},p}}{N_{\mathrm{cl,tot}}}
\end{equation}
Consequently, in calculating the covariance matrix using these samples, we weight each sample by $w_p = N_{\mathrm{cl},p}/N_\mathrm{cl,tot}$.

For each of the $N_\mathrm{reg}$ stacks, we measure a data vector $\boldsymbol{C}_m=C_m(r_i)$ representing the multipole profile divided into radial bins $r_i$, where $i\in(1,N_{\mathrm{bin}})$. We combine the $\boldsymbol{C}_m$ from all regions into a matrix $X$ of size $N_\mathrm{bin}\times N_\mathrm{reg}$. We subtract the weighted average (with weights $w_N$) of each bin across all regions (i.e., the row mean), creating the modified matrix $M$:
\begin{equation} \label{eq:mean_subtracted_matrix}
    M_{ip} = X_{ip}-\frac{\sum_{p'=1}^{N_{\mathrm{reg}}}X_{ip'}
    w_{p'}}{\sum_{p'=1}^{N_{\mathrm{reg}}} w_{p'}}.
\end{equation}
Finally, the covariance matrix element between two bins $r_i$ and $r_j$ is
\begin{equation} \label{eq:covmat_regions}
    \Sigma_{ij} = \left[ \frac{ \sum_{p'=1}^{N_{\mathrm{reg}}}w_{p'}}{(\sum_{p'=1}^{N_{\mathrm{reg}}}w_{p'})^2-\sum_{p'=1}^{N_{\mathrm{reg}}}w_{p'}^2}\right] \sum_{p'=1}^{N_{\mathrm{reg}}} (w_{p'} M_{ip'} M_{jp'}).
\end{equation}

We apply this method equivalently to Buzzard and Cardinal as to ACT$\times$DES. Buzzard and Cardinal generally have far fewer clusters at any given richness level, and thus there are fewer stacking locations in our simulated samples; the errorbars therefore tend to be larger than those for ACT$\times$DES despite the absence of instrumental noise in the simulation. We \textit{consider} rescaling the simulated errorbars by $\sqrt{{N_{\mathrm{sim}}/N_{\mathrm{DES}}}}$ to estimate the level of errors if the simulated cluster sample were to be the same size, assuming random noise. However, this scaling would artificially misrepresent the Buzzard or Cardinal results as having smaller cosmic variance than they truly do, and therefore we do not perform this rescaling.

In addition, the ACT $y$ map has spatially-varying noise. We test whether it is important to incorporate this in the weights, such that the total weight per region would be $w_{\mathrm{s},p}\times w_p$ where $w_{\mathrm{s},p}$ represents the spatially-varying noise in the $p^\mathrm{th}$ region. To calculate each $w_\mathrm{s}$, we smooth the $y$ map with a top-hat beam of angular radius $\theta_\mathrm{TH}$. We create several maps with different $\theta_\mathrm{TH}$ filters that represent the angular size corresponding to the constant-comoving-size binning of 7.5~Mpc used for the final analysis. $\theta_\mathrm{TH}$ is smallest for the highest-redshift stacks and largest for the lowest redshifts. For each smoothed map, we calculate the variance $\sigma^2_p | \theta_\mathrm{TH}$ in each region, approximating the \texttt{kmeans\_radec} regions as rectangular. This is an appropriate way to estimate the noise in each region because the $y$ map is noise-dominated. For each filter we normalize by the variance $\sigma^2_{\mathrm{map}}|\theta_\mathrm{TH}$ of the entire map, and finally set $w_{\mathrm{s},p} = \langle ( \sigma^2_{\mathrm{map}} | \theta_\mathrm{TH} )/( \sigma^2_{p} | \theta_\mathrm{TH}) \rangle$ averaged over all $\theta_\mathrm{TH}$. However, we find that $w_\mathrm{s}$ has negligible impact (few-percent-level) on the total signal and the errors, and hereafter use only $w_p$ for weighting.
\subsection{Binning}
We choose to radially bin the data in five bins of 7.5 Mpc each. This binning scheme divides evenly into the 40 Mpc extent of each profile and places the second bin near the peak of the profile in most higher-order moments, capturing the rise and fall of those profiles. Finer binning of the Compton $y$ data vectors results in mean profiles which fluctuate rapidly, obscuring the underlying smoother trend. The relatively small sample size (24) from which we calculate all covariances and uncertainties also motivates a small number of bins.

We note that the signal-to-noise ratios calculated for the complete data vectors in the results section (Sec.~\ref{sec:res_indiv}) have some dependency on bin size and placement. We therefore clearly state how each signal-to-noise ratio is calculated and emphasize that it is always done for five bins. The following section will present several null tests and other tests of robustness; these are all conducted with the same five bins for consistency. To avoid bin dependency, a more optimal approach would be to fit a model to each data vector and calculate all statistics for the continuous model. However, this is beyond the scope of the current work.

\section{Tests of robustness}\label{sec:tests}
In this section we perform several tests to verify the validity of our covariance estimation, the impact of the CIB, and search for survey-dependent orientation effects. For all tests, we calculate a $p$-value and reject the null hypothesis if $p<0.05$.

\textbf{Gaussian uncertainties:}
Our method assumes that the stack result from each map sub-region is sampled from an underlying distribution which is consistent across the entire map. If this is true, the distribution of results across all regions should be Gaussian-distributed around the mean (i.e., around the full-map stack), with the standard deviation derived from the variance of the signal across the per-region stacks. However, in the case of either DES or ACT, some of the spatially distinct regions may be impacted by unique survey effects (anisotropic noise, masking, etc.), which may cause the errors to be non-Gaussian.

\begin{figure*}
\begin{tabular}{c}
    \includegraphics[width=0.9\textwidth, trim={2.6cm 1cm 6.3cm 1.7cm }, clip]{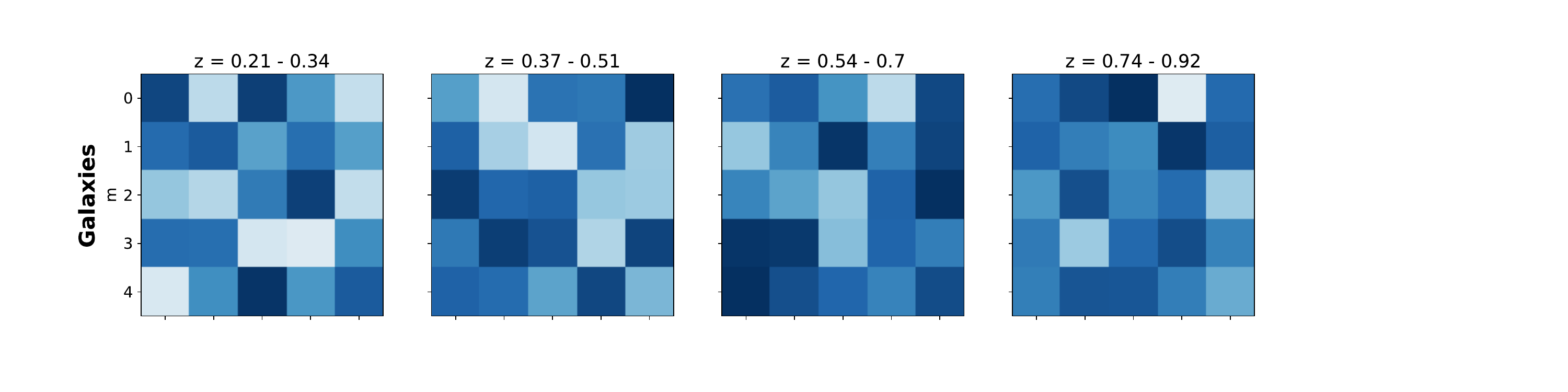}  \\
     \vspace{0.0mm}
     \includegraphics[width=0.9\textwidth, trim={2.6cm 1cm 6.3cm 1.7cm }, clip]{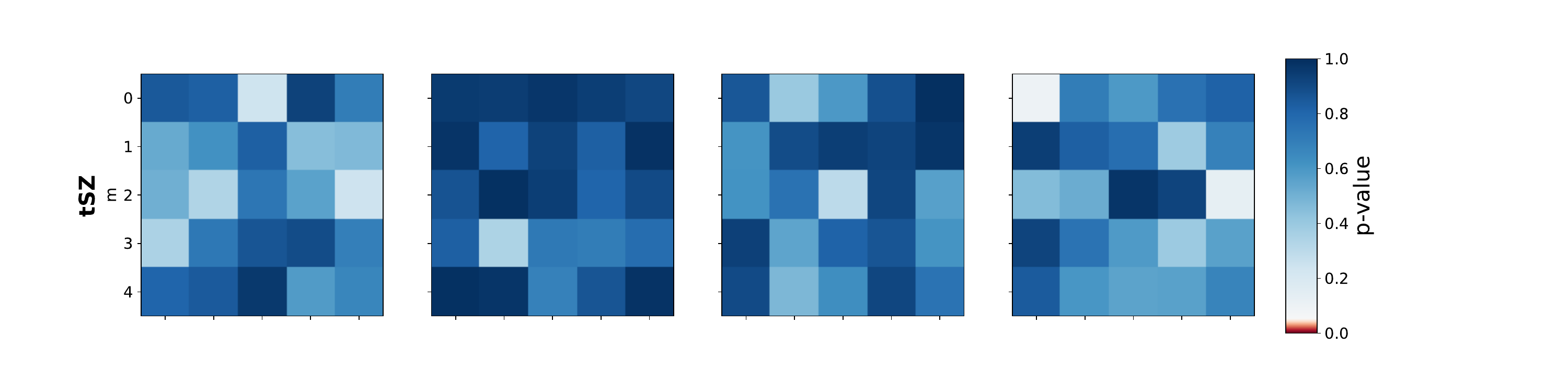} \\
    \vspace{0.0mm}
    \includegraphics[width=0.9\textwidth, trim={2.6cm 1cm 6.3cm 1.7cm }, clip]{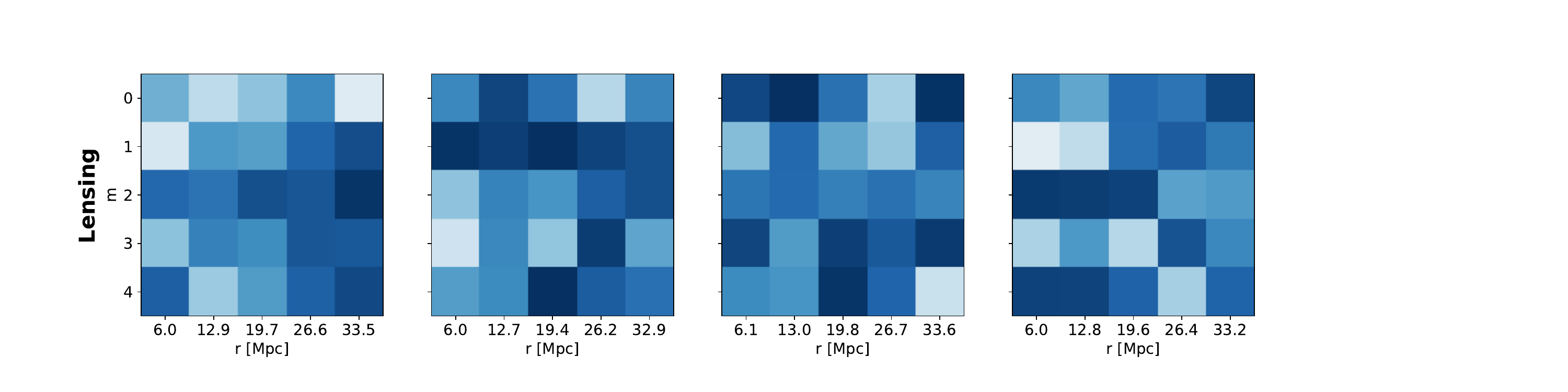}
\end{tabular}
  \caption{The Gaussianity KS-test $p$-value results for each $m$ value (rows) and each radial bin in the profile (columns) for each successive redshift bin. The results for galaxies, tSZ, and lensing are presented from top to bottom. All tests pass, as indicated by the absence of red squares, demonstrating that the results are consistent with being from Gaussian distributions.}
     \label{fig:ks-tests-gaussian}
\end{figure*}

We test whether the distribution of data vectors across the regions deviates from a Gaussian distribution using the following prescription. For each redshift bin, we examine each multipole profile of the galaxy density stack in individual radial bins. Each bin contains $N_{\rm{reg}}$ samples. Using the one-sample Kolmogorov-Smirnov (KS) test, we compare the samples to the continuous Gaussian distribution with that bin's mean and standard deviation. If $p<0.05$, we reject the null hypothesis that the samples are indeed drawn from a Gaussian distribution; otherwise, the test passes. We repeat this test with stacks of the $y$ map and $\kappa$ map. The results of these tests are presented in Figure~\ref{fig:ks-tests-gaussian}. The threshold defined by the red color in that figure is $p=0.05$ (representing a failure); the lack of red indicates that no individual bin is significantly inconsistent with a Gaussian distribution, as the lowest p-value we find is 0.17. The p-value distribution across all tests skews somewhat high, indicating that the data is even more Gaussian than expected from random chance; visual comparison with randomly-generated Gaussian data confirms this. This may be a consequence of binning, and indicate that the binning is wider than necessary; however, further exploration of bin size variations is beyond the scope of this work.

\begin{figure}
    \centering
    \includegraphics[width=\columnwidth]{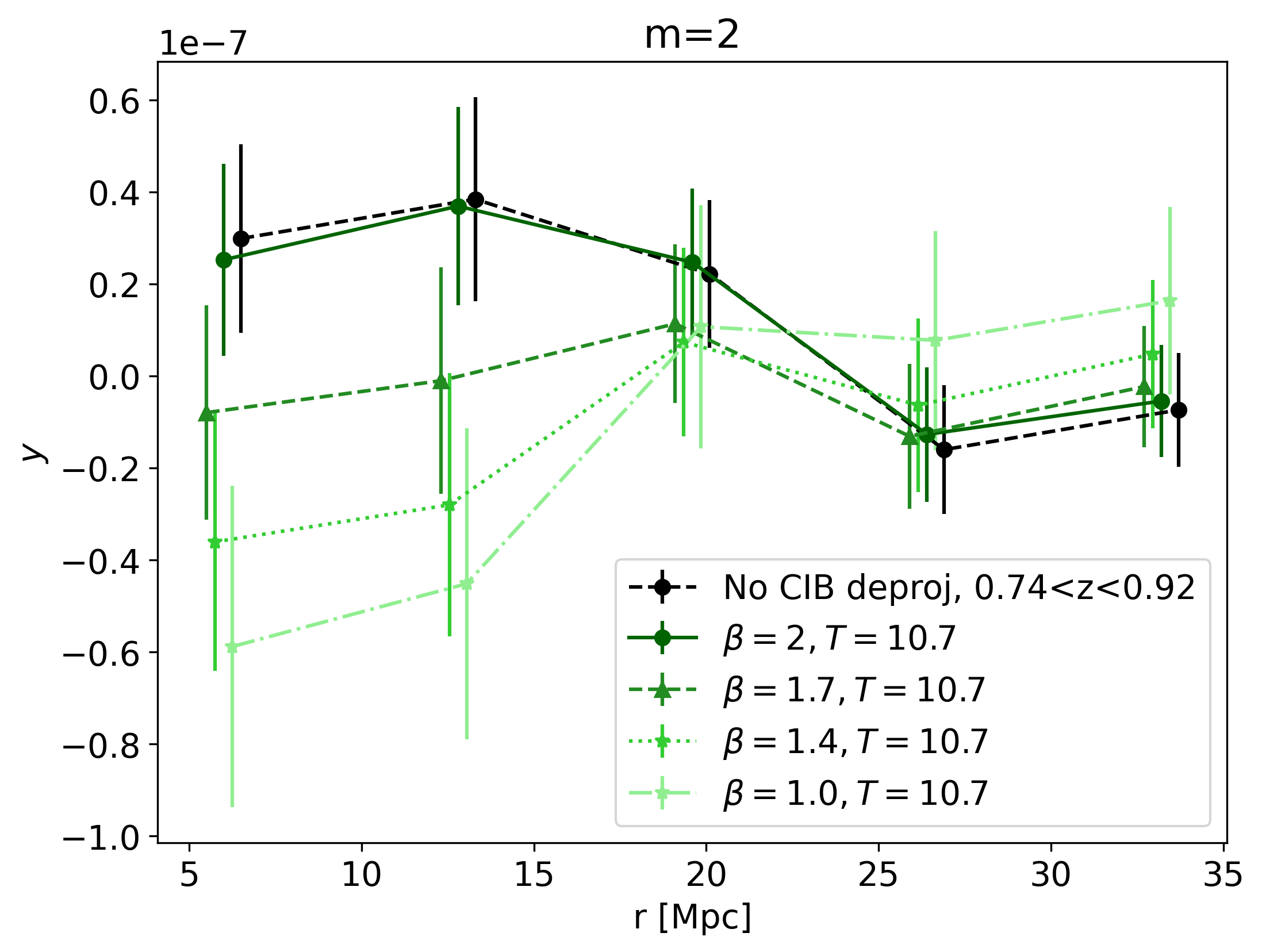}
    \caption{The impact of different CIB deprojections with fixed $T$ and varying $\beta$ ($\boldsymbol{C}^{y,\mathrm{{deCIB}}}_{2}$), shown as green binned profiles (connected for visual aid) compared to the black profile from the map with nothing deprojected ($\boldsymbol{C}^{y}_{2}$). These data represent the most extreme impacts of CIB deprojection, as seen in the cos(2$\theta$) moment of the highest $z$ bin, $0.74<z<0.92$.}
    \label{fig:CIB-dependence-bin4}
\end{figure}

\textbf{CIB deprojection:} Next, we examine the differences between the Compton-$y$ results using CIB-deprojected maps and the fiducial ILC map (with no components deprojected). The CIB-deprojected maps created for ACT DR6 aim to deproject a simple two-parameter model for the CIB frequency dependence that includes parameters $\beta$ and $T_{\rm{CIB}}^{\rm{eff}}$ \citep[see][for details]{Coulton2023arXiv230701258C}. The available maps sample the two parameters along and somewhat beyond the range given by the highly degenerate constraints placed by \citet{McCarthy2024PhRvD.109b3528M} using \textit{Planck} data (their Figure 3). The deprojection comes at a cost of increased noise. To test whether the deprojection procedure impacts the signal in a manner dependent on the chosen parameter values, we run our stacking procedure on four maps with a fixed $T_{\rm{CIB}}$ but different values of $\beta$. If the impact from the deprojection were $\beta$-independent, we would expect the multipole profiles from all CIB-deprojected stacks to scatter randomly around a shared mean. Instead, we find a correlation between the $\beta$ parameter and the observed $y$ signal, which is (often, but not always) monotonic for a single radial bin of a multipole profile at a given redshift. For example, Figure~\ref{fig:CIB-dependence-bin4} displays the quadrupole moment in the highest redshift bin, which is most impacted; the $y$ signal decreases as $\beta$ decreases for $T_{\rm{CIB}}=10.7$~K. The largest difference is a 2.5-fold depletion in $y$ signal for the first bin in the light-green dot-dashed curve ($\beta=1.0$).

This monotonic dependence on $\beta$ suggest that there is true CIB impact, yet it is unclear which---if any---of the parameter choices produces the correct deprojection. A similar effect was found in preliminary results from ACT$\times$unWISE (Kusiak et al, in prep) shown in \citet[][their Fig. 16]{Coulton2023arXiv230701258C}. In the discussion therein, the authors demonstrate that maps with an additional term deprojected---specifically, the derivative of the CIB frequency-dependence equation with respect to $\beta$ or $T_{\rm{CIB}}$---is more stable to the choice of parameters. However, as that approach increases the noise even further due to the additional deprojection, we instead elect to report our lower-noise results from the non-deprojected map and, when reporting results, provide context for which moments and redshifts may be severely biased.

To gain insight into this, we consider whether the binned radial profile for each $\cos(m)$ moment of each CIB-deprojected map ($\boldsymbol{C}^{y,\mathrm{{deCIB}}}_{m}$) is \textit{significantly} in tension with that of the fiducial map ($\boldsymbol{C}^y_{m}$). Our null hypothesis is that they are consistent with each other, so we compute the $\chi^2$ statistic between the two:
\begin{equation}
    \chi^2_m = (\boldsymbol{C}^{y,\mathrm{{deCIB}}}_{m} - \boldsymbol{C}^y_{m})^\mathrm{T} \boldsymbol{\Sigma}^{-1}_{\rm{comb},m}(\boldsymbol{C}^{y,\mathrm{{deCIB}}}_{m}-\boldsymbol{C}^y_{m})\big.
\end{equation}
In a case in which the two stacks were completely uncorrelated, $\boldsymbol{\Sigma}_{\rm{comb}}=\boldsymbol{\Sigma}_m+\boldsymbol{\Sigma}_{m}^\mathrm{deCIB}$, the sum of the two covariance matrices. Assuming this minimal correlation limit, we determine the probability that a $\chi^2$ as extreme as that of the CIB-deprojected signal would be observed if the null hypothesis were true, i.e. the $p$-value (or PTE, probability-to-exceed).
We find that the highest redshift bin is impacted the most by any of the CIB deprojections (not shown in any figure), and that deprojecting the model with $(\beta=1.0,\; T_\mathrm{CIB}=10.7$~K) yields results that deviate the most from fiducial. Under the uncorrelated assumption, the impacts are only significant ($p<0.05$) in a single moment of the highest redshift bin for that map. However, in reality the uncorrelated assumption is untrue: the fluctuations about the mean from true $y$ signal are the same in the two stacks, but the amount of contamination and of atmospheric and instrumental noise varies due to the distinct ILC weights. The Pearson correlation coefficient between the two datavectors ranges from $\sim0.8-0.999$, indicating a high degree of correlation. This means that the CIB contamination is likely far more significant than the lower limit indicates.

In summary, there is evidence that the CIB causes significant impacts in the multipole profiles of the stacked signals which depend on $m$ and redshift bin. The highest $z$ bin is impacted worse than the others for most $m$; besides that, there does not appear to be a trend in $m$ or $z$. Clearly, advancement in observations and theory are needed to better constrain the CIB. In this work, given that it is unknown which deprojected model (if any) properly removes the CIB contamination, and due to the large noise penalties of deprojecting, we elect to report the stacked $y$ values from the fiducial map without CIB deprojection and with statistical uncertainties only. To give the reader insight into which results should be considered to have additional systematic uncertainties and/or bias, we include in the main $y$ result figure (introduced in Sec.~\ref{sec:res_multiple}) profiles from the most impactful model ($\beta=1.0,\; T_\mathrm{CIB}=10.7$~K). This provides a pessimistic estimate into the possible magnitude of the systematic bias, and also shows the direction of the possible bias as a function of radial moment, bin, and redshift.

\textbf{Mask effects:} Next, we test for impacts of the DES mask on orientation, as introduced in Sec.~\ref{subsec:mask_holes}. Having excluded clusters near masked regions, orientations should not be biased by the mask shape (in other words, by the shape of the footprint or by holes in the overdensity map where data has been masked). If the sub-selection of clusters has been successful, an oriented stack of the mask itself, used as a binary map, should be isotropic. We thus stack cutouts of the DES Y3 galaxy sample mask, converted to binary format, centered on locations of our constrained cluster sample and test whether the higher-order moments of the stack are consistent with zero. 

For each stack of the mask on clusters from each redshift bin, we calculate the multipole profiles for every $m$ and determine the $\chi^2$ with respect to a null vector ($\chi^2_{\rm{true}}$). We compute the $p$-value from the $\chi^2$ CDF to determine whether the $\chi^2_{\rm{true}}$ value is extreme with respect to the distribution from the null hypothesis.

\begin{figure}
    \centering
    \includegraphics[width=.7\columnwidth]{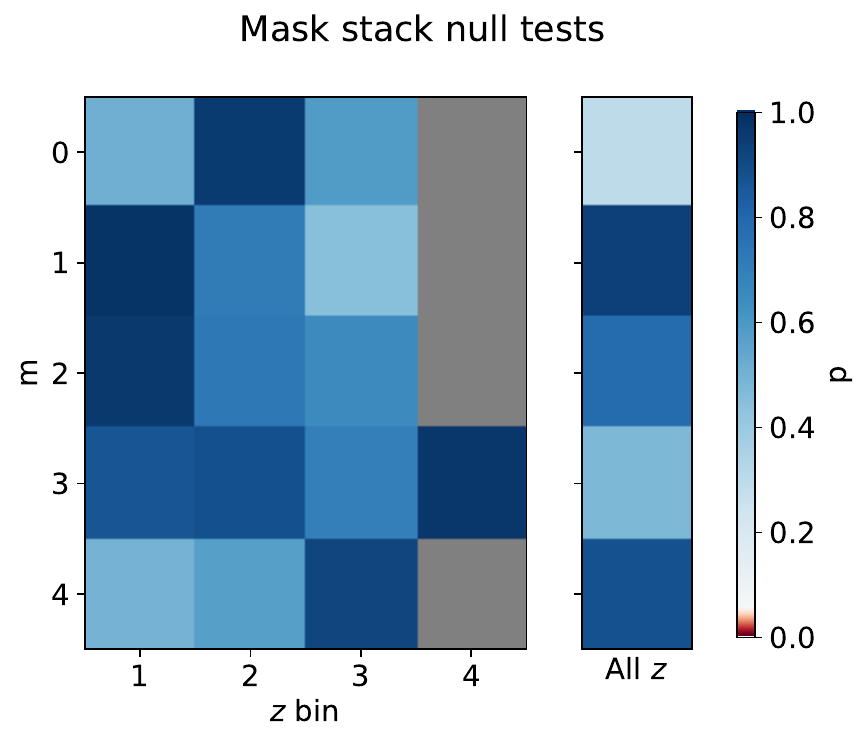}\\
    \caption{The null test results for each $m$ profile (rows) and each cluster redshift bin (columns) for stacks of the DES mask. The final column shows a combined stack using the full sample of constrained cluster locations at all redshifts. Red colors with $p<0.05$ would represent a failure, but with the minimum $p$-value being 0.3 (light blue, upper-right), all tests pass. Grey squares represent cases in which the covariance matrix could not be inverted to compute the $\chi^2$. The success of these null tests indicates that the shape of the survey footprint, including many small holes where data were masked, does not impact our results given the careful treatment done in Sec.~\ref{subsec:mask_holes}.}
    \label{fig:null-test-mask}
\end{figure}

The mask test results are shown in Fig~\ref{fig:null-test-mask}. For three of the four moments of stacks on cluster locations from the highest redshift bin, the covariance matrix is not positive-definite so a $\chi^2$ cannot be calculated; we represent those with grey squares. As shown visually by the blue colors, the profiles for all moments for which the test can be conducted are consistent with null, passing the test. Furthermore, when performing the stack for the entire sample of supercluster centers, regardless of redshift (at right), all moments pass the test. In particular, the passing of the tests in the $m>0$ moments which are sensitive to orientation demonstrates that the avoidance of mask holes described in Sec.~\ref{subsec:mask_holes} was successful.

\begin{figure*}[htbp!]
\includegraphics[width=.7\textwidth, trim={0 0 0 1cm},clip]{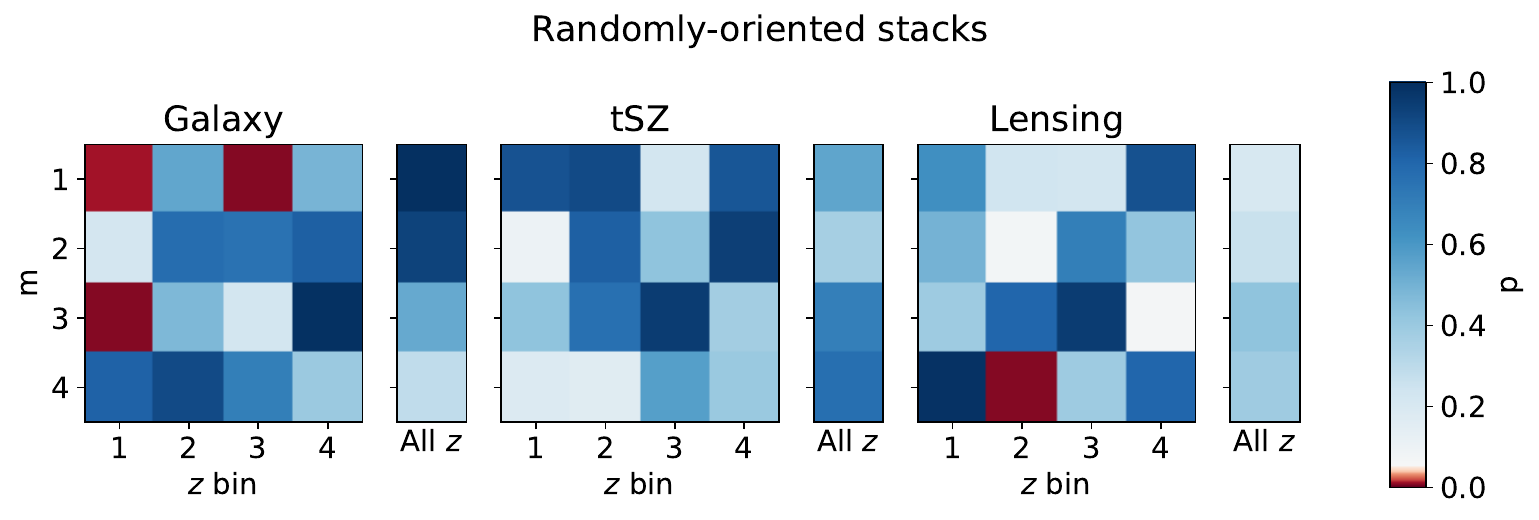}
\includegraphics[width=.3\textwidth]{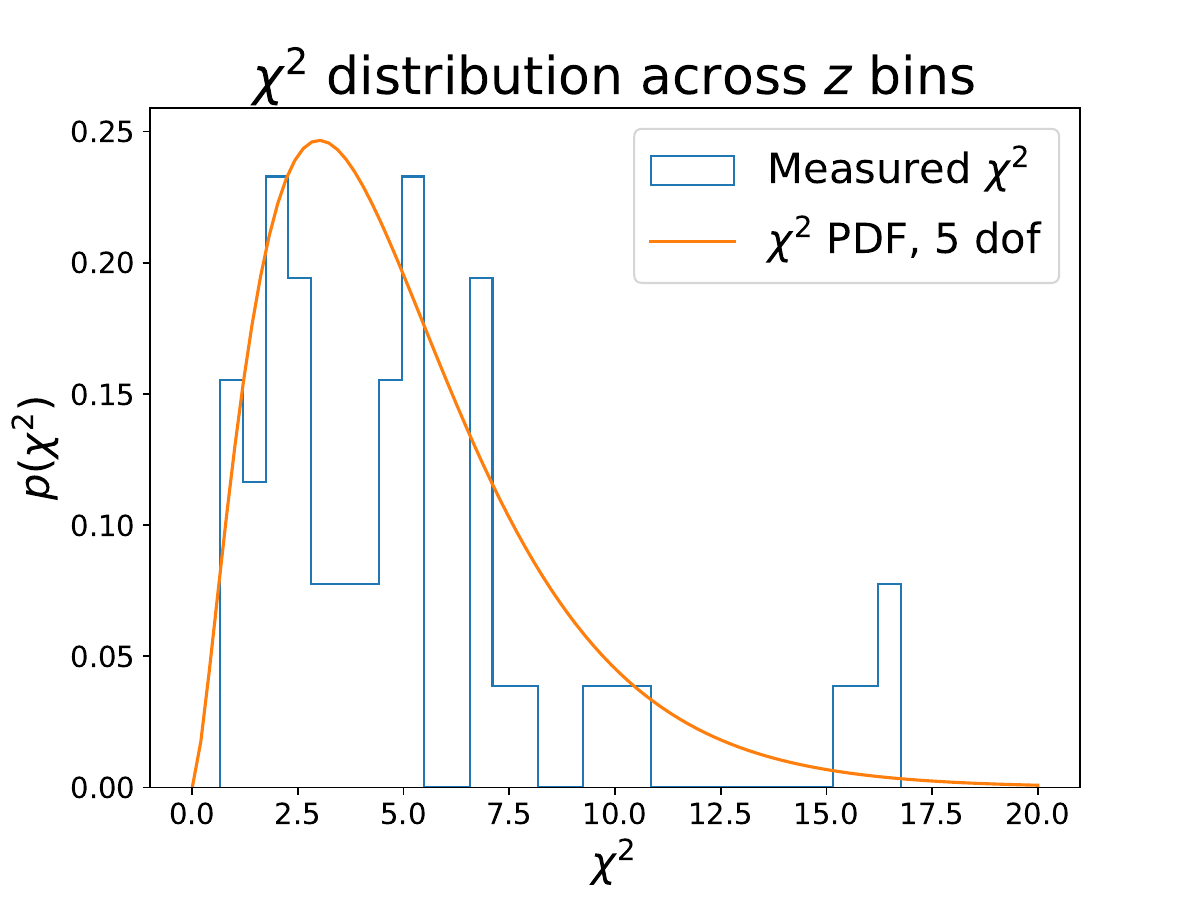}
    \caption{Null tests for higher-order moments of randomly-oriented stacks. A randomly-oriented stack has the same (non-null) $m=0$ signal as an oriented stack, but in the limit of stacking infinitely many cutouts is expected to be isotropic and thus have null signal for $m>0$. Red colors indicate that there is significant evidence ($p<0.05$) for the $m$ profile in the particular $z$ bin to be non-null, which occurs for several cases. However, when the stacks are combined across bins (``all $z$" columns), there is no evidence for anisotropic signal, fulfilling the expectation of increasing isotropy with more stacked cutouts. The $\chi^2$ distribution of the individual bin case is roughly consistent with the expected (right), and the PTE distribution is consistent with uniform (not shown).}
    \label{fig:rand_orientation_null}
\end{figure*}
\textbf{Random orientations:} Next, we test whether stacks in which each cutout is given a random orientation, i.e., with no association to large-scale structure, result in a signal that is consistent with null in all higher-order moments. This serves as a further test of the covariance matrix estimation: if the uncertainties were under-estimated, the small deviations from radial symmetry caused by random noise in a stack of a discrete number of images could masquerade as a signal. The results are shown in Figure~\ref{fig:rand_orientation_null}. Several moments fail ($p<0.05$) for several redshift bins. We run the test again with different random angles (not shown), finding that the failing cases occur for different bins and moments, indicating their random nature. Furthermore, when the stacks are combined across redshifts, there are no failures: the random anisotropies in individual bins fall in different directions and cancel when combined. This suggests that the cause of failures is the relatively small number of stacked cutouts in individual redshift bins (e.g., 500 in bin 1). To demonstrate that the opposite occurs when performing oriented stacking for truly correlated LSS, we show all-$z$-combined stack results in Sec.~\ref{sec:res_indiv}, finding in general that combining redshift bins enhances rather than diminishes the anisotropy as expected.

Nevertheless, it is important to note that the test suggests that significant residual anisotropies occur $\sim8\%$ of the time in the divided redshift bins. The distribution of $\chi^2$ is shown at right, and we note that the bump around 16 would register as a significance of $4\sigma$ in the rough signal-to-noise estimates we will present in Sec.~\ref{sec:res_indiv}. To distinguish results from true correlations in individual redshift bins, we will use 4.5$\sigma$ as a significance threshold.

\textbf{Non-physical structure:} Finally, we perform several approximate tests to search for systematic/non-physical extended coherent structure in the DES galaxy data which could bias orientations. For example, there is some excess in the magnitude limit of the DES Y3 survey along curves of constant R.A. (\citet{SevillaNoarbe2021ApJS..254...24S}, their Fig E.3). The  systematic weights developed for \maglim are intended to correct for such effects, and in fact \citet{RodriguezMonroy2022MNRAS.511.2665R, Porredon2021_cosmoconstraints} used various tests to demonstrate that the weights properly correct for dependence of the galaxy number density on survey properties. Nevertheless, for robustness especially given the unique sensitivity of our analysis to any form of local anisotropy, we proceed with our own test.

If any survey-dependent effects were to have propagated into the \maglim~catalog, they could masquerade as true large-scale structure and dominate the orientations, leading to false signal in the galaxy density stacks and potentially lensing stacks (but not the $y$-stacks as ACT has entirely different survey systematics). If this were the case, we would expect certain spatial regions to be associated with an excess of orientation in a special direction, and the distribution of orientation angles would therefore deviate from a uniform distribution.

We first examine the distribution of orientation angles at locations of the cluster sample, finding no evidence to reject the null hypothesis of a uniform distribution by using the KS test ($p=0.4$). Next, we divide the DES footprint into four approximately-equal sized quadrants, finding the same result for all ($p=0.7, 0.1, 0.4, 0.8$). We examine strips of constant-declination at the top and bottom of the footprint, again finding consistency with uniformity $(p=0.6, 0.97)$. Finally, we test strips of constant-R.A. around R.A.=5\degree, R.A.=30\degree, R.A.=60\degree \citep[those most visible in Figure E.3 of][]{SevillaNoarbe2021ApJS..254...24S} which are also consistent with uniform $p=(0.2, 0.6, 0.7$). In all the above tests, the parity transformation is also consistent with either direction being equally likely. Combined with the visual inspections of orientations near masked regions as described in Section~\ref{subsec:mask_holes}, we find no evidence to suggest survey effects are impacting the orientations determined using DES data. 

\section{Results for individual tracers}\label{sec:res_indiv}

\subsection{Comparisons with Paper I}
\begin{figure*}
    \centering
    \includegraphics[width=\textwidth]{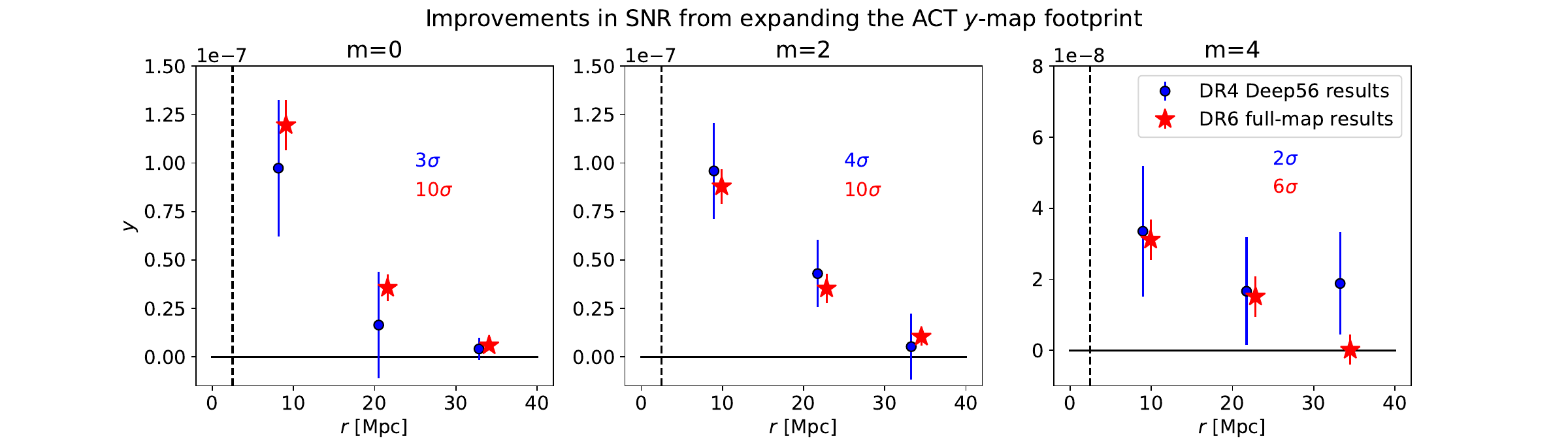}
    \caption{The first three even moments from the $y$ stacks combined over redshifts from $z\sim0.2$ to $z\sim0.7$. The blue points show results from Paper I using the DR4 $y$ map covering the Deep56 sky region, which have large errorbars due to the small sky area coverage. Red stars show the much more precise results from this work, in which the broad-coverage DR6 $y$ map is stacked on identical cluster locations and oriented by identical galaxies.  The blue points average over redshifts $0.25<z<0.72$, while the red points have a marginally different redshift range from $0.26<z<0.7$. The diminishing error bars from blue to red show the improvements, mostly due to the larger ACT $y$ footprint which increased the overlap with DES by $\sim9\times$, and partially also due to reduced DR6 noise. The anisotropic signal in both $m=2$ and $m=4$ is now significant.}
    \label{fig:pI_pII_y}
\end{figure*}
Figure~\ref{fig:pI_pII_y} quantitatively displays the gains in signal-to-noise compared to Paper I. Comparing the even cosine moments from the first to the third stack originally presented in Fig.~\ref{fig:ystack_impvts}, the shrinking errorbars from blue points to red stars display the improvement from using the lower-noise DR6 $y$ map and near-full use of the ACT-DES overlap region ($\sim9\times$ more sky area than before). Both stacks were processed without the gradient-flip aspect, preserving left-right and up-down symmetry and thus only inducing signal in the even cosine moments. The radial binning exactly matches that done in Paper I to enable direct comparison. In both profiles, we remove the inner 2.5~Mpc to avoid issues from miscentering, as the focus of this work is on the large-scale structure, not the central stacked clusters. 

The most robust way to compute the significance of the results would be to fit a model to the data and measure the significance of the model compared to null. However, without a readily available model for our results, we leave this for future work. To estimate the significance accounting for the correlation between neighboring radial bins, we use the covariance-weighted, multi-dimensional Mahalanobis distance \citep{Mahalanobis} from a null vector:
\begin{equation}
    d = \sqrt{\chi^2} = \sqrt{\boldsymbol{O}^\mathrm{T} \boldsymbol{\Sigma}^{-1}_O \boldsymbol{O}},
\end{equation}
where $\boldsymbol{O}$ is the observational data vector and $\Sigma_O$ is the covariance matrix representing radial bin-to-bin correlations. For Gaussian distributions (which we expect our data adheres to, given the tests of Gaussianity in Sec.~\ref{sec:tests}), $d^2$ is $\chi^2$-distributed and thus commonly referred to as $\chi^2$. We present the result as a number of standard deviations $d\sigma$, but note that it is susceptible to changes in the number and placement of the radial bins. So while it is useful as a point of comparison between results, it should be considered only as an approximation of true significance given the lack of a proper model fit.

The significances for the DR6 results are $\sim3-4\times$ higher than for DR4, mostly due to the expanded footprint. We find $\sim10\sigma$ for the m=0 and m=2 moments, showing evidence (respectively) of $y$ overdensity in the far-field beyond the stacked clusters and a large-scale quadrupole.  Importantly, there is now a $\sim6\sigma$ detection of a non-zero $m=4$ profile, which previously was non-significant. In Paper I, we demonstrated that a signature in $m=4$ does not appear in simulations of Gaussian random fields with the same power spectrum and cross-correlation as the galaxy-$y$ data. We also showed how the $m=4$ signal grows at late times in the Websky simulations, and interpreted it therefore as a signature of late-time non-Gaussianity indicative of the emergence of filamentary structure.\footnote{Our results on non-Gaussianity were consistent with prior analytic work by \citet{Blazek2011} which showed (in Appendix A therein) that only a $\cos{(2\theta)}$ term (i.e., $m=2$) appears as angular dependence in the projected galaxy alignment correlation function $w_g(r_p,\theta)$ under a linear intrinsic alignment model for a Gaussian random field.}
\begin{figure*}
    \centering
    \includegraphics[width=1\textwidth, trim={3.5cm 0cm 0cm 0.5cm},clip]{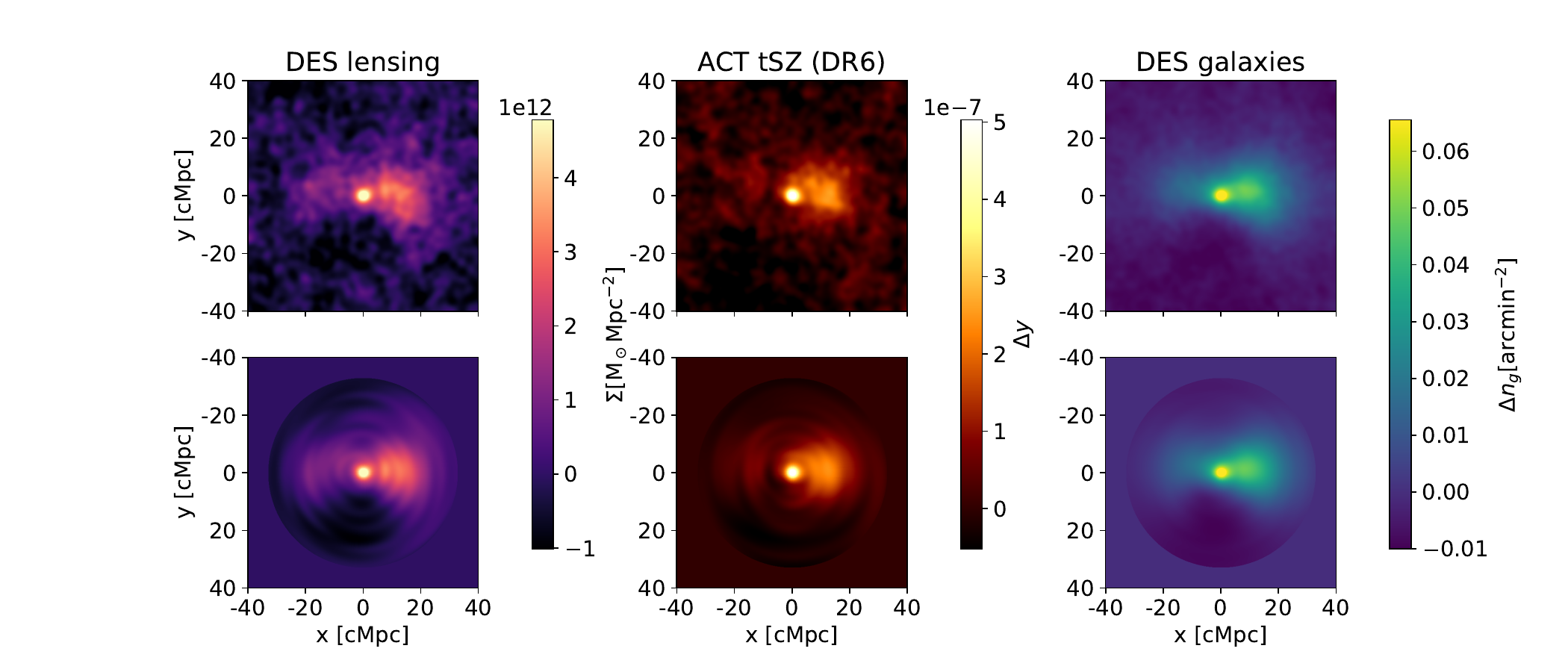}
    \caption{From left to right: the stacked surface mass excess density $\Sigma$, the excess $y$, and excess $n_\mathrm{g}$ combined over the full redshift range of viable overlapping DES and ACT DR6 data, $0.2<z<0.92$ (4648 selected clusters). The upper panel shows the stack filtered with a Gaussian beam (FWHM=2.6 Mpc) for visual purposes; the lower panel shows a version filtered even further by including only selected low-$m$ moments (see text for details). The reconstruction is shown for a circle extending out to $\sim35$~Mpc in radius, and beyond that the lower images are set to zero.}
    \label{fig:sigma_y_ng_stacks}
\end{figure*}

\subsection{All tracers} \label{subsec:all_tracers}
Next, we analyse the stacked results for $y$, lensing, and galaxy number density using the full updated set of methods described in Sec.~\ref{sec:methods}. These differ from Paper I by stacking only on $\lambda>20$ clusters, including asymmetry, smoothing the galaxy maps with a 20~Mpc Gaussian filter (slightly larger than the scale of the preceding comparison), and using 75\% of \maglim data for orientation rather than the full \redmagic sample. With these updates, Fig.~\ref{fig:sigma_y_ng_stacks} shows the stacks for $\Delta y$, $\Sigma$, and $\Delta n_g$ across the full redshift range of available data, $0.2<z<0.92$ (here the four broad redshift bins are stacked together). With smoothing applied in post-processing, the stacked cluster and extended feature are visually clear, representing a strong detection of the anisotropic signal in all three maps. The shape of the signal is remarkably similar between the three, showing that they are tracing the same structure.

There is clear reflective asymmetry about the $y$-axis (left-right) of each image, and more subtle asymmetry about the $x$-axis (up-down), both having been induced by the $+x$ and $+y$ gradient-based flips (Sec.~\ref{sec:methods_overview}). It is unsurprising that the stronger asymmetry is left-right: in the simple picture where a cluster is connected to only one straight filament that terminates at a neighboring cluster, the filament will be aligned along the $+x$ axis after the imposed rotation and flip. This natural asymmetry in the long axis of the eigenvector basis extends to cases of multiple filaments with varied sizes, odd-numbered connectivity, and varying shapes. Meanwhile, asymmetry in the short-axis (directed towards $+y$ in the stacked image) arises less commonly; e.g., it does not exist in the simple straight inter-cluster filament picture. However, elements such as curved structures, odd numbers of connected filaments, and uneven angular offsets between neighboring connected filaments (which are all further complicated by projection effects) can all contribute to the excess toward $+y$ seen in Fig.~\ref{fig:sigma_y_ng_stacks}.

When examining the $C_r(m)$ and $S_r(m)$ coefficients of Fig.~\ref{fig:sigma_y_ng_stacks} (with Eq.~\ref{eq:image_composition_series}), we find that the largest contributions come from the monopole ($m=0$) and the $\cos{(m\theta)}$ moments with $m=1$, 2, and 4. We note that integration with $\cos{(4\theta)}$ maximally picks up positive signal at $\theta=0$, 90, 180, and 270 degrees; however, the contributions in this case are likely to stem from concentrated signal in a small opening angle around the horizontal axis (0 and 180\degree) as there is no visual evidence for excess signal about the vertical axis. There is substantially lower $y$ signal in all sine components compared to the respective cosine components, and the sine radial profiles are noisier in all cases. The largest contributions in sine come from $m=1$, which captures the excess in the upper-half of the stacks compared to the lower. We find that the $S_1(r)$ is about $\sim60\%$ as large as $C_1(r)$, the cosine equivalent capturing the left-right asymmetry. The next-largest sine contributions are from $m=3$. Given this information, we reconstruct a filtered version of each stacked image in the lower panel of Fig.~\ref{fig:sigma_y_ng_stacks} by adding all cosine moments from $m=0$ up to (and including) $m=4$, and including also $S_1(r)$ and $S_3(r)$. The faithful reconstruction filters out small-scale noise and demonstrates that most of the signal is encompassed in these first five cosine and two sine moments.

Despite the presence of signal in some of the $\sin(m\theta)$ components, due to their noisy nature we will focus the following sections on comparing only the cosine $C_m(r)$ profiles with the respective profiles from the simulations. To optimize this in future work, one could consider combining the sine and cosine components in cases where the SNR is high. There may also be more optimized ways to handle $m=1$ moments through changes to the methodology. By considering the way that large-scale structure moves towards ``great attractors", one could incorporate the constraint of a dipole as the first step of the oriented stacking process and use the anisotropic tensor information secondarily. These explorations are beyond the scope of this work.

\subsubsection{Galaxy stack moments}
\begin{figure*}
    \centering
    \includegraphics[width=\textwidth]{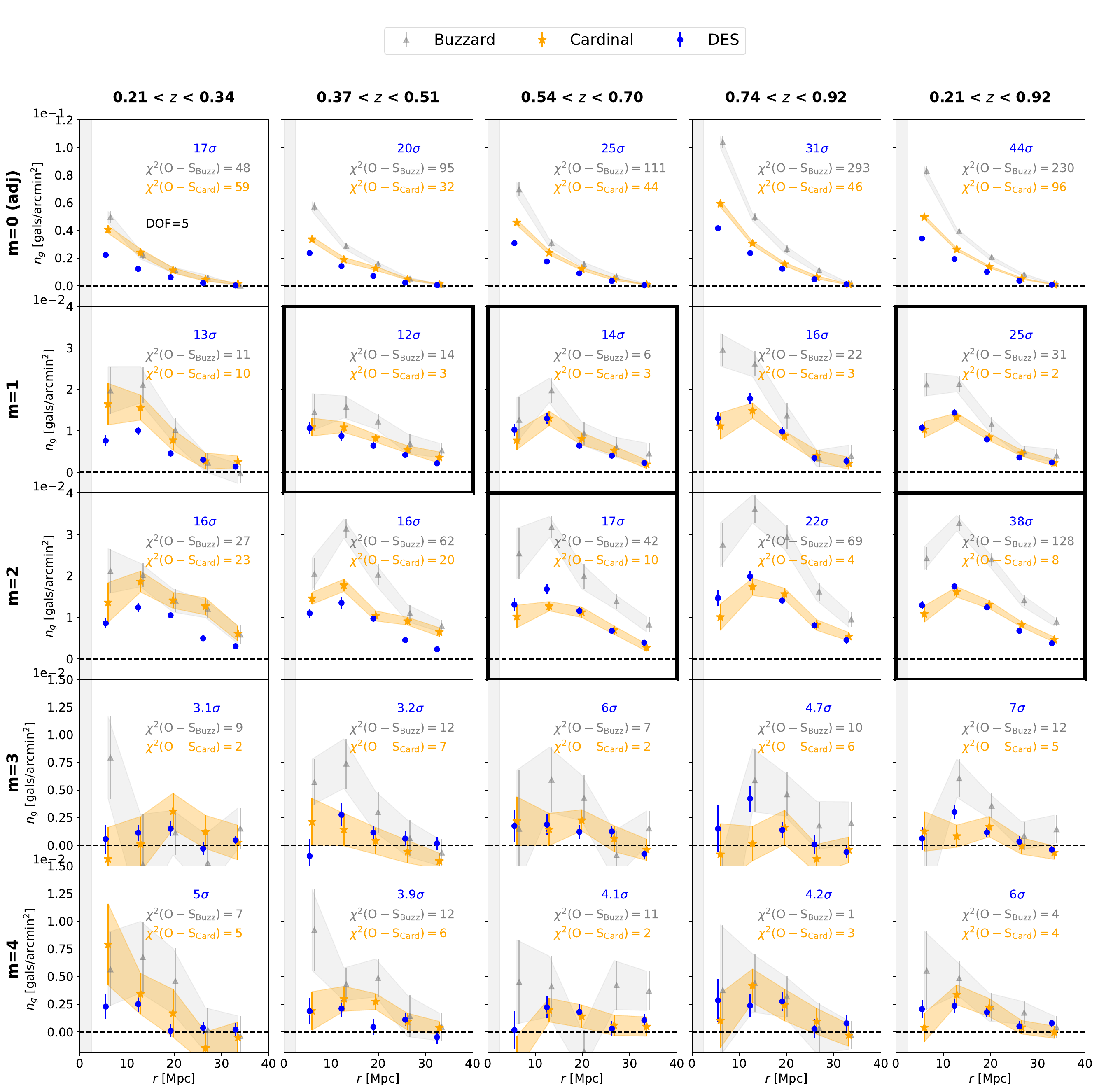}
    \caption{The radial profiles of the \maglim $n_\mathrm{g}$ oriented stacks per cosine moment (rows) and redshift bin (columns). The final column shows the multipole decomposition of a combined stack using data from all redshifts. Blue points show DES results, light gray triangles with a shaded band shows Buzzard predictions, and orange stars and band show Cardinal, all evaluated with the same $r$ binning and shown with small offsets in $r$ for visual purposes only. There are more \redmapper clusters in each bin in DES than in either Buzzard or Cardinal, causing the larger errorbars in the simulated results. The Cardinal simulations show a large improvement over Buzzard; nevertheless there are significant discrepancies in the monopole ($m=0$). Of the cases where the results are consistent between Cardinal and DES, those that are also significant in the ACT$\times$DES Compton-$y$ results (presented later) are highlighted in bold borders to facilitate comparison with Fig.~\ref{fig:giant_result_figure_y}.}
    \label{fig:giant_result_figure_g}
\end{figure*}

We begin the quantitative discussion with the galaxy-only results, as they provide a measure of how faithful the simulations are in simulating the dark matter and galaxy distribution in the cosmic web, as observed by DES, without the added complication of gas modelling. In Fig.~\ref{fig:giant_result_figure_g}, we show the multipole decomposition (up to $m=4$, with each $m$ per row) of each galaxy stack in each of the four redshift bins (columns). The final column shows the stack combined over all redshifts. The blue points represent the observational DES results. The gray shows the multipole information in Buzzard stacks, while the orange shows the same stack on Cardinal. We emphasize that Buzzard and Cardinal were run on the same underlying halo catalog, and therefore the differences stem from only the updated galaxy prescriptions in Cardinal.

In addition, we show the squared Mahalanobis distances, or $\chi^2$ values, between the observations and Buzzard (gray) or Cardinal (orange) simulations. For a simulated data vector $\boldsymbol{S}$, this is calculated according to:
\begin{equation}
    \chi^2(\boldsymbol{O}-\boldsymbol{S}) = (\boldsymbol{O} - \boldsymbol{S})^\mathrm{T} \boldsymbol{\Sigma}^{-1}_{O+S} (\boldsymbol{O}-\boldsymbol{S}),
\end{equation}
where $\boldsymbol{\Sigma}^{-1}_{O+S}$ is the inverse of the covariance matrix that adds the simulated and observed covariances. When divided by the number of degrees of freedom (DOF), equal to the number of radial bins (5), the result should be $\sim1$ when the observation and theory results are statistically consistent.

The $m=0$ profiles in DES are strongly detected in each redshift bin, showing the slow decrease in isotropic density of the average supercluster region as one probes outwards from the central stacked \redmapper clusters (whose signal is removed as indicated by the vertical gray band at low $r$). We caution against attempting to read the physical evolution with redshift from the figure, because the $y$-axes are in angular density, the \maglim selection function is redshift-dependent, and there is some further redshift dependence of the selected cluster samples particularly in $\lambda$ (see Fig.~\ref{fig:multiscale_props}). In all redshift bins, the $m=0$ moments are all highly inconsistent between both Buzzard and Cardinal when compared to DES in low radial bins. This is likely due to difficulty in reproducing the bias of \maglim in the constrained regions, especially near \redmapper clusters. There is better agreement in the outskirts. Treating each profile as a whole, the reduced-$\chi^2$ compared to simulations is poor: always greater than six. Nevertheless, a strong improvement has been made from Buzzard to Cardinal in all but the lowest redshift bin. The improvement largely comes from a drastically updated treatment of color-dependent clustering affecting the mock \redmapper sample, as described in \citet{To2024ApJ...961...59T}.

For the anisotropic moments on which this work is focused, the signal-to-noise ratio of DES galaxies is high in $m=1$ and $m=2$, ranging from $\sim13-22\sigma$ in the separate $z$ bins. For the noisier $m=3$ and $m=4$ moments, we will conservatively interpret the significances using a $4.5\sigma$ threshold for detection, given that some unoriented stacks produced $\sim4\sigma$ signatures of residual anisotropy in the individual redshift bins (Sec.~\ref{sec:tests}).  We also note that there is $6-7\sigma$ signal in the combined-$z$ stack for $m=3$ and $m=4$, which was consistently null in the unoriented tests, suggesting that the binned redshift results are produced by true correlations rather than randomness. There is $m=3$ detection in the highest two redshift bins, of interest because this moment is expected to relate to the shapes and connectivity of filaments as discussed previously in this section. In $m=4$, there is a significant detection in the lowest redshift bin, probing the non-Gaussianity of the late universe. There are tentative signals of $m=4$ at the other redshifts. As with $m=0$, Cardinal presents an improvement in consistency over Buzzard for nearly all bins in the anisotropic moments. The quadrupole profiles ($m=2$) in the lowest two redshift bins are the most discrepant.

We highlight in Figure~\ref{fig:giant_result_figure_g} (using a bold border) the cases in which the Cardinal prediction is consistent with the observational result at the 95\% level ($p>0.05$), \textit{and} the Compton-$y$ observational result presented later in this section is non-null at a significant ($>4.5\sigma)$ level. The purpose is to identify cases in which the galaxy structure, which defines the selection and orientation process, is well-matched by the simulations. Such cases are highlighted as the best to search for discrepancies in the stacked $y$ signal alone, as these are less likely to be due to differences in the galaxy/cluster distribution and more likely to be attributable to inaccuracies in the gas modeling.

\subsubsection{Normalizing by the monopole} \label{subsec:monopole_norm}
The $n_\mathrm{g}$ analysis helps to identify regions where the underlying structure is similar between simulations and data. However, as the difference between Buzzard and Cardinal results from Fig.~\ref{fig:giant_result_figure_g} shows, the same exact underlying halo structure can produce drastically different galaxy density results in all moments when the galaxy bias and/or color-dependent clustering are simulated inaccurately. To understand more about how well both Buzzard and Cardinal reproduce anisotropic structure in the underlying matter distribution, we can attempt to remove the large-scale projected cluster-galaxy bias. We begin by assuming that the relationship of the projected galaxy density $\delta_\mathrm{g,p}(r,\theta)$ to the projected total matter distribution $\delta_\mathrm{p}(r,\theta)$ depends linearly on the response function $R_\mathrm{p}(r)$ which is a function \textit{only} of radial position from the cluster, and not the angle: 
\begin{equation} 
\delta_\mathrm{g,p}(r,\theta) = R_\mathrm{p}(r) \delta_\mathrm{p} (r,\theta).  
\label{eq:theta-independent-response}
\end{equation}
Here, the $p$ subscripts indicate that these are all projected quantities. If the assumption of a $\theta$-independent response is true, then the ratio of any moment $m$ to another, $m'$, removes the response function of the underlying matter anisotropy. Computing the ratios of multipole moments $C_m(r)$ for a galaxy stack, we combine Eq.~\ref{eq:theta-independent-response} with Eq.~\ref{eq:multipole_moments}:
\begin{align}
    \frac{C_{\mathrm{g},m}(r)}{C_{\mathrm{g},m'}(r)} &= \frac{\frac{1}{2\pi}\int_0^{2\pi}  \langle R_\mathrm{p}(r) \delta_\mathrm{p} (r,\theta) | \xi \rangle  \cos{(m \theta)}d\theta}{\frac{1}{2\pi}\int_0^{2\pi}  \langle R_\mathrm{p}(r) \delta_\mathrm{p} (r,\theta) | \xi \rangle  \cos{(m' \theta)}d\theta}  \nonumber \\
    &= \frac{\int_0^{2\pi} \langle \delta_\mathrm{p}(r,\theta)| \xi \rangle \cos{(m\theta)} d\theta}{\int_0^{2\pi} \langle \delta_\mathrm{p}(r,\theta)| \xi \rangle \cos{(m'\theta)} d\theta},
\label{eq:bias_removal}
\end{align}
for the case where neither $m$ nor $m'$ is 0 (otherwise, a factor of 2 appears).
In other words, a $\theta$-independent response simply cancels because it appears equally in all moments at a given radial distance $r$. The concept of studying matter anisotropies through multipole ratios has been applied previously in studies of the weak lensing maps and galaxy catalogues \citep{Gouin2017A&A...605A..27G, Gouin2022}, but without orientation or the emphasis on bias removal.

Several recent studies have found the galaxy or halo bias to be dependent on the anisotropy of the tidal field, or (similarly) the filamentary environment, in simulations \citep{Paranjape2018, Obuljen2019JCAP...10..020O} and data \citep{Alam2024MNRAS.527.3771A, Obuljen2020JCAP...10..058O}. In other words, halos and therefore galaxies form and cluster differently depending on environment, and therefore the bias of some galaxy types is in fact expected to be somewhat orientation-dependent. This would imply that the response function is \textit{not} $\theta$-independent, at least for small scales and sufficiently high precision. Such dependency in three dimensions should propagate into projected measures that preserve some information about the 3D tidal field, as in our work, meaning that testing whether or not Eq.~\ref{eq:bias_removal} holds true in data, requiring the assumption in Eq.~\ref{eq:theta-independent-response} to be true, is one way to search for the effects found in the previous works. However, if the response is indeed $\theta$-dependent but is subtle enough to lie within our errorbars or is washed out by our relatively large (7.5~Mpc) binning size, Eq.~\ref{eq:bias_removal} will appear to hold true.

\begin{figure*}
\includegraphics[width=\textwidth]{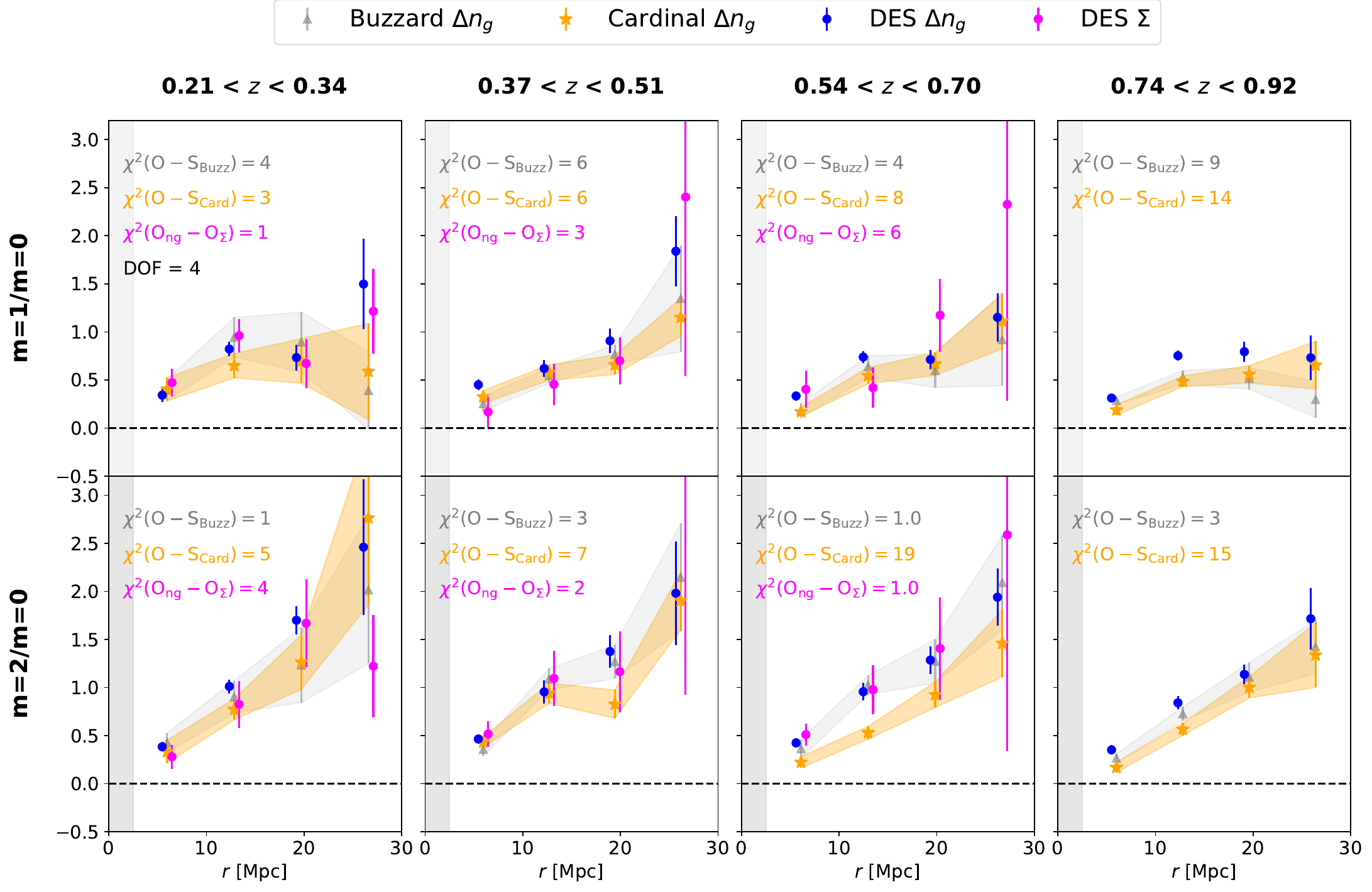}
    \caption{Ratios of the $m=1$ and $m=2$ profiles by the $m=0$ profile of the same stack. \maglim galaxies are shown in blue, Buzzard galaxy predictions in gray, Cardinal galaxies in orange, and DES surface mass density in magenta. The division by $m=0$ removes the galaxy bias under the assumption of a linear large-scale bias which is isotropic around the central stacked cluster sample. This brings the Cardinal and Buzzard $\Delta n_g$ results largely into agreement, and both into agreement with the data in most cases. If galaxy bias has been fully removed, both blue and magenta points are expected to be consistent as they measure the shapes of the same underlying matter distribution. The magenta $\chi^2$ values are low, showing consistency between them and no evidence for a $\theta$-dependent and/or nonlinear bias.}
    \label{fig:giant_result_figure_mratio}
\end{figure*}
To perform this test, Fig.~\ref{fig:giant_result_figure_mratio} compares the DES, Buzzard, and Cardinal \maglim multipole profiles after division of the adjusted $m=0$ moment. Only the normalized $m=1$ and $m=2$ profiles are shown, as the higher-order ratios are noisier. If Eq.~\ref{eq:bias_removal} were true for both the Buzzard and Cardinal simulations, the two simulation results would come into perfect agreement with each other as their galaxies are pasted on the same $\delta_m$ matter distribution. Indeed, out of the shape measurements in all 32 radial bins shown in the plot, only three are over $1\sigma$ offset, showing that the approximation of $\theta$-independent large-scale galaxy bias for these simulations is valid at most redshifts and radii. Additionally, for this normalized shape measure, the Buzzard measurements are far more consistent with DES than in Fig.~\ref{fig:giant_result_figure_g}. In other words, while the large-scale cluster-galaxy bias is inaccurate in Buzzard, causing amplitude offsets in every un-normalized moment, the relative amount of matter distributed in the dipole and quadrupole is much more similar to the reality inferred from DES measurements, demonstrating that most of the Buzzard inaccuracies enter at the level of the galaxy prescription. This statement is consistent with the fact that Cardinal shows such a large improvement over Buzzard in Fig.~\ref{fig:giant_result_figure_g}. The only point for which neither simulation matches the DES results well is at $r\sim13$~Mpc of $m=1$, bin $z_4$: the distribution of matter into the dipole predicted by the simulations is 2.5-3.5$\sigma$ lower than the real data. If this discrepancy is not a statistical fluke, the cause could be (1) a nonlinear and/or $\theta$-dependent galaxy bias in the observed \maglim data, or (2) a difference in the shape of underlying dark matter structure, at this scale and redshift for our selected sample.

The simple concept of dividing out a $\theta$-independent bias provides an interesting additional application. Eq.~\ref{eq:bias_removal} can directly be compared to the same ratio for the stack on weak lensing, a less biased tracer of the same underlying matter. If they are consistent, there is no evidence to reject the hypothesis that the response is linear and isotropic (Eq.~\ref{eq:theta-independent-response}). An \textit{inconsistent} signal between galaxies and lensing would provide evidence for a directionally-dependent and/or nonlinear bias.

Fig.~\ref{fig:giant_result_figure_mratio} shows the lensing results in magenta overlaid on the same plot as the normalized galaxy number density shape measures. The magenta numbers indicate the $\chi^2$ value measuring consistency for the observed normalized $n_\mathrm{g}$ and $\Sigma$ profiles, calculated in the same manner as the others. Note that in this case, unlike in the observation-simulation consistency tests, the two data vectors are highly correlated by virtue of tracing the same physical structure. The reduced $\chi^2$ values are all $\lesssim1$, indicating no evidence for anisotropic galaxy bias and showing, as expected in that case, strong coherence from the two methods of tracing the same underlying density field. In the future, higher precision results from a more optimized treatment of the lensing data and/or future galaxy survey data may make this test a useful and novel way to search for deviations from a simple linear and isotropic bias prescription.

\begin{figure*}
\includegraphics[width=\textwidth]{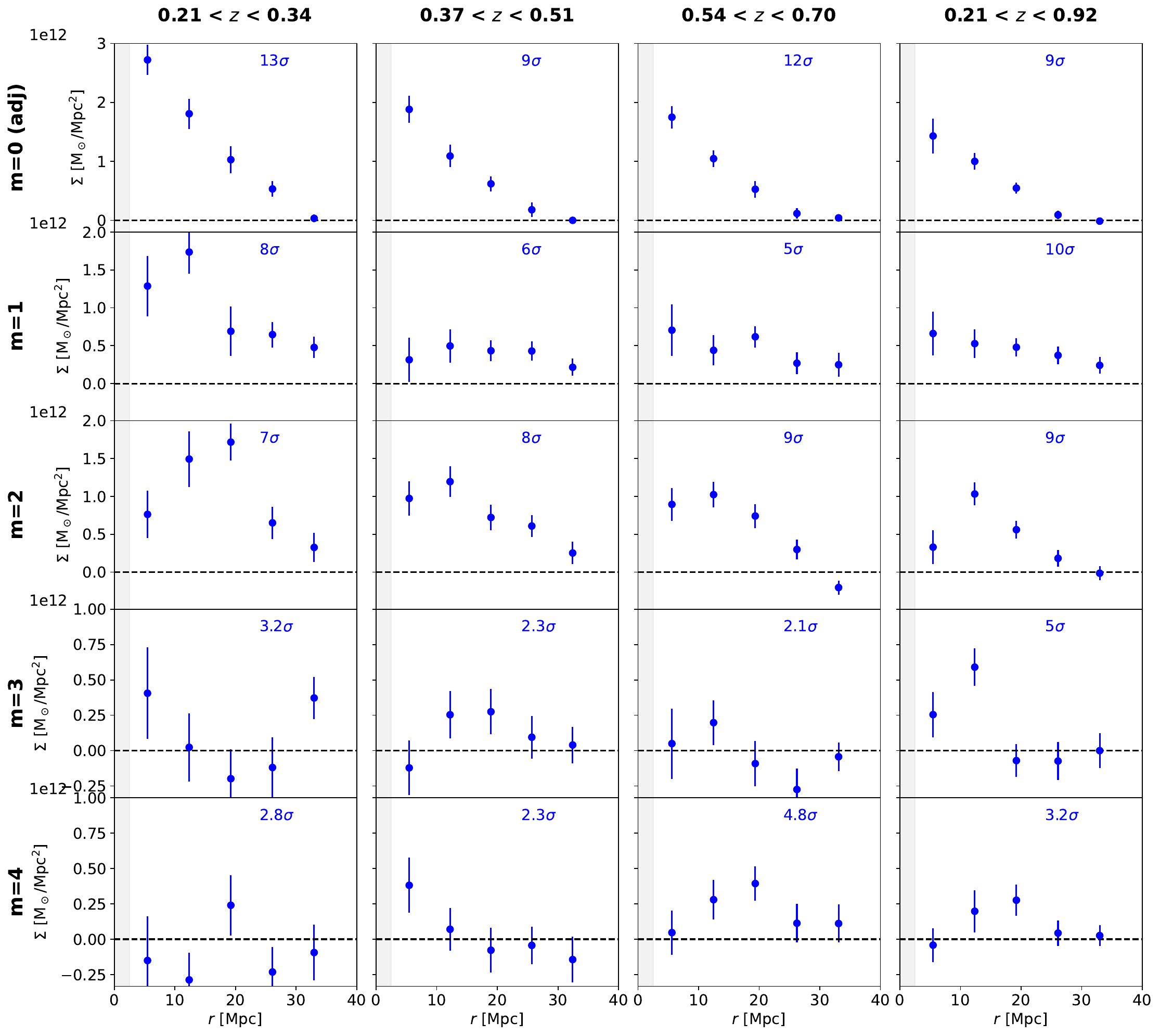}
    \caption{$\kappa$ map stacks converted to $\Sigma$ for the DES data. We show only the first three redshift bins as the fourth overlaps with the source bin for the $\kappa$ map and therefore is very noisy. The right-most column shows the combined stack over all available redshifts (including the fourth bin) to enable direct comparison with the same all-$z$ results in $y$ and $n_\mathrm{g}$. Note that the $y$-axes differ in each row. The lensing kernel for the single $\kappa$ map we use, sourced from galaxies at high redshifts (Fig.~\ref{fig:lensing_kernel}), peaks in our second-lowest redshift bin. The values in text show the significance of detection over a null vector in $\Sigma$. The $m=0$ profiles have been adjusted by subtracting an annulus from $30-40$~Mpc. There is detection of anisotropy in $m=1$ and $m=2$ for all three redshift bins, but detection of $m=4$ only in the third bin. The all-$z$ stacks have significant signal in all moments up to $m=3$ and tentative signal in $m=4$.}
    \label{fig:giant_result_figure_sigma}
\end{figure*}

\subsubsection{Lensing stack moments}
Next, in Figure~\ref{fig:giant_result_figure_sigma}, we report the un-normalized results for the stacked surface mass density $\Sigma$, from stacking the lensing convergence maps and using the conversions detailed in Sec.~\ref{subsec:lensing_conversions}. We show only the first three redshift bins, as the highest bin overlaps significantly with the source distribution from which the $\kappa$ map was created, and therefore has low signal-to-noise. We also show the combined result over all $z$. As there are no readily-available Buzzard/Cardinal $\kappa$ mock maps, we show only the observed results from DES. For a point of comparison and validation of the $\kappa$ to $\Sigma$ conversions in Sec.~\ref{subsec:lensing_conversions}, we turn to the study of \citet{Yang2020}. The authors stacked $\Sigma$ maps at locations of luminous red galaxy pairs from SDSS, separated by $\sim8\hMpc$ (probing structure at scales similar, but $\sim20\%$ smaller, than our study) at an effective redshift of $z\sim0.5$. They find a maximum value of $\sim2.5\times10^{12}$\Msun Mpc$^{-2}$ in the intervening bridge region in the stacks. For similar redshifts -- in our second redshift bin -- we can make a comparison by combining the signal from the $m=0$, $m=2$, and $m=4$ moments (using only moments that are symmetric about the $x$ and $y$ axes because the LRG stacks were symmetric in that manner). The total is $\sim3\times10^{12}$~\Msun Mpc$^{-2}$, quite similar to the \citet{Yang2020} result.

The $\Sigma$ results are generally far noisier than the galaxy data, yet there are significant detections of signal throughout the first three moments. 
From $m=0$, we observe that the large-scale mass overdensity in the selected supercluster regions is largest at the lowest redshifts, perhaps a consequence of the slightly higher-richness-skewed cluster sample (see Fig.~\ref{fig:multiscale_props}). The adjusted $m=0$ profile also illustrates why annulus-subtraction at $30-40$~Mpc is not ideal; the profiles do not move asymptotically to zero but rather appear to still have downward slopes near the highest radial bin, indicating that large-scale structure above the mean-field still exists at the outskirts and has artificially been removed by our annulus subtraction. Larger cutouts would better allow us to measure the mean-field beyond the large overdensities ($\nu>2$ at $\sim15$~Mpc scales) being stacked, but larger cutouts come with technical issues from edge effects as discussed in Section~\ref{subsec:mask_holes}. The resulting depletion of the sample would not be worthwhile given that the $m=0$ moment is not the main focus of this work. Turning to the anisotropic moments, there is strong detection of anisotropy from $m=1$ and $m=2$, also peaking in the lowest redshift bin. There is also significant detection of $m=3$ signal in the redshift-combined stack (right-most column) and of $m=4$ signal in bin $z_3$. 

The lensing results are a useful proof-of-concept that the method is applicable to a wider variety of tracers than explored in Paper I, but we note that they could be improved by using an ideal combination of $\kappa$ maps created from all DES source bins. This is beyond the scope of this work, but both this optimization, as well as the translation of the Cardinal shear data into mass maps analogous to those produced for the DES observational data, would be worthwhile in the future to gain more insights from locally-anisotropic cross-correlation measurements.

\subsubsection{Compton-y stack moments}

\begin{figure*}
    \centering
    \includegraphics[width=\textwidth]{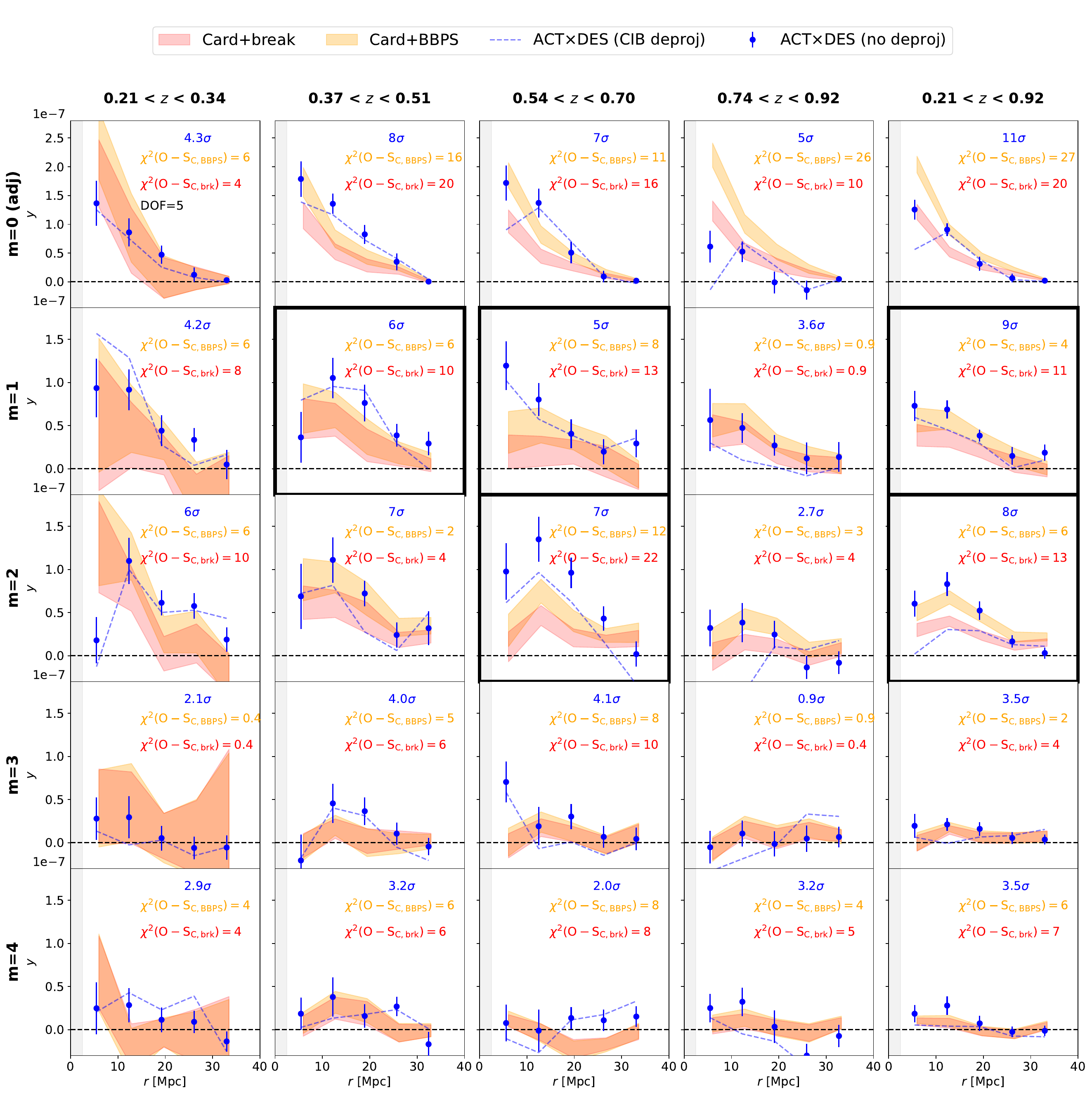}
    \caption{$y$ map stacks from ACT$\times$DES data (blue) and their comparison with Cardinal. Blue points show the results with the fiducial $y$ map and dashed lines show the same profile from the most-impacted CIB-deprojected map as first introuced in Sec.~\ref{sec:tests}. The yellow-orange shaded bar shows the results for the fiducial BBPS model $y$ map, while the red-orange bar show the same results from the alternative `break' model map. Orange and red text display the reduced $\chi^2$ comparison between the data and the simulations for the BBPS and \textit{break} models respectively. A $\chi^2>11$ is statistically inconsistent at the 95\% level given 5 degrees of freedom; thus there is consistency between data and simulations in most higher-order moments. The bolded subplots are those which have a significant detection of a non-zero $y$ profile (using a threshold of SNR$>$4.5 as motivated by the unoriented-stack null tests, Sec. \ref{sec:tests}) \textit{and} a galaxy stack result that is consistent between Cardinal and DES.  The \textit{break model} is less consistent with the fiducial $y$ data in all highlighted cases. The selected CIB deprojection alleviates the tension with the \textit{break model} especially in the highest redshift bin.}
\label{fig:giant_result_figure_y}
\end{figure*}

Lastly, we present the central focus of this work, the oriented stacks on Compton-$y$, in Figure~\ref{fig:giant_result_figure_y}. This figure includes the observational ACT$\times$DES results in blue, including points with errorbars showing results from the fiducial map (with no components deprojected) and the dashed line showing results from the CIB-deprojected map with the largest difference from fiducial, out of those we studied ($\beta=1.0, T=10.7$~K). Errorbars are not shown for the dashed blue line (which connects discrete points for visual purposes only, and is not a fit) but are $\sim1.5-2\times$ larger than those from the fiducial map. In addition, we show stacks on the `Cardinal' mock $y$ map, with pasted gas profiles, in a shaded bands: yellow-orange for the fiducial BBPS model and red for the \textit{break model} (Eq. \ref{eq:break_model}). The \textit{break model} profiles always have lower signal than the BBPS model, due to the suppression in pressure in low-mass halos. Although we do not show it in the figure, we also note that stacking the mock $y$ map while selecting and orientating the cutouts using Buzzard galaxies and clusters, rather than Cardinal, makes a significant difference in some of the mock $y$ profiles. This demonstrates that accurate galaxy modeling is important to the oriented $y$ measurements.

The deprojected observational results using the fiducial DR6 map mark a large improvement from the previous DR4 work, with $>4\sigma$ significance in the first three redshift bins in both $m=1$ and $m=2$. In $m=3$ and $m=4$, none of the individual $z$ bin profiles pass the 4.5$\sigma$ threshold motivated from the unoriented null tests, but the all-$z$ combined results are detected at a level ($3.5\sigma$) that was not seen in the respective null tests. Interestingly, this is lower than the significance at which we detected $m=4$ for the stacks shown in Sec.~\ref{sec:res_indiv}. This could be due to stacking on more massive halos ($\lambda>20$ rather than $\lambda>10$), pointing to a difference in the environments of halos at higher masses, something worth investigating for future work. Meanwhile, the CIB deprojection depletes the signal in most, but not all, cases. It is especially impactful in the lowest radial bin in the monopole and in the highest redshift bin for all moments. The CIB-deprojected signal-to-noise is lower throughout due partly to the depletion but mostly to the larger errorbars (not shown); in $m=1$ and $m=2$ it ranges from 3.5-4.5$\sigma$ (not written in the figure).

Examining qualitatively the redshift dependence of the signal, the largest evolution in the observational data occurs between bins $z_4$ and $z_3$ for most moments. In $m=2$, for example, the signal at the peak increases by $\sim250\%$ from $z_4$ to $z_3$ in the observed profile. Since this increase is not seen as strongly in the lensing data between $z_4$ (not shown) and $z_3$, it may be indicative of evolution in how the gas pressure responds to matter at these redshifts. Interestingly, the Cardinal models both predict a smaller increase in the signal between $z_4$ and $z_3$; this could indicate inaccuracies in the model of the evolution of the thermal energy response between these redshifts.

For a more quantitative comparison, we examine the $\chi^2$ statistic when comparing the fiducial observations to the BBPS-model simulation $\boldsymbol{S_{\rm{BBPS}}}$ as well as that compared to the \textit{break model} $\boldsymbol{S_{\rm{br}}}$, to assess consistency between observational and simulated profiles. Considering $\chi^2>11$ as evidence of inconsistency (given 5 degrees of freedom, $p<0.05$), both simulations are consistent for most redshift bin and multipole combinations. The BBPS predictions are generally in better agreement with the observational results than the \textit{break model} predictions. The largest discrepancies for the BBPS profile are in the monopole of bins $z_2$ and $z_4$ as well as the all-$z$-combined result. When the CIB is deprojected, there is no clear preference between the two models in the first three redshift bins. The highly-impacted bin $z_4$ tends to prefer the \textit{break model} and, driven by the inclusion of these high redshift results, the all-$z$-combined CIB-deprojected results are also more consistent with the \textit{break model} (driven by $m=0$ and $m=2$).

Of those redshift and multipole combinations in which there is a significant signal in ACT$\times$DES, we highlight (with bold borders) the same cases as in Fig.~\ref{fig:giant_result_figure_g}, where there is consistency between the Cardinal and DES result in the \textit{\maglim galaxies}. Assuming that this consistency means that the Cardinal sims provide the statistically-correct prediction both for the multi-scale constrained sample and for the large-scale orientation in those regions, then it is reasonable to attribute any remaining differences to the halo-based gas pressure modelling. Of the highlighted cases, the BBPS model tends to be successful in predicting the extended signal in the fiducial results; only $m=2$ in bin $z_3$ shows statistical discrepancy at the 95\% confidence level due to an under-prediction of signal in the first 3 radial bins. That discrepancy is alleviated with the CIB deprojection. Combining over all redshifts (rightmost column), the fiducial $y$ result prefers the BBPS model in both bolded cases, but we caution again that the inclusion of high-redshift data makes the CIB impacts especially uncertain here.

Besides the bold-bordered plots, we also note that in the $z_2$ and $z_3$ monopole $(m=0)$ moments, the galaxy results were inconsistent because they were highly \textit{over}-predicted in Cardinal---yet, the gas pressure results are \textit{under}-predicted for several radial bins. This indicates that the modeled relationship between gas and galaxies is low for these redshifts.

With the reminder that we have not attempted to vary parameters to infer best-fit values, we draw tentative conclusions from the comparison with the two fixed-parameter models. Overall, the discrepant cases from the fiducial observations suggest that when modeling extended structure, the uncompensated \textit{break model} overly depletes the thermal energy. This could be from an excess in depletion internal to the $M<2\times10^{14}$~\hMsun\, halos (done as a proxy for AGN feedback), or it could indicate that the thermal energy lost from these halos needs to be compensated elsewhere by explicitly moving it to the halo outskirts and/or by modeling filaments. This latter idea is consistent with past conclusions drawn from hydrodynamic simulations with AGN and from studies of filaments and inter-cluster bridges, in which gas far beyond massive halos can be hot and dense enough to contribute significantly to an oriented tSZ signal \citep{Tanimura2019, deGraaff2019, Hincks2022, Lokken2023MNRAS.523.1346L}. Meanwhile, despite the approximations made by assuming fully isotropic gas profiles and applying a model originally fit to only high-mass halos, the BBPS model shows broad consistency with both the fiducial and CIB-deprojected observations for redshifts $z<0.7$ and radii $r>10$. This indicates that, even if BBPS over-predicts the tSZ signal for small halos as shown in previous works, the way in which the signal is smeared out into the large radial bins mimicks the true extended, large-scale tSZ signal.

\section{Comparison of multiple tracers} \label{sec:res_multiple}
Because the orientation procedure is applied identically to all three datasets (tSZ, lensing, and galaxy positions), it is appealing to examine how the extended structure revealed through each tracer relates to the others. The relationships are sensitive to the bias of the different tracers, or in other terminology can be thought of as response functions (as first discussed in Sec.~\ref{subsec:monopole_norm}) measuring how one signal responds to the presence of another. In particular, we will analyze the ratio of $y$/$n_\mathrm{g}$ and $y$/$\Sigma$ for several binned multipole profiles. How $y$, the integrated gas pressure, responds to the presence of matter (traced by $\Sigma$) and galaxies ($n_\mathrm{g}$) is expected to depend not only on the initial collapse of gas into filaments and halos, but also on the cycles that redistribute baryons in the late universe.

\subsection{Motivation to only examine higher-order moments}\label{subsec:discard_monopole}
We limit this study only to the ratios of the higher-order moments between the different tracers. Appendix~\ref{appdx:ratios_tracers} details the analytic forms of the stacked signals and demonstrates why working with mean-zero maps is important. To do so, we perform the annulus subtraction as presented in Sec.~\ref{sec:methods}; however, this may still cause complications when comparing the outskirts of the $m=0$ moments between tracers. If the annulus average of $y$ only contained contributions from the long-wavelength CMB, removing it would yield a pure $\Delta y$ measurement which would be directly comparable to the other tracers. However, if there is some true $y$ structure that is also removed by the annulus subtraction, comparing the annulus-subtracted $y$ stack with the galaxies and lensing will inaccurately represent the relationship at the outskirts. The same is true if $n_\mathrm{g}$ or $\Sigma$ stacks contain some unknown sources of additive bias. This subtlety can have a large impact where the denominator of a ratio is approaching zero: for example, consider the following ratio of annulus-subtracted stacks on $y$ ($I_\mathrm{y}$) and on $n_\mathrm{g}$ ($I_\mathrm{g})$: 
\begin{equation} \label{eq:stack_ratio}
\frac{I'_\mathrm{y}(r,\theta)}{ I'_\mathrm{g}(r,\theta} = \frac{I_\mathrm{y}(r,\theta)-\langle I_\mathrm{y,ann} \rangle}{I_\mathrm{g}(r,\theta) - \langle I_\mathrm{g,ann} \rangle} + \tilde{N},
\end{equation}
where $\langle I_{\rm{ann} \rangle}$ is the annulus-average. When $r$ is small (within or nearby the central stacked cluster), $I_\mathrm{y}(r)\gg\langle I_\mathrm{y,ann} \rangle$ and $I_\mathrm{g}(r)\gg\langle I_\mathrm{g,ann} \rangle$ so the left-hand side of Eq.~\ref{eq:stack_ratio} is equivalent to $\sim I_\mathrm{y}(r)/I_\mathrm{g}(r)$. However, in the regime of large $r$ far afield of the central cluster, both numerator and denominator approach zero, making the annulus-subtracted ratio very sensitive to small fluctuations and thus very noisy. Subtleties in the mean removal become important in this regime and are challenging to fully understand for each tracer. 

Therefore, we focus here on the comparative analysis of the higher-order moments of each stack. These moments are by definition insensitive to the mean, an advantageous feature that makes them insensitive to any additive isotropic biases.

\subsection{Propagation of errors for ratios} \label{subsec:ratio_errs}
For the ratios of the tracers, we propagate the covariances presented in Sec.~\ref{sec:uncertainties}. The derivation for this propagation is described in Appendix~\ref{appdx:errors_ratios_tracers}. To simplify notation, let us call the radially binned multipole moment from a stack on a map of one tracer, previously referred to as $\boldsymbol{C}_{m}$, simply $\boldsymbol{X}$. If we divide $\boldsymbol{X}$ by a profile $\boldsymbol{Y}$ from a second (correlated) tracer (for example, $
y$ divided by $\Sigma$), finding the ratio $\boldsymbol{R}=\boldsymbol{X}/\boldsymbol{Y}$, then the covariance $\mathrm{Cov}(R_i,R_j)$ between radial bins $i$ and $j$ is: 
\begin{multline}
    \mathrm{Cov}(R_i,R_j) = \frac{1}{Y_i Y_j} \bigg[\mathrm{Cov}(X_i,X_j)-\frac{X_i}{Y_i}\mathrm{Cov}(X_i,Y_j) \\ -\frac{X_j}{Y_j}\mathrm{Cov}(Y_i,X_j) + \frac{X_i X_j}{Y_i Y_j}\mathrm{Cov}(Y_i,Y_j)\bigg].
\end{multline}
We use this formula to calculate the uncertainties in the following results.

\subsection{Results of tracer ratios}

\begin{figure*}
    \centering
    \includegraphics[width=.8\textwidth]{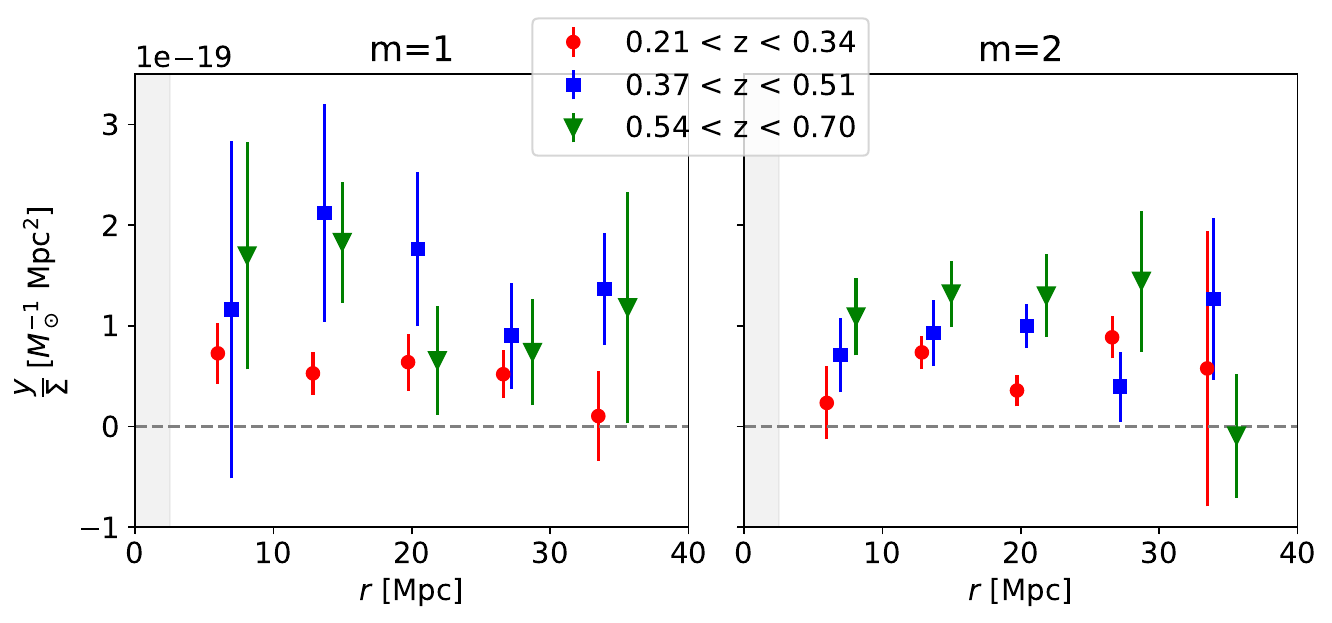}
    \caption{The ratio of $y$ to $\Sigma$ in the binned profiles for $m=1$ and $m=2$, first three redshift bins. The ratio is fairly consistent in all radial bins, showing no strong trends with distance from the central cluster (as usual, the inner 2.5~Mpc has been removed). The relationship is also consistent between the two moments (left and right): every radial bin is within 1$\sigma$ for $m=1$ and $m=2$ for a given redshift. There is evidence $(p<0.05$ for 5 degrees of freedom) that the lowest redshift bin has a significantly lower ratio, averaging across radial bins, than the higher redshifts.}
    \label{fig:y_Sigma_ratio}
\end{figure*}
We begin by presenting the ratio between $y$ and $\Sigma$ in the $m=1$ and $m=2$ anisotropic moments in Fig.~\ref{fig:y_Sigma_ratio}. We show only the results for the fiducial $y$ with no CIB deprojection. The figure shows only these lower multipoles as the higher orders are very noisy and thus challenging to interpret. As in previous results, the bins begin at $r=2.5$~Mpc, removing the central stacked clusters from the analysis, and are each 7.5~Mpc in width thus effectively smoothing out fluctuations on smaller scales. We find $y/\Sigma$ to be constant with radius within the uncertainties, at all redshifts. This is quite remarkable given that the results span a wide range of $r$. This does not mean that the amount of anisotropy in either profile is constant, but rather, that their profiles fluctuate together; both the $y$ and lensing stacks have a similar shape as a function of radius. Second, the $m=1$ and $m=2$ results are fairly similar in value despite the fact that these multipoles probe different aspects of the structural shape. This suggests the relationship does not change drastically as a function of the large-scale anisotropic superclustering environment, although analysis of the higher multipoles, especially $m=4$, with more precision would help to confirm or deny this point. Both of these features indicate that the gas pressure is tracing the matter on large scales. In other words, there is large-scale coherence in the shapes and extents of the ionized gas and the dark matter. Finally, the higher redshift bins have significantly larger $y$/$\Sigma$ responses than the lowest redshift bin, indicating some redshift dependence of this relationship.

For completeness, we note that the $y$-axis values in Fig.~\ref{fig:y_Sigma_ratio} are related to the value of the halo-bias-weighted average electron pressure $\langle b P_\mathrm{e} \rangle$, as studied in \citet{Vikram2017, Pandey2021, Pandey2022PhRvD.105l3526P, Yan2021, Chen2023arXiv230916323C}. The measurements presented here are distinct as they are local, i.e., a function of $r$ and $\theta$, and measured under the multi-scale constraints $\xi$. Relating the projected ratio to the gas pressure bias, defined in 3D, would require making several approximations to deproject the integrated signals and convert the excess matter density to an overdensity. As too many assumptions are involved in that case to find a reliable value, the better option is to fully forward model the signal in the simulations, which is challenging as there are no readily-available Cardinal $\kappa$ maps. Such an approach to translating these measurements to $\langle b P_\mathrm{e} \rangle$ are beyond the scope of this work, but Sec.~\ref{sec:energy_ratio} presents an alternative physical interpretation of the ratio values.

\begin{figure*}
    \centering
    \includegraphics[width=\textwidth]{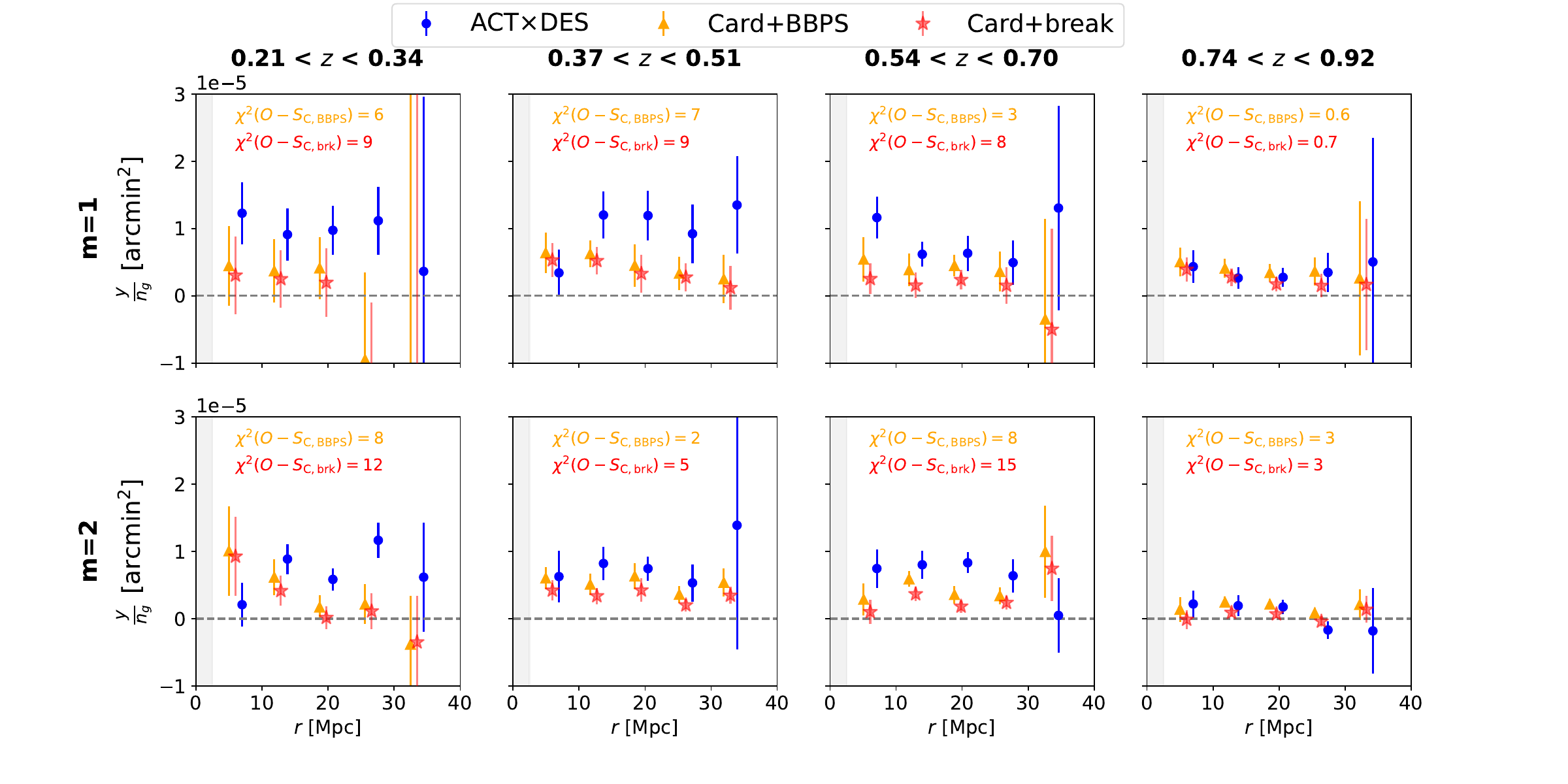}
    \caption{The ratio between the first two moments of stacked $y$ and stacked $n_\mathrm{g}$ as a function of radial distance $r$. The relationship is consistent as a function of $m$ and $r$. Qualitatively, the Cardinal results appear to slightly underestimate the relationship, but statistically, the simulated ratio using the BBPS model is consistent with the observed. The \textit{break model} is only marginally less consistent.}
    \label{fig:y_ng_relationship}
\end{figure*}

Finally, we study the relationship between $\Delta y$ and $\Delta n_g$ in Fig.~\ref{fig:y_ng_relationship}. We compare the observational result to the Cardinal simulations, in which this ratio---the amount of excess electron pressure per excess galaxy (in projection)---depends on the prescriptions for the \maglim galaxies and gas pressure profiles applied to the underlying halo distribution. The figure shows comparisons between ACT$\times$DES, the Cardinal simulation with a BBPS pressure prescription, and the simulation with the standard \textit{break model} values applied. As before, only $m=1$ and $m=2$ results are shown, as $m=3$ and $m=4$ have such large uncertainties that all bins are statistically consistent. Taken as a whole with their bin-to-bin correlations considered, the $y/n_g$ profiles are statistically consistent between the Cardinal+BBPS simulations and data, as shown by all $\chi^2$ values being $<11$ (therefore $p>0.05$). The \textit{break model} is marginally less consistent in each redshift bin, given that it always suppresses the large-scale-averaged $y$ signal, but the difference from observations is only significant in the first and third redshift bins of $m=2$.

It is interesting to see that both the relationship predicted by Cardinal and that observed in the data is fairly consistent as a function of $r$ and $m$. Interpreted through the lens of Cardinal, which only contains a halo model prescription, this suggests that each 7.5~Mpc transverse radial bin encompasses a similar distribution of halo masses. If the distribution had radial dependence on these scales, the different dependencies on mass for $y$ vs. galaxies in the halo models would cause fluctuations in this relationship. In future work, this could be further explored by stacking maps of the halo mass distribution itself.

With the current levels of signal-to-noise, the $y$-$n_\mathrm{g}$ response function is also independent of $m$. With future data that will improve the SNR, this measurement will be interesting to repeat as it probes the relative distribution of the tracers in their filamentary environments, which may cause an $m$-dependent effect measurable at sufficiently high precision. This concept would also be interesting to study in future work with hydrodynamic simulations with varying astrophysical feedback mechanisms, as processes which eject and heat gas further out would modulate the shape of the gas pressure distributed through filaments (by puffing it out), while presumably having smaller effects on the galaxy distribution.

\input{thermal_energy}
\section{Discussion and Conclusions} \label{sec:conclusion}

This is the second paper in a series studying anisotropic large-scale superclustering in observational data with novel techniques. In this work, we identified high-superclustering regions in DES projected photometric galaxy data, measured the LSS orientation around clusters using the same data, and stacked cutouts of an ACT Compton-$y$ map, DES weak lensing mass map, and DES galaxy number density maps with orientation. In doing so, we produced high significance measurements of extended anisotropic structure in four distinct redshift bins. This process included several refinements of the techniques first introduced in Paper I \citep{Lokken2022PaperI}, and we broadly expanded the application by incorporating multiple matter tracers and analyzing their relationships.

For the $y$ data, we found that usage of the full ACT DR6 sky footprint, overlapping nearly completely with DES, yielded a 3-4$\times$ improvement in SNR in the stacked profiles. What began as a promising but tentative signal, only 3.5-4$\sigma$ evidence for a non-zero $m=2$ moment in Paper I, has grown to $8-10\sigma$ with the current data (when all redshifts are combined). Moreover, we found $\sim6\sigma$ detection of the $m=4$ moment in the $y$ data when stacking on $\lambda>10$ clusters and $\sim3.5\sigma$ detection when stacking on more massive $\lambda>20$ clusters, whereas with the previous 450 sq. deg. $y$ map the $m=4$ moment was not detectable. This demonstrated our sensitivity to late time non-Gaussianity: we showed with simulations in Paper I that the same methodology produces an $m=4$ signal only for non-Gaussian fields, and null $m=4$ signal when applied to a Gaussian random field. The addition of the gradient-based flip to the stacking procedure additionally yielded a strong detection of a large-scale dipole in the thermal energy distribution ($m=1$), interpreted as being sourced from the typical asymmetry of LSS along the supercluster axis to either side of each selected cluster. Due to the SNR improvements, we were also able to split the data into four distinct redshift bins and measure significant $y$ signal up to $m=2$ in most of the bins.

We expanded the application of the techniques to DES \maglim galaxy number density and a single weak lensing convergence map sourced from the highest-redshift galaxies in DES. The visual similarity of the stacks qualitatively demonstrated that they probe the same underlying matter distribution. In the abundant galaxy number density data, the errorbars are small and we find significant detections of all cosine multipoles up to $m=4$ in combined redshift data, as well as in some individual redshift bins. The lensing data is noisier, but tells a similar story.

Using the Buzzard simulation of the DES galaxy and cluster populations as well as its successor, Cardinal, we proceeded to analyze the consistency between simulations and DES data using a multipole decomposition of the galaxy stacks. There is very poor agreement with the Buzzard results due to the inaccurate large-scale cluster-galaxy bias in the simulations; this is greatly improved in Cardinal. When normalized by the $m=0$ moments, the large-scale bias is largely removed; this leads to the observed galaxy data, both simulations, and the observed surface mass density from weak lensing (tracing the same matter) all coming into broad agreement. These results demonstrate the impact of the galaxy bias on the oriented stack observables, a method to mitigate it, and the underlying consistency in the shape of structure between the observations and simulations.

Although the non-normalized Cardinal galaxy results still show many inconsistencies with the observed galaxy distribution, we found that in particular redshift bins and multipole moments, the stacked galaxy signal in the simulations predicts the observed signal well. In those cases --- where the Cardinal large-scale structure mimics that of our real universe, despite whatever cosmological and survey differences exist between the DES and Cardinal skies---we highlighted the comparison between the stacks of ACT vs. Cardinal Compton-$y$. By creating $y$ maps for Cardinal with two different models--- BBPS, which gives a constant $Y-M$ power-law relationship, and the \textit{break model}, which depletes $Y$ at masses below $M_\mathrm{br}=2\times10^{14}$~\hMsun\ (both with a particular set of fixed parameters) ---we tested how two different feedback models hold up against extended $y$ measurements. In all highlighted cases, BBPS is a better fit, and is even significantly too low in one case. However, when swapping the observational $y$ data for the CIB-deprojected $y$ map with the strongest impact on the results, the model preference is washed out for most redshifts by a decrease in $y$ signal and enlarged errorbars. In that case, there is a preference for the \textit{break model} at high $z$.

Lastly, we made use of these multi-tracer data products to study the anisotropic responses of the gas pressure to the galaxies and total matter content. We measured the ratios of $y$ to galaxies and $y$ to lensing as a function of multipole moment and distance from the stacked clusters. The results, although noisy, are remarkably consistent over $r$ and between $m=1$ and $m=2$, for both ratios. This demonstrates that when oriented on $\sim15$~Mpc scales, the anisotropic superclustering of matter in galaxies, hot gas, and dark matter is coherent. In other words, all of the forms of matter have similar large-scale shapes and extents; they trace each other. As the Cardinal results for the $y$/$n_\mathrm{g}$ ratio are similarly consistent, we observe that this is expected under the halo model. Of the two halo models applied, the \textit{break model} is less consistent with the ficudial $y$ data for $z<0.7$, and has no preference under the chosen CIB deprojection.

The \textit{break model} was found to fit similar tSZ data well in previous works \citep[see][and references therein]{Pandey2022PhRvD.105l3526P}, and also fits low-mass halos better than BBPS in some simulations with AGN feedback \citep{LeBrun2017MNRAS.466.4442L}. Even when taking the non-CIB-deprojected results as observed truth, the inconsistency found with the \textit{break model} doesn't necessarily go against those findings. We find this inconsistency when \textit{explicitly} observing extended structure, rather than extracting signal mostly from the interiors of halos. In this case, it is possible that the uncompensated \textit{break model} (the one we implemented) does poorly because the thermal energy removed from the low-mass halos is not accounted for beyond those halos, and thus the large-scale binned signal is overly depleted in the model. Recent works such as \citet{Amodeo2021}, \citet{Schaan2021}, and \citet{Hadzhiyskafeedback2024} have found evidence for highly extended gas profiles, which is consistent with this picture.

To advance these probes to a stage in which the measurements can be used for cosmological or gas physics constraints, it is clear that further work is also needed on the modelling side. In modern hydrodynamic simulations, feedback redistributes and heats gas well beyond halo boundaries and tidal forces shock-heat gas in an anisotropic manner. Depending on how extreme the feedback is, for some simulations it is unlikely that the halo model framework can properly capture the tSZ signal. For example, in \theth cluster simulations, gas beyond $2R_{200c}$ of halos contributes $\sim25\%$ of the tSZ signal \citep{Lokken2023MNRAS.523.1346L}. This effect is not encapsulated by simply radially extending the BBPS profiles with fixed parameter values, as the outskirts have too small $y$ signal (Lokken et al., in prep.), but could potentially be fit by also varying the parameters to modulate the profile shapes for a certain range of halo masses. The \textit{compensated} \textit{break model} is a related way to rapidly simulate the effects of the pushed-out gas by feedback, as modeled in \citet{Horowitz2017MNRAS.469..394H} and studied in \citet{Hill2018}. However, any halo model encounters the challenge of avoiding over-counting from neighboring overlapping halos. Mixed halo-filament rapid modelling which incorporates the tidal field, e.g. extending the type of anisotropic modeling studied in \citet{Paranjape2021MNRAS.502.5210P}, may be a way forward. More ideally, future attempts to place cosmological and/or gastrophysics constraints with anisotropic methods like oriented stacking would be most informative via comparison to large-volume suites of hydrodynamical simulations like \textsc{Flamingo} \citep{Schaye2023MNRAS.526.4978S}, which contains both cosmology and gas variations. The challenge in those cases is to post-process the simulations with survey selection effects, or marginalize over them.

In future  work, while pursuing improvements in the modelling, we will also apply similar oriented techniques to data from ACT in combination with galaxy data from the Dark Energy Spectroscopic Instrument, DESI \citep{DESI2019}. The DESI survey, currently taking data, offers the advantage of spectroscopic redshifts, which may improve the accuracy of LSS orientation. However, it also has lower galaxy number density, a disadvantage. Given this tradeoff, as done here with DES, we will explore different configurations of the smoothing scale and superclustering constraints with DESI mocks to optimize signal-to-noise in the oriented cross-correlations. The work will aim to develop an inference framework with the goal of placing constraints on the gas physics.

This series of works presents a novel approach to measuring a beyond-two-point cross-correlation statistic of cosmological matter tracer fields which is flexibly applicable to both clean and noisy data. If isotropic statistics continue to confirm $\Lambda$CDM with greater precision in forthcoming cosmological data, we hope that ultimately beyond-2-point statistics like this one will prove useful in searches for evidence of beyond-$\Lambda$CDM physics. While Paper I presented some of the conceptual motivations behind focusing on localized anisotropies, future simulation-based work will explore the ways in which primordial non-Gaussianity and non-standard dark energy impacts local anisotropic structure to further motivate this concept. Before the cosmological applications are possible, however, we must understand the physics that distributes baryons throughout the cosmos; our study presents a step toward that goal by testing halo-model approaches for the baryons in a unique way. 
\input{acknowledgments}

\appendix

\section{Comparing the \maglim vs \redmagic catalogs for orientation} \label{appdx:maglim_v_redmagic}

To test whether the trade-off between the higher number density and poorer redshift precision of \maglim is worthwhile, compared to \redmagic, we use the Buzzard mocks \citep{DeRose2019}. Both the \redmagic algorithm and the DNF \citep{devicente2016MNRAS.459.3078D} algorithm were run on Buzzard galaxy catalogs to produce mocks of \redmagic and \maglim. The mock catalogues reproduce the number density $n(z)$ and photometric uncertainties $\sigma_z$ reasonably well \citep{DeRose2019, Porredon2021_cosmoconstraints}, although the Buzzard \maglim sample is somewhat over-optimistic in $\sigma_z$ at $z\sim0.4$ due to the fact that the simulations do not reproduce a degeneracy between magnitudes and photo-$z$s feature that appears in the real data.

For a test sample between $z=0.3$ and $z=0.6$, we apply our orientation methods (described in further detail in Sec.~\ref{sec:orientation_methods}) separately to each catalogue, by mapping out galaxies in thin redshift slices using their position and photometric redshift information. We determine the orientation of LSS surrounding locations of Buzzard mock \redmapper clusters for the Gaussian- smoothed galaxy maps, using a FWHM of 20 Mpc. To test which catalogue produces more accurate orientations, a reference for the `ideal' orientation is needed. Because the \maglim sample has higher number densities, for the ideal comparison we apply the same methods to the \maglim catalogue using the \textit{true} redshifts of the galaxies from the simulation. Finally, we measure the offset in the orientation angle, defined with respect to the R.A.axis, determined at each location between the photo-$z$ result (\maglim or \redmagic) and the ideal angle from true-$z$s. We find that, in both cases, the distribution of angle offsets is skewed toward 0\degree  with a tail extending to 90\degree. For \maglim, 50$\%$ of orientation angles become offset by less than 10\degree from the ideal angles. For \redmapper, that percentage is only 40$\%$; the \redmapper sample performs slightly worse. This test indicates that the increase in photo-$z$ accuracy with \redmapper does not compensate for the loss in number density when determining orientation.

We test how the orientation angle accuracy degrades as the orientation fraction decreases. We find $<6$\degree average angle offset when including $3/4$ of the \maglim data for orientation, which increases to $\sim10$\degree when decreasing the sample to $1/2$ and $\sim15$\degree for $1/4$. We consider a $<6$\degree average offset to be acceptable and choose to keep $3/4$ of the data for orientation.

\section{Tests of bin size and p(z) with Websky}\label{appdx:binsize_tests}
To seek the highest orientation accuracy given broad photometric redshift uncertainties in the galaxy data, we test several variations to the redshift treatment. We begin with variations to the redshift bin size of the projected galaxy data. The crux of the question is, what redshift slice width is ideal? Too thin, and the photometric redshift uncertainties introduce too much noise. Too large, and projection of uncorrelated foreground and background structure dominates. The latter can be problematic because when the orientation of a cluster is determined by uncorrelated structure in the galaxy field, the stacks on any correlated tracer (the galaxies themselves, tSZ, or lensing) will also stack that structure. Thus, the resulting stack contains signal along the orientation axis, but this signal does not correspond to that from true correlated filaments and superclusters. In such a case, the signal would reflect the projected alignment correlation function for that redshift bin -- a fine measurement in itself, but unlikely to reveal insights into filaments.

We take $z=0.4$ as a test case for all tests in this section, where the $\sigma_z/(1+z)$ for the Maglim galaxy sample is poorest \citep{Porredon2021}. For a point of comparison, we first set up the simulation to yield `ideal' orientation, defined as follows. Taking all Websky halos with $M>1.5\times10^{12}$ Msun, and a satellite fraction of 0.15 (similar to the satellite fraction of Maglim at this redshift \citep{Zacahregkas2022}, we make mass-weighted maps of the halo density in bins of 200 Mpc. Due to photometric redshift uncertainties, we do not explore the possibility of a narrower bin than 200 Mpc. Although 200 Mpc is large compared to the scales of interest (the inter-cluster bridge scale, $\sim10-20$ Mpc on either side of a cluster so $\lesssim40$ Mpc in total), and thus introduces some projection effects into orientation, this width is motivated by both the \maglim and \redmapper cluster photo-$z$ uncertainties. For \maglim, the  average uncertainty is  typically $0.02<\sigma_z/(1+z)<0.05$ \citep{Porredon2021}, so the comoving distance spanned by the 68\% confidence interval $(z-\sigma_z, z+\sigma_z)$ is $\geq200$~Mpc. For \redmapper, the average uncertainty is $0.01<\sigma_z/(1+z)<0.02$ \citep{McClintock2019}, corresponding to $\sim100-200$~Mpc. Thus determining the orientation from structure restricted to bins smaller than 200~Mpc would increase the mismatch in physical space between clusters and surrounding galaxy data in the angular size.

For the `ideal' case, we create overlapping 200 Mpc bins, where each bin center is offset by 100 Mpc compared to the previous one. We determine the orientation for clusters which lie within 50 Mpc of the bin center. Thus, the clusters (if their redshifts were perfectly determined) are situated in the midst of a larger redshift bin used to provide large scale structure information. For testing purposes, we do not introduce photometric uncertainty into the simulated cluster $z$, only the galaxy $z$, choosing to focus on the impacts of the (much larger) galaxy photometric redshift uncertiainties.

For photo-$z$ testing, we update the catalog of Websky mock Maglim galaxies to one in which the galaxy photo-$z$ is randomly drawn from a Gaussian distribution around the true-$z$, with a sigma of $0.05\times(1+z)$ \citep[the largest $\sigma_z$ in Maglim,][] {Porredon2021}. We then create three approaches to orientation with the mock photo-$z$-smeared data for comparison with the ideal orientation.

In the first, we combine all galaxies in a bin with $\Delta z$ = 0.15, a typical Maglim bin size. We make a smoothed $\delta_\mathrm{g}$ map in which all galaxies are weighted equally and measure the orientations at positions of $M>10^{14} \Msun$ halos (approximately $\lambda>20$ for \redmapper). For a fair comparison, we choose a bin with this approximate width (specifically, $0.31<z<0.45$), to line up with the clusters in three consecutive 200 Mpc bins for a fair, one-to-one comparison of the orientations.

In the second, we repeat the 200 Mpc-wide bin method, with half-offset bins, in exactly the same manner as the ideal case but using the mock photo-$z$ galaxy catalog.

In the third, we attempt to account for the fact that every galaxy's position in redshift space is an extended probability function along the line-of-sight direction. Rather than representing each galaxy as just a single galaxy at its mean photometric redshift, we create 10 samples of each galaxy, spaced out in varying increments to reflect a Gaussian distribution, i.e., the $p(z)$. This is similar to integrating over the probability between two redshifts and applying a weight to the galaxy to reflect how likely it is to truly lie in that redshift range. However, integration is computationally slow, while the Gaussian-spacing method is much faster as the spacing of the $z$ samples can be rapidly stretched or compressed for a given $\sigma_z$.

For the stacking points, we take Websky halos with $M>10^{14}$ corresponding approximately to $\lambda>20$ for \redmapper. We smear their redshifts by taking a random number from a Gaussian distribution with $\sigma_z/(1+z)=0.015$. We then split the clusters into 100 Mpc bins, where we will find the orientation for bins in each cluster by the number density map of the projected galaxy field from the surrounding 200 Mpc. 

We find that compared to the ideal case, orienting the clusters by the unweighted photo-$z$ orientation maps with 200 Mpc bin width results in the closest-matching orientations, with an average $\Delta \theta$ of $\sim30^{\circ}$ (method 2). The map in which galaxies are projected over a bin with $\Delta z$=0.15 (method 1) overly incorporates projection effects, such that the orientation angles are nearly randomly distributed in comparison to the ideal case. 
Meanwhile, the maps with weighted photo-$z$ treatment (method 3) do not improve the orientation accuracy over the projected 200 Mpc bin approach.

It is unclear why the $p(z)$ approach was ineffective. The $z_{\mathrm{mean}}$ determined by a given algorithm like \redmagic is not the true $z_{\mathrm{mean}}$ of the cluster, so the estimated $p(z)$ distribution is always shifted from the true. Perhaps there is a better way to represent $p(z)_{estimate}$ given the likelihood for $z_\mathrm{mean} | z_{\mathrm{true}}$. This would be an interesting avenue for future work given the abundance of existing photometric data from DES and forthcoming from LSST.

\section{Redshift isolation, mean-zero maps, and their impacts on tracer ratios} \label{appdx:ratios_tracers}
Comparing different observational tracers of matter is subtle due to their different relationships to the underlying quantities, redshift dependence, and observational treatment. These subtleties become more important when computing ratios in regimes of low signal-to-noise. For clarity, in this appendix we describe analytically what is probed by the ratios of oriented stacks in Compton-$y$, the convergence $\kappa$, and the galaxy number density $n_\mathrm{g}$. The main purposes of this appendix are to detail the assumptions about the redshift distribution of the signals we are sensitive to via oriented stacks, describe why it is necessary to use mean-zero maps in order to take ratios of maps with different amounts of line-of-sight projection, and highlight where correlated noise enters into the resulting measurements. This contextualizes the measurements in Sec.~\ref{sec:res_multiple}.

For each $i^{\rm{th}}$ cluster at sky coordinate $\boldsymbol{\hat{n}}_i=$(R.A.,dec), in the basis defined by the LSS eigenvalues, the surrounding $y$ field can be described in terms of a distance from the cluster $r$ and an angle $\theta$ from the long-axis. Let us say $y'=n_\mathrm{e} \frac{k_\mathrm{B} T_\mathrm{e}}{m_\mathrm{e} c^2} \sigma_\mathrm{T}$, in other words, refer to the integrand of the Compton-$y$ parameter (Eq. \ref{eq:tSZ}) as $y'$, related to the 3D electron pressure. The measured $y$ signal is:
\begin{equation}
    \Delta y(r, \theta)|\boldsymbol{\hat{n}}_i = \int_{0}^{\infty}d\ell\, y'(r,\theta,\ell)|\boldsymbol{\hat{n}}_i -  \frac{1}{4\pi f_{\rm{sky,A}}} \int_{\rm{A}}\,d\Omega \int_{0}^{\infty} d\ell\, y'(\boldsymbol{\hat{n}}, \ell) + \tilde{N}_{\rm{y}},
\end{equation}
where the second term is the sky-averaged $y$ signal over the ACT footprint covering $f_{\rm{sky,A}}$, $\langle y \rangle$, and $\tilde{N}_{\rm{y}}$ stands for noise. Here, `noise' refers to any non-signal component with mean 0.

Consider a large-scale structure, such as a chain of several clusters, that is highly correlated with a cluster at redshift $z_i$. Generally, the $y$ integrals can be divided up arbitrarily, so let us write them as an addition of the component coming from all redshifts at which there is correlated structure, and the component containing all other redshifts. The same can be done for the averages $\langle y \rangle$. Then we have,

\begin{align*}
    \Delta y(r, \theta)|\boldsymbol{\hat{n}}_i &= y^{\rm{corr}(z_i)}+y^{\rm{uncorr}} - \langle y^{\rm{corr}(z_i)} \rangle - \langle y \rangle^{\rm{uncorr}} + \tilde{N}_y\\
    &\approx \Delta y(r,\theta, z_i\pm \delta z_i)|\boldsymbol{\hat{n}}_i + \tilde{N}_{y},
\end{align*}
where in the second step we subsumed $y^{\rm{uncorr}}-\langle y \rangle^{\rm{uncorr}}$ into the noise $\tilde{N}$ because it is a mean-zero variable: at any coordinate $(r,\theta)$ centred at any $\boldsymbol{\hat{n}}_i$, it is equally likely to be above or below the mean, which is not true for the correlated redshifts. In the second step we also made the approximation that the correlated structure spans a redshift range $\pm \delta z_i$ around the cluster at $z_i$, such that all the $\Delta y$ signal comes from integrating over those redshifts, an assumption we'll further discuss later. Next, we stack (with orientation) $N$ clusters and their large-scale surroundings, limited to a (narrow) redshift slice centered at $z_{\rm{slice}}$:

\begin{align} \label{eq:y_stack}
    S_y \equiv \langle \Delta y(r, \theta) \rangle |\xi &= \frac{1}{N} \sum_{i=1}^{N} \big(\Delta y(r,\theta, z\pm \delta z_i)|\boldsymbol{\hat{n}}_i + \tilde{N}_y\big) \nonumber\\
    &\approx \Delta y(r, \theta, z_\mathrm{slice} \pm \delta z)|\xi + \tilde{N}_y,
\end{align}
where $\xi$ are the constraints placed to select the sample of $\boldsymbol{\hat{n}}_i$ coordinates. In the first line, we define the stack $S_y$ as the average over the integrated $y$ signal from many different structures in the narrow redshift bin. In the second, we make the assumption that this is equivalent to integrating over the extent of a single, representative, average structure which is centered at the midpoint of the bin and has redshift extent $\delta z$, the average of the $\delta z_i$. The narrow width of not only each of our cluster bins (100~Mpc) but also the galaxy bins (200~Mpc), which identify correlated structure well beyond the clusters through the orientation procedure, help to justify these approximations.

Ideally, we would convert $S_y$ into a measure of a $y$ overdensity by dividing by $\langle y(z_{\rm{slice}}) \rangle$, the sky-averaged mean $y$ at the cluster redshift. However, as there are only theoretical predictions for this value, including the theory prediction would introduce a model-dependence to the measurements that we wish to avoid.

Let us now construct the analogous expression for the galaxy distribution, which measures the fluctuations of the projected number density $n_\mathrm{g}$ around the mean for a given narrow redshift range. Given the knowledge of galaxy photometric redshifts, we have more control over the $z$ distribution. To begin, we divide the galaxy sample into broad bins of width $\Delta z_{\rm{g}}=z_{\rm{g,2}}-z_{\rm{g,1}} \gg \Delta z_{\rm{slice}}$, where $\Delta z_{\rm{slice}}$ represents the 
effective width of the galaxy-cluster slices used for selection and orientation (between 100 and 200~Mpc). We make maps of the projected 2D number density fluctuations $n_\mathrm{g}$ in each broad bin:

\begin{equation}
    n_{\rm{g}}(\boldsymbol{\hat{n}}) = \int_{\Delta z_{\rm{g}}} dz \, n_{\rm{g,3D}}(\boldsymbol{\hat{n}}, z).
\end{equation}
In order to make the same arguments about correlated and uncorrelated components as were presented for $y$, this map must be made mean-zero. To do so, the sky-average of $n_g(\boldsymbol{\hat{n}})$ could be subtracted at this stage. In our pipeline, we elect to instead later subtract the average over an annulus measured on the stack at large $r$, as we perform the same to the $y$ data to mitigate large-scale contamination (and do the same to $\Sigma$). The annulus average is an estimator of the sky-averaged signal, although it will also capture any offsets from structure correlated at very large scales ($30-40$~Mpc) with the constrained region centers. Concerns with this detail are discussed in Sec.~\ref{subsec:discard_monopole}, and ultimately lead us to focus only on the ratios of non-zero moments.

After mean-subtraction, the same argument about correlated and uncorrelated structure holds for galaxy number density, but now the uncorrelated structure is reduced to smaller ranges ($z_{\rm{g,1}}, z_1$) and ($z_2, z_{\rm{g,2}}$) instead of including the contributions of structure over the full \maglim redshift range, thus reducing the noise. The galaxy stack can be approximated by
\begin{align}
    S_g \equiv \langle \Delta n_g(r,\theta) \rangle | \xi
    &\approx \Delta n_g(r,\theta, z_\mathrm{slice} \pm \delta z)|\xi +\tilde{N}_g,
\end{align}
analogous to the $y$ measurement.

Thus the relationship between gas pressure and galaxy number at a given redshift, under constraints $\xi$, can be probed by dividing the stack of $y$ by the stack of $n_\mathrm{g}$,
\begin{equation}
    \frac{S_y}{S_g}=\frac{\Delta y(r,\theta,z_{\rm{slice}} \pm \delta z)|\xi +\tilde{N}_y}{\Delta n_g(r,\theta,(r,\theta,z_{\rm{slice}} \pm \delta z))|\xi + \tilde{N}_g},
\end{equation}
where $\tilde{N}_y$ and $\tilde{N}_g$ are correlated noise terms. The analogous relationship holds for the relationship between a stack on $y$ and $\Sigma$. The noise terms are correlated because they contain information about the foreground and background structure which appears in both tracers, in addition to the instrumental noise which differs between ACT and DES. Assuming these noise terms are Gaussian (which we test in Sec.~\ref{sec:tests}, finding no evidence to the contrary), we can propagate errors for each ratio as long as all covariances are taken into account (as derived in Appendix.~\ref{appdx:errors_ratios_tracers}).

A major assumption in this discussion is that the correlated structure comes from a limited redshift range around each cluster. In reality, the 3D structures contributing superclustering signal span various true lengths and extend to different projected lengths along the line-of-sight, depending on their 3D orientations. In some cases, they can contribute signal beyond the narrow redshift range of the cluster or galaxy bin. Additionally and more importantly, as our oriented stacking approach uses photometric cluster and galaxy data, there is a great deal of mixing between truly correlated and uncorrelated structures along the line-of-sight direction. Therefore, the interpretation of the signal as an average over structures with limited l.o.s. extents, such as inter-cluster bridges, is likely to be inaccurate. Nevertheless, as we combine the results from five neighboring redshift slices into a single larger bin for each of the four redshift bins that enter our final analysis, it is more accurate to assume the signal in the larger redshift bin is truly coming from within that bin. Future work with spectroscopic data will reveal how much the photo-$z$ uncertainties play a role in the results.

\section{Uncertainties on ratios of tracers} \label{appdx:errors_ratios_tracers}

In Sec.~\ref{sec:res_multiple} we compute the ratio of multipole measurements in Compton $y$ to projected galaxy density and surface mass density. Here we derive the covariance matrix for those ratios. Given the measured moments of the first stacked tracer (e.g., $y$) in a radius bin $r_k$, generically $ X_k \equiv  X(r_k)$, and the moments of a different tracer at the same radius, $ Y \equiv  Y(r_k)$, we find the ratio $R(r_k)$. In general, we find that at radial bins $i$ and $j$ the measured $X_i$ and $Y_{j}$ moments are correlated, as the different signatures trace the same or similar underlying structure, which is itself correlated on large scales (i.e., between radial bins). We expect this covariance to lead to a reduction of the uncertainty in the ratio $ R_k$, compared with a case where the two measurements are uncorrelated.  Furthermore, in order to compare with models we would like to compute the covariance of the ratio between radial bins.  Propagating errors, we have
\begin{equation}
     \mathrm{Cov}( R_k,  R_l) = \begin{bmatrix} \frac{\partial R_k}{\partial X_i} & \frac{\partial R_l}{\partial Y_i}  \end{bmatrix}
     \begin{bmatrix} \mathrm{Cov}( X_i,  X_j) & \mathrm{Cov}( Y_i,  X_j) \\ \mathrm{Cov}( X_i,  Y_j) & \mathrm{Cov}( Y_i,  Y_j)  
     \end{bmatrix}
     \begin{bmatrix} \frac{\partial R_l}{\partial X_i} \\ \frac{\partial R_l}{\partial Y_i}
     \end{bmatrix},
\end{equation}
with summation implied over the repeated indices $i$ and $j$.  Given that $\frac{\partial R_k}{\partial Y_i} = \frac{1}{Y_k} \delta_{ik} $    and $\frac{\partial R_k}{\partial Y_i} = -\frac{ X_k}{ Y_k^2} \delta_{ik} $, this becomes 
\begin{equation}
     \mathrm{Cov}( R_k,  R_l) = \frac{1}{ Y_k  Y_l} \left (  \mathrm{Cov}( X_k,  X_l) - \frac{ X_l} {  Y_l}\mathrm{Cov}( X_k, Y_l)  - \frac{ X_k }{  Y_k}\mathrm{Cov}( Y_k, X_l) + \frac{ X_k }{  Y_k}\frac{ X_l }{  Y_l} \mathrm{Cov}( X_k, X_l) \right ).
\end{equation}
We use this expression for both the ratio $ y /  n_g$ and $ y /  \Sigma$ analyzed in Sec. \ref{sec:res_multiple}.

\bibliography{biblio, thesis_biblio}
\section*{Affiliations}
\input{affil_at_end}

\end{document}

%% file: authors_numbered.tex
\author[0000-0001-5917-955X]{M.~Lokken$^{1,2,3,4}$}\thanks{E-mail: mlokken@ifae.es}

\author[0000-0002-3495-158X]{A.~van~Engelen$^5$}

\author{M.~Aguena$^6$}

\author[0000-0002-7069-7857]{S.~S.~Allam$^7$}
  
\author[0000-0003-3312-909X]{D. Anbajagane$^{8,9}$}

\author{D.~Bacon$^{10}$}

\author{E.~Baxter$^{11}$}

\author[0000-0002-4687-4657]{J.~Blazek$^{12}$}

\author[0000-0002-4900-805X]{S.~Bocquet$^{13}$}

\author[0000-0003-2358-9949]{J.~R.~Bond$^3$}

\author[0000-0002-8458-5047]{D.~Brooks$^{14}$}

\author{E. Calabrese$^{15}$}

\author[0000-0003-3044-5150]{A.~Carnero~Rosell$^{16,6}$}

\author[0000-0002-3130-0204]{J.~Carretero$^1$}

\author{M.~Costanzi$^{17,18,19}$}

\author{L.~N.~da Costa$^6$}

\author{W. R.~Coulton$^{20,21}$}

\author[0000-0001-8318-6813]{J.~De~Vicente$^{22}$}

\author[0000-0002-0466-3288]{S.~Desai$^{23}$}

\author{P.~Doel$^{14}$}

\author{C.~Doux$^{24,25}$}

\author[0000-0003-2856-2382]{A. J.~Duivenvoorden$^{26,27}$}

\author{J.~Dunkley$^{27,28}$}

\author{Z.~Huang$^{29}$}

\author{S.~Everett$^{30}$}

\author{I.~Ferrero$^{31}$}

\author[0000-0003-4079-3263]{J.~Frieman$^{7,9}$}

\author[0000-0002-9370-8360]{J.~Garc\'ia-Bellido$^{32}$}

\author{M.~Gatti$^{24}$}
\author[0000-0001-9632-0815]{E.~Gaztanaga$^{33,10,34}$}

\author[0000-0002-3730-1750]{G.~Giannini$^{1,9}$}

\author[0000-0002-3589-8637]{V.~Gluscevic$^{35}$}

\author[0000-0003-3270-7644]{D.~Gruen$^{13}$}
\author{R.~A.~Gruendl$^{36,37}$}

\author[0000-0002-1697-3080]{Y.~Guan$^4$}

\author[0000-0003-0825-0517]{G.~Gutierrez$^7$}

\author{S.~R.~Hinton$^{38}$}

\author[0000-0002-0965-7864]{R.~Hlo\v zek$^{2,4}$}

\author{D.~L.~Hollowood$^{39}$}

\author[0000-0002-6550-2023]{K.~Honscheid$^{40,41}$}

\author[0000-0001-5160-4486]{D.~J.~James$^{42}$}

\author[0000-0003-0120-0808]{K.~Kuehn$^{43,44}$}

\author[0000-0002-1134-9035]{O.~Lahav$^{14}$}

\author{S.~Lee$^{45}$}

\author[0000-0002-0309-9750]{Z.~Li$^{46,47,3}$}

\author[0000-0001-6740-5350]{M.~Madhavacheril$^{24}$}

\author[0000-0002-8571-8876]{G. A. Marques$^{7,9}$}

\author[0000-0003-0710-9474]{J.~L.~Marshall$^{48}$}

\author[0000-0001-9497-7266]{J. Mena-Fern{\'a}ndez$^{49}$}

\author[0000-0002-1372-2534]{F.~Menanteau$^{36,37}$}

\author[0000-0002-6610-4836]{R.~Miquel$^{50,1}$}

\author{J.~Myles$^{28}$}

\author[0000-0001-7125-3580]{M. D. Niemack$^{51,52}$}

\author{S.~Pandey$^{53}$}

\author{M.~E.~S.~Pereira$^{54}$}

\author[0000-0001-9186-6042]{A.~Pieres$^{6,55}$}

\author[0000-0002-2598-0514]{A.~A.~Plazas~Malag\'on$^{56,57}$}

\author{A.~Porredon$^{22,58}$}

\author[0000-0001-6163-1058]{M.~Rodr\'iguez-Monroy$^{32}$}

\author[0000-0001-5326-3486]{A.~Roodman$^{56,57}$}
\author{S.~Samuroff$^{12}$}
\author[0000-0002-9646-8198]{E.~Sanchez$^{22}$}
\author[0000-0003-3054-7907]{D.~Sanchez Cid$^{22}$}
\author{B.~Santiago$^{59,6}$}
\author[0000-0001-9504-2059]{M.~Schubnell$^{60}$}
\author[0000-0002-1831-1953]{I.~Sevilla-Noarbe$^{22}$}

\author[0000-0002-8149-1352]{C.~Sif\'on}

\author[0000-0002-3321-1432]{M.~Smith$^{62}$}

\author[0000-0002-7020-7301]{S.~T.~Staggs$^{27}$}

\author[0000-0002-7047-9358]{E.~Suchyta$^{63}$}

\author{M.~E.~C.~Swanson$^{36}$}

\author[0000-0003-1704-0781]{G.~Tarle$^{60}$}

\author[0000-0001-7836-2261]{C-H. To$^{40,41,64}$}

\author{N.~Weaverdyck$^{65,47}$}

\author{P.~Wiseman$^{66}$}

\author[0000-0002-7567-4451]{E. J. Wollack$^{67}$}


%% file: thermal_energy.tex
\subsection{Thermal energy per baryon} \label{sec:energy_ratio}


To derive more meaning from the physical values in Fig.~\ref{fig:y_Sigma_ratio}, in this section, we will use several approximations to compute the fraction of thermal energy compared to baryon rest-mass energy in the supercluster regions.

If we assume there is no anisotropic bias or nonlinear bias in the lensing signal or gas pressure, and that they trace exactly the same structure, then we expect that the $y$-to-$\Sigma$ ratio of each cosine and sine component is equal to the total signal ratio. In other words, we go beyond the definition and decomposition from Eq.~\ref{eq:image_composition_series}, 
\begin{equation}
\langle I_y(r,\theta) \rangle = \sum_m \big (  C_{y,m}(r) \cos(m \theta)+S_{y,m}(r) \sin(m \theta) \big),\nonumber
\end{equation}
and the equivalent for $\Sigma$, to make the assumption that the following should be theoretically true:
\begin{align}
    I_y(r,\theta) /  I_\Sigma(r,\theta) \rangle &= C_{\mathrm{0},y}(r) / C_{\mathrm{0},\Sigma}(r) \nonumber \\
    &= C_{\mathrm{1},y}(r) / C_{\mathrm{1},\Sigma}(r) \nonumber \\
    &= S_{\mathrm{1},y}(r) / S_{\mathrm{1},\Sigma}(r) \nonumber \\
    &=...
\end{align}
and so on for the higher moments. I.e., we assume the ratio to be a function of $r$ only, and thus equivalent for each moment; in the previous sections we have observed this to be roughly true. Therefore, we can use $m=1$ and $m=2$ results as a proxy for the relationship in the full stack and avoid the need for more careful attention to the $m=0$ annulus subtraction.

Because we observe the relationship to be more or less constant over $r$, let us take the average over $m=1$ and $m=2$, as well as $r$, as $\langle y/\Sigma \rangle$. The second assumption we will make is to assume the relationship does not evolve over the line-of-sight extent of our average stacked structure in a given redshift bin ($\Delta z \sim 0.15$). If we do so, we can interpret the $y/\Sigma$ relationship in terms of a ratio of thermal energy density to the energy density of combined dark matter plus baryons. The argument follows.

Assuming a fully-ionized gas consisting of Hydrogen and Helium, we can compute the fraction of number of electrons to total number of particles in the system. We will assume negligible heavier elements, and take a Helium gas mass fraction of 0.24 and Hydrogen 0.76. The number fraction of Helium is then $N_\mathrm{He}=0.24/4=0.06$ for the typical 4-nucleon Helium atom and $N_\mathrm{proton}=0.76$ for Hydrogen. With two electrons per Helium and one per Hydrogen, $N_\mathrm{e}=0.76+0.24/2=0.88$. The total number fraction including electrons is $N_\mathrm{T}= N_\mathrm{He}+N_\mathrm{proton}+N_\mathrm{e}=1.7$. Then, electron pressure $P_\mathrm{e}$ is related to the total thermal pressure $P_{\rm{th}}$ by:
\begin{equation}
    P_\mathrm{e} = \bigg[\frac{N_\mathrm{e}}{N_T}\bigg]P_\mathrm{th}.
\end{equation}
The factor multiplying $P_\mathrm{th}$ is the same that appears, e.g., in \citet{Pandey2022PhRvD.105l3526P} as [$(4-2Y_\mathrm{He})/(8-5Y_\mathrm{He})]$, where $Y_\mathrm{He}$ is the helium mass fraction.
The electron pressure is related to $y$ as
\begin{equation}
    y = \int \frac{\sigma_T}{m_e c^2} P_\mathrm{e} d\ell .
\end{equation}
Because pressure is equivalent to energy per volume, we can use these relations to relate our observations to the total thermal energy enclosed in a volume $V$, assuming an ideal monatomic gas,
\begin{align}
    E_\mathrm{th}&= \int_V \frac{3}{2} n k_B T \nonumber \\
    &= \int_V \frac{3}{2} P_\mathrm{th} \nonumber \\
    &= \int_V \frac{3}{2}\frac{N_\mathrm{T}}{{N_\mathrm{e}}} P_\mathrm{e} \nonumber \\
    &= \frac{3}{2} \frac{N_\mathrm{T}}{{N_\mathrm{e}}}  \int_A \int_{L_1}^{L_2}  P_\mathrm{e} \dv{\ell} \nonumber \\
    &= \frac{3}{2} \frac{N_\mathrm{T}}{{N_\mathrm{e}}} \int_A \frac{y\, m_e c^2}{\sigma_T}  ,
\end{align}
where in the last lines we have divided a cylindrical volume integral into an area integral and line-of-sight integral, and assumed that the $y$ signal comes only from a limited extent from $L_1$ to $L_2$ in physical coordinates. Meanwhile, the total matter (combined dark matter and baryon) rest mass energy in the same physical volume is:
\begin{align}
    {E_\mathrm{DM+B}} &= \int_A \int_{L_1}^{L_2}  \rho c^2 \dv{\ell} \nonumber \\
    &\approx \int_A \int_{L_1}^{L_2}  (\rho - \bar{\rho}) c^2 \nonumber \\
    &= \int_A \Sigma c^2.
\end{align}
The ratio is:
\begin{align} \label{eq:frac_therm_energy}
\frac{E_\mathrm{th}}{E_\mathrm{DM+B}} &= \frac{\int_A \frac{3}{2}\frac{N_\mathrm{T}}{{N_\mathrm{e}}} \frac{y c^2 m_e}{\sigma_T} }{\int_A \Sigma c^2\,} \nonumber \\
&\approx \frac{\bar{y}}{\bar{\Sigma}} \frac{3N_\mathrm{T} m_e}{2 N_\mathrm{e} \sigma_T},
\end{align}
where the averages in the last line are over area, following the assumption that both $y$ and $\Sigma$ maintain the same projected relationship throughout the underlying matter distribution.  The result is a dimensionless fraction. Using the $\frac{y}{\Sigma}$ results from only the first redshift bin due to its lower noise, we take the inverse-variance-weighted average of the first 4 radial bins of the $m=1$ and $m=2$ moments, getting a value of $y/\Sigma\sim6\times 10^{-20}$~M$_\odot^{-1}$~Mpc$^2$ and computing the fractional thermal energy as $\frac{E_\mathrm{th}}{E_\mathrm{DM+B}}\sim1.1\times10^{-6}$.

The meaning of this can be interpreted by assuming the matter is distributed as per the cosmic baryon to dark matter ratio determined by \citet{Planck2018}, i.e., $\rho_\mathrm{DM}\sim5 \rho_\mathrm{B}$. Then,
\begin{align}
\frac{E_\mathrm{th}}{E_\mathrm{DM+B}} 
&\approx \frac{E_\mathrm{th}}{5E_\mathrm{B}+E_\mathrm{B}} \nonumber \\
\frac{E_\mathrm{th}}{E_\mathrm{B}} &\approx \frac{E_\mathrm{th}}{E_\mathrm{DM+B}} \times 6
\end{align}
with which, inputting the computed value from Eq.~\ref{eq:frac_therm_energy}, we conclude that the thermal energy per baryon is $7\times10^{-6}$ times the baryon's rest mass energy in the studied regions.

For reference, thermal energy at room temperature is roughly $\sim0.025$eV, which compared to a proton's rest-mass energy is a fractional amount of $\sim 10^{-11}$. The fraction in our high-superclustering regions is $\sim10^5\times$ higher.

For another point of reference, we can estimate a gas temperature for the gas responsible for our measurements by considering the average thermal energy per baryon (assuming it is all coming from electrons) compared to the average rest mass energy per baryon as follows:
\begin{equation}
    \frac{3 Y_\mathrm{T} k_B T_e}{2 Y_\mathrm{e} m_p c^2} = 7\times10^{-6}.
\end{equation}
Considering a DM-to-baryon fraction range of 5--5.5, and the systematic uncertainty range from CIB contamination, this yields $T_e\sim(2$--7)$\times10^7$~K, or $\sim2$--6 keV, suggesting that the dominant contributions to the signal are from gas at the typical temperature of relatively massive clusters \citep[see][for the relationship between temperature and DES cluster richnesses]{Rykoff2016}. This finding suggests that every radial bin along the superclustering axis encompasses some massive halos, which dominate the signal in both $y$ and $\Sigma$ and therefore dominate the relationship. The filament gas, which should be at lower temperatures and thus contribute a distinct $y/\Sigma$ relationship, is likely being overpowered and would require much finer binning and/or some masking scheme to draw out.

In future, the approximations made in producing these estimates can be refined by using simulations to better understand what fraction of halos and filament gas contribute and dominate the measurement, and whether it is valid to apply the cosmic baryon fraction in that regime. Despite the caveats, this section serves to illustrate the relationship of our measurements to physical quantities and the utility of pursuing such measurements.


%% file: acknowledgments.tex
\section*{Acknowledgments}

Support for ACT was through the U.S.~National Science Foundation through awards AST-0408698, AST-0965625, and AST-1440226 for the ACT project, as well as awards PHY-0355328, PHY-0855887 and PHY-1214379. Funding was also provided by Princeton University, the University of Pennsylvania, and a Canada Foundation for Innovation (CFI) award to UBC. ACT operated in the Parque Astron\'omico Atacama in northern Chile under the auspices of the Agencia Nacional de Investigaci\'on y Desarrollo (ANID). The development of multichroic detectors and lenses was supported by NASA grants NNX13AE56G and NNX14AB58G. Detector research at NIST was supported by the NIST Innovations in Measurement Science program. Computing for ACT was performed using the Princeton Research Computing resources at Princeton University, the National Energy Research Scientific Computing Center (NERSC), and the Niagara supercomputer at the SciNet HPC Consortium. SciNet is funded by the CFI under the auspices of Compute Canada, the Government of Ontario, the Ontario Research Fund–Research Excellence, and the University of Toronto. We thank the Republic of Chile for hosting ACT in the northern Atacama, and the local indigenous Licanantay communities whom we follow in observing and learning from the night sky.

Funding for the DES Projects has been provided by the U.S. Department of Energy, the U.S. National Science Foundation, the Ministry of Science and Education of Spain, 
the Science and Technology Facilities Council of the United Kingdom, the Higher Education Funding Council for England, the National Center for Supercomputing 
Applications at the University of Illinois at Urbana-Champaign, the Kavli Institute of Cosmological Physics at the University of Chicago, 
the Center for Cosmology and Astro-Particle Physics at the Ohio State University,
the Mitchell Institute for Fundamental Physics and Astronomy at Texas A\&M University, Financiadora de Estudos e Projetos, 
Funda{\c c}{\~a}o Carlos Chagas Filho de Amparo {\`a} Pesquisa do Estado do Rio de Janeiro, Conselho Nacional de Desenvolvimento Cient{\'i}fico e Tecnol{\'o}gico and 
the Minist{\'e}rio da Ci{\^e}ncia, Tecnologia e Inova{\c c}{\~a}o, the Deutsche Forschungsgemeinschaft and the Collaborating Institutions in the Dark Energy Survey. 

The Collaborating Institutions are Argonne National Laboratory, the University of California at Santa Cruz, the University of Cambridge, Centro de Investigaciones Energ{\'e}ticas, 
Medioambientales y Tecnol{\'o}gicas-Madrid, the University of Chicago, University College London, the DES-Brazil Consortium, the University of Edinburgh, 
the Eidgen{\"o}ssische Technische Hochschule (ETH) Z{\"u}rich, 
Fermi National Accelerator Laboratory, the University of Illinois at Urbana-Champaign, the Institut de Ci{\`e}ncies de l'Espai (IEEC/CSIC), 
the Institut de F{\'i}sica d'Altes Energies, Lawrence Berkeley National Laboratory, the Ludwig-Maximilians Universit{\"a}t M{\"u}nchen and the associated Excellence Cluster Universe, 
the University of Michigan, NSF NOIRLab, the University of Nottingham, The Ohio State University, the University of Pennsylvania, the University of Portsmouth, 
SLAC National Accelerator Laboratory, Stanford University, the University of Sussex, Texas A\&M University, and the OzDES Membership Consortium.

Based in part on observations at NSF Cerro Tololo Inter-American Observatory at NSF NOIRLab (NOIRLab Prop. ID 2012B-0001; PI: J. Frieman), which is managed by the Association of Universities for Research in Astronomy (AURA) under a cooperative agreement with the National Science Foundation.

The DES data management system is supported by the National Science Foundation under Grant Numbers AST-1138766 and AST-1536171.
The DES participants from Spanish institutions are partially supported by MICINN under grants PID2021-123012, PID2021-128989 PID2022-141079, SEV-2016-0588, CEX2020-001058-M and CEX2020-001007-S, some of which include ERDF funds from the European Union. IFAE is partially funded by the CERCA program of the Generalitat de Catalunya.

We acknowledge support from the Brazilian Instituto Nacional de Ci\^encia
e Tecnologia (INCT) do e-Universo (CNPq grant 465376/2014-2).

This manuscript has been authored by Fermi Research Alliance, LLC under Contract No. DE-AC02-07CH11359 with the U.S. Department of Energy, Office of Science, Office of High Energy Physics.

Canadian co-authors acknowledge support from the Natural Sciences and Engineering Research Council of Canada. Websky computations were performed on the SciNet supercomputer at the SciNet HPC Consortium. SciNet is funded by: the Canada Foundation for Innovation; the Government of Ontario; Ontario Research Fund - Research Excellence; and the University of Toronto.

IFAE is partially funded by the CERCA program of the Generalitat de Catalunya.

R. H. acknowledges support from CIFAR, the Azrieli and Alfred. P. Sloan foundations, and the Connaught Fund.

GAM is part of Fermi Research Alliance, LLC under Contract No. DE-AC02-07CH11359 with the U.S. Department of Energy, Office of Science, Office of High Energy Physics.

EC acknowledges support from the European Research Council (ERC) under the European Union’s Horizon 2020 research and innovation programme (Grant agreement No. 849169).

CHT received support from the United States Department of Energy, Office of High Energy Physics under Award Number DE-SC-0011726.

The Flatiron Institute is supported by the Simons Foundation.

CS acknowledges support from the Agencia Nacional de Investigaci\'on y Desarrollo (ANID) through Basal project FB210003.

This work received support from the U.S. Department of Energy under contract number DE-AC02-76SF00515 at SLAC National Accelerator Laboratory. This research used computing resources at SLAC National Accelerator Laboratory and at the National Energy Research Scientific Computing Center (NERSC), a U.S. Department of Energy Office of Science User Facility located at Lawrence Berkeley National Laboratory, operated under Contract No. DE-AC02-05CH11231.

%% file: affil_at_end.tex
\noindent$^1$Institut de F\'{\i}sica d'Altes Energies (IFAE), The Barcelona Institute of Science and Technology, Campus UAB, 08193 Bellaterra (Barcelona) Spain\\
$^2$David A. Dunlap Department of Astronomy \& Astrophysics, University of Toronto, 50 St. George St., Toronto, ON M5S 3H4, Canada\\
$^3$Canadian Institute for Theoretical Astrophysics, University of Toronto, 60 St. George St., Toronto, ON M5S 3H4, Canada\\
$^4$Dunlap Institute of Astronomy \& Astrophysics, University of Toronto, 50 St. George St., Toronto, ON M5S 3H4, Canada\\
$^5$School of Earth and Space Exploration, Arizona State University, Tempe, AZ 85287, USA\\
$^6$Laborat\'orio Interinstitucional de e-Astronomia - LIneA, Rua Gal. Jos\'e Cristino 77, Rio de Janeiro, RJ - 20921-400, Brazil\\
$^7$Fermi National Accelerator Laboratory, P. O. Box 500, Batavia, IL 60510, USA\\
$^8$Department of Astronomy and Astrophysics, University of Chicago, Chicago, IL 60637, USA\\
$^9$Kavli Institute for Cosmological Physics, University of Chicago, Chicago, IL 60637, USA\\
$^{10}$Institute of Cosmology and Gravitation, University of Portsmouth, Portsmouth, PO1 3FX, UK\\
$^{11}$Institute for Astronomy, University of Hawai‘i, 2680 Woodlawn Drive, Honolulu, HI 96822,
USA\\
$^{12}$Department of Physics, Northeastern University, Boston, MA 02115, USA\\
$^{13}$University Observatory, Faculty of Physics, Ludwig-Maximilians-Universit\"at, Scheinerstr. 1, 81679 Munich, Germany\\
$^{14}$Department of Physics \& Astronomy, University College London, Gower Street, London, WC1E 6BT, UK\\
$^{15}$School of Physics and Astronomy, Cardiff University, The Parade, Cardiff, Wales CF24 3AA, UK\\
$^{16}$Instituto de Astrofisica de Canarias, E-38205 La Laguna, Tenerife, Spain\\
$^{17}$Astronomy Unit, Department of Physics, University of Trieste, via Tiepolo 11, I-34131 Trieste, Italy\\
$^{18}$INAF-Osservatorio Astronomico di Trieste, via G. B. Tiepolo 11, I-34143 Trieste, Italy\\
$^{19}$Institute for Fundamental Physics of the Universe, Via Beirut 2, 34014 Trieste, Italy\\
$^{20}$Kavli Institute for Cosmology Cambridge, Madingley Road, Cambridge CB3 0HA, UK\\
$^{21}$DAMTP, Centre for Mathematical Sciences, University of Cambridge, Cambridge CB3 OWA, UK\\
$^{22}$Centro de Investigaciones Energ\'eticas, Medioambientales y Tecnol\'ogicas (CIEMAT), Madrid, Spain\\
$^{23}$Department of Physics, IIT Hyderabad, Kandi, Telangana 502285, India\\
$^{24}$Department of Physics and Astronomy, University of Pennsylvania, Philadelphia, PA 19104, USA\\
$^{25}$Universit\'e Grenoble Alpes, CNRS, LPSC-IN2P3, 38000 Grenoble, France\\
$^{26}$Center for Computational Astrophysics, Flatiron Institute, New York, NY 10010, USA\\
$^{27}$Joseph Henry Laboratories of Physics, Jadwin Hall, Princeton University, Princeton, NJ, USA 08544\\
$^{28}$Department of Astrophysical Sciences, Princeton University, Peyton Hall, Princeton, NJ 08544, USA\\
$^{29}$School of Physics and Astronomy, Sun Yat-sen University, 2 Daxue Road, Zhuhai, 519082, China\\
$^{30}$California Institute of Technology, 1200 East California Blvd, MC 249-17, Pasadena, CA 91125, USA\\
$^{31}$Institute of Theoretical Astrophysics, University of Oslo. P.O. Box 1029 Blindern, NO-0315 Oslo, Norway\\
$^{32}$Instituto de Fisica Teorica UAM/CSIC, Universidad Autonoma de Madrid, 28049 Madrid, Spain\\
$^{33}$Institut d'Estudis Espacials de Catalunya (IEEC), 08034 Barcelona, Spain\\
$^{34}$Institute of Space Sciences (ICE, CSIC),  Campus UAB, Carrer de Can Magrans, s/n,  08193 Barcelona, Spain\\
$^{35}$Department of Physics and Astronomy,  University of Southern California, Los Angeles, CA, 90007, USA\\
$^{36}$Center for Astrophysical Surveys, National Center for Supercomputing Applications, 1205 West Clark St., Urbana, IL 61801, USA\\
$^{37}$Department of Astronomy, University of Illinois at Urbana-Champaign, 1002 W. Green Street, Urbana, IL 61801, USA\\
$^{38}$School of Mathematics and Physics, University of Queensland,  Brisbane, QLD 4072, Australia\\
$^{39}$Santa Cruz Institute for Particle Physics, Santa Cruz, CA 95064, USA\\
$^{40}$Center for Cosmology and AstroParticle Physics (CCAPP), The Ohio State University, Columbus, OH 43210, USA\\
$^{41}$Department of Physics, The Ohio State University, Columbus, OH 43210, USA\\
$^{42}$Center for Astrophysics $\vert$ Harvard \& Smithsonian, 60 Garden Street, Cambridge, MA 02138, USA\\
$^{43}$Australian Astronomical Optics, Macquarie University, North Ryde, NSW 2113, Australia\\
$^{44}$Lowell Observatory, 1400 Mars Hill Rd, Flagstaff, AZ 86001, USA\\
$^{45}$Jet Propulsion Laboratory, California Institute of Technology, 4800 Oak Grove Dr., Pasadena, CA 91109, USA\\
$^{46}$Berkeley Center for Cosmological Physics, University of California, Berkeley, CA 94720, USA\\
$^{47}$Lawrence Berkeley National Laboratory,  1 Cyclotron Road, Berkeley, CA 94720, USA\\
$^{48}$George P. and Cynthia Woods Mitchell Institute for Fundamental Physics and Astronomy, and Department of Physics and Astronomy, Texas A\&M University, College Station, TX 77843,  USA\\
$^{49}$LPSC Grenoble - 53, Avenue des Martyrs 38026 Grenoble, France\\
$^{50}$Instituci\'o Catalana de Recerca i Estudis Avan\c{c}ats, E-08010 Barcelona, Spain\\
$^{51}$Department of Physics, Cornell University, Ithaca, NY 14853,USA\\
$^{52}$Department of Astronomy, Cornell University, Ithaca, NY 14853, USA\\
$^{53}$Department of Physics, Columbia University, 538 West 120th Street, New York, NY, USA 10027, USA\\
$^{54}$Hamburger Sternwarte, Universit\"{a}t Hamburg, Gojenbergsweg 112, 21029 Hamburg, Germany\\
$^{55}$Observat\'orio Nacional, Rua Gal. Jos\'e Cristino 77, Rio de Janeiro, RJ - 20921-400, Brazil\\
$^{56}$Kavli Institute for Particle Astrophysics \& Cosmology, P. O. Box 2450, Stanford University, Stanford, CA 94305, USA\\
$^{57}$SLAC National Accelerator Laboratory, Menlo Park, CA 94025, USA\\
$^{58}$Ruhr University Bochum, Faculty of Physics and Astronomy, Astronomical Institute, German Centre for Cosmological Lensing, 44780 Bochum, Germany\\
$^{59}$Instituto de F\'\i sica, UFRGS, Caixa Postal 15051, Porto Alegre, RS - 91501-970, Brazil\\
$^{60}$Department of Physics, University of Michigan, Ann Arbor, MI 48109, USA\\
$^{61}$Instituto de F\'isica, Pontificia Universidad Cat\'olica de Valpara\'iso, Casilla 4059, Valpara\'iso, Chile\\
$^{62}$Physics Department, Lancaster University, Lancaster, LA1 4YB, UK\\
$^{63}$Computer Science and Mathematics Division, Oak Ridge National Laboratory, Oak Ridge, TN 37831\\
$^{64}$Department of Astronomy, The Ohio State University, Columbus, OH 43210, USA\\
$^{65}$Department of Astronomy, University of California, Berkeley,  501 Campbell Hall, Berkeley, CA 94720, USA\\
$^{66}$School of Physics and Astronomy, University of Southampton,  Southampton, SO17 1BJ, UK\\
$^{67}$NASA Goddard Spaceflight Center, 8800 Greenbelt Rd, Greenbelt, MD 20771, USA\\